\PassOptionsToPackage{final}{graphicx}
\PassOptionsToPackage{nosumlimits,nonamelimits}{amsmath}
\PassOptionsToPackage{capitalize,noabbrev,nameinlink}{cleveref}
\PassOptionsToPackage{bookmarks,unicode,colorlinks=true,linkcolor=blue,citecolor=blue}{hyperref}%

\documentclass[a4paper,UKenglish,cleveref,thm-restate,numberwithinsect,final]{lipics-v2021}
 \usepackage{microtype}

\nolinenumbers

\usepackage{enumitem}
\usepackage{xspace}
\usepackage{xcolor}
\usepackage{soul}
\usepackage[utf8]{inputenc}
\usepackage{amsmath}
\usepackage{mathrsfs}
\usepackage{stmaryrd}
\usepackage{wasysym}
\usepackage{framed}
\usepackage{mathpartir}

\usepackage{etex,etoolbox}

\usepackage[marginclue,footnote,final]{fixme}
\FXRegisterAuthor{hs}{ahs}{HS}
\FXRegisterAuthor{ls}{als}{LS}
\FXRegisterAuthor{uc}{auc}{UC}

\newif\iffullversion
\fullversiontrue

\iffullversion\hideLIPIcs\fi

\newcommand{\Br}{\mathsf{bar}}
\newcommand{\Perm}{\mathsf{Perm}(\Names)}
\newcommand{\trc}{\mathsf{tr}}
\newcommand{\pre}{\mathsf{pre}}
\newcommand{\abs}{\mathsf{abs}}
\newcommand{\res}{\mathsf{res}}
\newcommand{\Alg}{\mathsf{Alg}}
\newcommand{\Term}{\mathsf{Term}}
\newcommand{\sig}{\Sigma}
\newcommand{\sigp}{\Sigma_0}
\newcommand{\sigf}{\Sigma_{\mathsf f}}
\newcommand{\sigb}{\Sigma_{\mathsf b}}
\newcommand{\centails}{\Rightarrow}
\newcommand{\game}{\mathsf{G}}

\newcommand{\loc}{\mathsf{loc}}
\newcommand{\drop}{\mathsf{drop}}

\newcommand{\fs}{\mathsf{fs}}
\newcommand{\ar}{\mathsf{ar}}
\newcommand{\Id}{\mathsf{Id}}
\newcommand{\FN}{\mathsf{FN}}

\newcommand{\Names}{\mathbb{A}}
\newcommand{\barNames}{\overline{\Names}}

\newcommand{\alphaeq}{\equiv_\alpha}
\newcommand{\powufs}{\pow_{\mathsf{ufs}}}
\newcommand{\pow}{\mathcal{P}}
\newcommand{\powfs}{\pow_{\fs}}
\newcommand{\powf}{\pow_{\omega}}

\newcommand{\mycomment}[1]{}
\newcommand{\fresh}{\mathbin{\#}}
\newcommand{\supp}{\mathsf{supp}}
\newcommand{\fix}{\mathop{\mathsf{fix}}}
\newcommand{\Fix}{\mathop{\mathsf{Fix}}}

\newcommand{\bind}[1]{\langle #1\rangle}

\newcommand{\mbar}{\overline{M}_1}

\newcommand{\resetCurThmBraces}{%
\gdef\curThmBraceOpen{(}%
\gdef\curThmBraceClose{)}}
\resetCurThmBraces
\newcommand{\removeThmBraces}{%
\gdef\curThmBraceOpen{}%
\gdef\curThmBraceClose{}}
\resetCurThmBraces

\usepackage{etoolbox}
\patchcmd{\thmhead}{(#3)}{\curThmBraceOpen #3\curThmBraceClose}{}{}

\theoremstyle{plain}

\newcommand{\midmid}{\hspace{0.2ex}{\rule[-0.1ex]{0.6pt}{1.65ex}}\hspace{0.2ex}}
\newcommand{\scriptmidmid}{\hspace{0.2ex}{\rule[-0.1ex]{0.6pt}{1.1ex}}\hspace{0.2ex}}

\newcommand{\newletter}[1]{{\midmid}#1}
\newcommand{\scriptnew}[1]{{\scriptmidmid}#1}

\newcommand{\id}{\mathit{id}}

\newcommand{\ub}{\mathsf{ub}}

\numberwithin{equation}{section}

\usepackage{algorithmicx}
    \usepackage[ruled]{algorithm}
\usepackage[noend]{algpseudocode}

\usepackage{tikz}
 \usetikzlibrary{trees}
 \usetikzlibrary{shapes}
 \usetikzlibrary{fit}
 \usetikzlibrary{shadows}
 \usetikzlibrary{backgrounds}
 \usetikzlibrary{arrows,automata}

\tikzset{
   n/.style= {circle,fill,inner sep=1.5pt,node distance=2cm}
  ,acc/.style={circle,draw,inner sep=3pt,node distance=2cm}
  ,phantom/.style={circle},
  ,arr/.style={->, >=stealth, semithick, shorten <= 3pt, shorten >= 3pt}
}

\usepackage{tikz-cd}

\newcommand{\sem}[1]{\llbracket #1 \rrbracket}

\makeatletter
\def\moverlay{\mathpalette\mov@rlay}
\def\mov@rlay#1#2{\leavevmode\vtop{%
   \baselineskip\z@skip \lineskiplimit-\maxdimen
   \ialign{\hfil$\m@th#1##$\hfil\cr#2\crcr}}}
\newcommand{\charfusion}[3][\mathord]{
    #1{\ifx#1\mathop\vphantom{#2}\fi
        \mathpalette\mov@rlay{#2\cr#3}
      }
    \ifx#1\mathop\expandafter\displaylimits\fi}
\makeatother

\newcommand{\Nat}{{\mathbb{N}}}
\newcommand{\monad}{{\mathbb{M}}}

\newcommand{\Set}{\mathbf{Set}}

\newcommand{\lrule}[3]{(\text{#1})\;\frac{#2}{#3}}

\newcommand{\Nom}{\mathbf{Nom}}

\newcommand{\BC}{\mathbf{C}}

\newcommand{\takeout}[1]{\empty}
\newcommand{\names}{\mathbb{A}}

\makeatletter
\newcommand{\mysubsec}[1]{%
  \par\medskip\noindent{\bfseries\sffamily #1}\hspace{3mm}%
  \@ifnextchar\par{\@gobble}{}% this eats a following \par if any
}
\makeatother

\theoremstyle{definition}
\newtheorem{defn}[theorem]{Definition}

\newcommand\PSpace{$\textsc{PSpace}$\xspace}

\newcommand{\Act}{\mathcal{A}}

\title{Graded Semantics of Nominal Systems}
\author{Hannes Schulze}{Friedrich-Alexander-Universit\"{a}t Erlangen-N\"{u}rnberg, Germany}{}{https://orcid.org/0009-0000-4713-4338}{}

\author{Lutz Schr\"{o}der}{Friedrich-Alexander-Universit\"{a}t Erlangen-N\"{u}rnberg, Germany}{}{https://orcid.org/0000-0002-3146-5906}{Funded by the Deutsche Forschungsgemeinschaft (DFG, German Research Foundation) -- project number 517924115.}%CoNaN

\author{\"{U}same Cengiz}{Friedrich-Alexander-Universit\"{a}t Erlangen-N\"{u}rnberg, Germany}{}{https://orcid.org/0009-0004-4092-7668}{Funded by the Deutsche Forschungsgemeinschaft (DFG, German Research Foundation) -- project number 419850228.}%CoMoC

\authorrunning{H. Schulze, L. Schr\"{o}der, and \"{U}. Cengiz}

\Copyright{Hannes Schulze, Lutz Schr\"{o}der, and \"{U}same Cengiz}

 %TODO mandatory: Please choose ACM 2012 classifications from https://dl.acm.org/ccs/ccs_flat.cfm

 %TODO mandatory; please add comma-separated list of keywords

\iffullversion
  \relatedversiondetails{Conference Version}{https://doi.org/10.4230/LIPIcs.CONCUR.2026.14}
\else
  \relatedversiondetails[cite={schulze2026gradedmonadssemanticsnominal}]{Full Version}{https://arxiv.org/abs/2510.24353}
\fi
\EventEditors{Ana Sokolova and Patrick Totzke}
\EventNoEds{2}
\EventLongTitle{37th International Conference on Concurrency Theory (CONCUR 2026)}
\EventShortTitle{CONCUR 2026}
\EventAcronym{CONCUR}
\EventYear{2026}
\EventDate{September 1--4, 2026}
\EventLocation{Liverpool, UK}
\EventLogo{}
\SeriesVolume{391}
\ArticleNo{14}

\begin{document}

\maketitle

\begin{abstract} Nominal automata models and transition systems serve
  as formalisms for languages and processes carrying data, and as such
  relate closely to classical register-based models. The paradigm of
  name allocation in nominal systems helps alleviate the pervasive
  computational hardness of register-based models in a tradeoff
  between expressiveness and computational tractability. For instance,
  regular nondeterministic nominal automata (RNNAs) correspond, under
  their \emph{local freshness} semantics, to a form of lossy register
  automata. Unlike the full register automaton model, RNNAs allow for
  inclusion checking in elementary complexity (parametrized
  $\PSpace$); similarly, trace inclusion in the underlying nominal
  transition systems is in parametrized $\PSpace$. In the present
  work, we develop a unified algebraic treatment of spectra of
  behavioural equivalences on nominal systems in the framework of
  graded monads, working in the setting of universal coalgebra. In
  particular, we extend the associated notion of graded algebraic
  theory to the nominal setting, and use this to give an algebraic
  axiomatization of the (linear-time) global and local freshness
  semantics of nominal systems with name allocation. As an
  illustration of the benefits of graded monads, we develop a nominal
  version of the generic game characterization of graded semantics,
  which we instantiate to obtain trace equivalence games for nominal
  transition systems under global and local freshness semantics.
\end{abstract}

\section{Introduction}\label{sec:intro}

In processes and formal languages, infinite alphabets serve as an
abstraction of data types that are either infinite or prohibitively
large. Typical examples include data values in XML
documents~\cite{NevenEA04}, object identities~\cite{GrigoreEA13},
channel names~\cite{Hennessy02}, process
identifiers~\cite{BolligEA14}, parameters of method
calls~\cite{HowarEA19}, or nonces in cryptographic
protocols~\cite{KurtzEA07}. Transition systems and automata models
processing infinite alphabets are conveniently described as systems
based on nominal sets~\cite{BojanczykEA14,ParrowEA21}, in which,
roughly speaking, the underlying set of names plays the role of the
infinite alphabet, and system states store a bounded number of names
in analogy to the standard register paradigm. For instance,
non-deterministic orbit-finite automata (NOFA)~\cite{BojanczykEA14}
are equivalent to register automata~\cite{KaminskiFrancez94} with
non-deterministic update~\cite{KaminskiZeitlin10}.

The nominal setting naturally supports explicit name-allocating
transitions, subject to $\alpha$-conversion of bound names; these are
featured, for instance, in \emph{nominal transition
  systems}~\cite{ParrowEA21} and in \emph{regular non-deterministic
  nominal automata (RNNA)}~\cite{SchroderEA17}, a nominal automata
model that is equivalent to a subclass of register automata
characterized by a lossiness condition called \emph{name
  dropping}. RNNA trade some of the expressiveness of register
automata for computational tractability; in particular, they allow
inclusion checking in elementary complexity (a problem that is
undecidable for the full class of register automata). Indeed,
name-allocating transitions support two distinct notions of freshness,
determined by the discipline imposed on renaming bound names: Under
\emph{global freshness}, as featured, for instance, in the
register-based model of \emph{session automata}~\cite{BolligEA14},
newly read names are required to be fresh w.r.t.~all previous
names. In the nominal setting, this is enforced by a discipline of
clean renaming (avoiding shadowing of bound names). On the other hand,
\emph{local freshness}, i.e.\ freshness w.r.t.\ (some of) the names
currently stored in the registers, is induced by unrestricted renaming
according to the usual rules of $\alpha$-equivalence familiar from
$\lambda$-calculus. These concepts have been extended to nominal
automata models processing trees~\cite{PruckerSchroder24} and infinite
words~\cite{UrbatEA21}, as well as to alternating nominal
automata~\cite{FrankEA25}.

Our main concern in the present work is to capture the various
semantics of nominal systems in the general categorical framework of
\emph{graded semantics}~\cite{MiliusEA15}, which provides a unified
treatment of spectra of semantics of automata models and concurrent
systems, such as the well-known linear-time / branching-time spectrum
of semantics of labelled transition systems~\cite{DorschEA19}. This is
achieved by combining the setting of universal
coalgebra~\cite{Rutten00} with additional algebraic laws introduced
via the choice of a \emph{graded monad}~\cite{Smirnov08}, a
categorical structure that abstractly represents a \emph{graded
  algebraic theory}. Here, the notion of \emph{grade} intuitively
corresponds to depth of look-ahead, measured in (exact) numbers of
transition steps. A typical example of a graded algebraic equation is
the equation $a(x+y)=a(x)+a(y)$, which occurs in the axiomatization of
the trace semantics of labelled transition systems~\cite{DorschEA19}
and says that taking an $a$-labelled step distributes over
nondeterministic choice~$+$. Here, one assigns depth~$1$ to the unary
operation $a(-)$ representing $a$-steps and depth~$0$ to the choice
operator~$+$, so that the overall depth of the equation is~$1$. The
notion of graded theory has originally been introduced for set-based
graded monads~\cite{Smirnov08}; it has been extended to graded monads
on the category of posets~\cite{FordEA21}, the category of metric
spaces~\cite{ForsterEA25a}, and more generally to graded monads on
categories of relational structures~\cite{FordThesis,ForsterEA25b},
while coverage of nominal systems is currently missing. Benefits of
casting a given semantics as a graded semantics include instantiation
of generic results obtained for graded semantics, such as logical
characterizations~\cite{MiliusEA15,DorschEA19,FordEA21,ForsterEA25a}
and game characterizations~\cite{FordEA22,ForsterEA25b}.

In this context, our main contributions in the present work are as
follows.\vspace{-0.5em}
\begin{itemize}[wide]
\item We introduce a notion of graded nominal algebra, based on a
  simplification of Gabbay and Mathijsen's framework of nominal
  algebra~\cite{GabbayMathijssen09}. We show that graded nominal
  theories induce graded monads on the category of nominal sets and,
  crucially, that graded nominal theories of depth~$1$ induce graded
  monads that are depth-$1$ in an abstract categorical
  sense~\cite{MiliusEA15,DorschEA19}, a key prerequisite for most
  technical results in graded semantics
  (e.g.~\cite{MiliusEA15,DorschEA19,FordEA22,ForsterEA25b}).
\item Besides more basic examples, we then present graded nominal
  theories capturing the global and local freshness semantics of name
  allocation in nominal transition systems and RNNA; in particular
  the case of local freshness turns out to be quite non-trivial even
  though it is characterized by a single simple additional axiom.
\item We transfer the game characterization of graded
  semantics~\cite{FordEA22} to nominal sets, obtaining as instances of
  this result Spoiler-Duplicator games for trace equivalence of
  nominal transition systems under both global and local freshness.
  Notably, establishing the technical conditions
  for correctness of the game in the case of local freshness requires a
  maybe unexpected twist to the setup of the corresponding graded
  nominal-algebraic theory, which as it turns out needs to include name
  restriction~\cite{Pitts13} as a depth-$0$ operation in order to
  unblock certain $\alpha$-renamings, in analogy to name-dropping
  constructions for RNNA~\cite{SchroderEA17}.
  % We conclude by transferring the notion of graded equivalence game,
  % which has previously been introduced for set-based graded
  % monads~\cite{FordEA22} and later extended to graded monads on
  % categories of relational structures~\cite{ForsterEA25b}, to the
  % nominal setting. Roughly speaking, graded games are
  % Spoiler-Duplicator games characterizing a given graded behavioural
  % equivalence as the winning region of Duplicator; they may be
  % viewed as playing out the existence of a graded algebraic proof of
  % the equivalence of given states. We instantiate the generic game
  % to games for trace equivalence of states in nominal transition
  % systems under both local and global freshness semantics.
\end{itemize}
\subparagraph*{Related work} We have already mentioned previous work
on graded
semantics~\cite{MiliusEA15,DorschEA19,FordEA21,FordEA22,ForsterEA25a,ForsterEA25b}
and nominal algebra~\cite{GabbayMathijssen09}; details on the
relationship with the latter are discussed in
\cref{rem:expressiveness}. We note that alternative systems of varying
expressiveness for universal algebra over nominal sets have been
proposed~\cite{CloustonPitts07,KurzPetrisan10}. Some of these
systems~\cite{KurzPetrisan10,KurzEA10} indeed use standard multisorted
universal algebra, which however needs to be restricted to
\emph{uniform} theories and terms, and then corresponds to a subclass
of (finitary) monads on nominal sets, the so-called
fb-monads~\cite{KurzEA10}. While many nominal-algebraic systems derive
freshness
judgements~\cite{Gabbay09,GabbayMathijssen09,CloustonPitts07},
sometimes in synchrony with equalities~\cite{CloustonPitts07}, our
system is entirely concerned with equalities, with freshness implicit
in the context of equalities. The global freshness semantics of RNNA
has been modelled coalgebraically using categorical distributive laws
(in what has been termed Eilenberg-Moore-type~\cite{AdamekEA12} and
Kleisli-type~\cite{HasuoEA07} coalgebraic trace semantics), which by
general results~\cite{MiliusEA15} entails a modelling as a graded
semantics, without however providing a syntactic algebraic
presentation~\cite{FrankEA22}. Categorical distributive laws have also
been used to obtain up-to techniques for proving language equivalence
of NOFAs~\cite{BonchiEA14}. % We mostly restrict the presentation to a
% variant of RNNA with all states final; these structures are
% particular \emph{nominal transition
%   systems}~\cite{ParrowEA21}. \emph{Fresh labelled transition
%   systems}~\cite{BandukaraTzevelekos25} combine global and local
% freshness in one system, in a similar way as in fresh register
% automata~\cite{Tzevelekos11}. Our use of name restriction in nominal
% theories of local freshness relates to the name dropping construction
% used in equivalence checking of RNNA~\cite{SchroderEA17}.

\section{Preliminaries}
\label{sec:prelims}
We assume basic familiarity with category theory
(e.g.~\cite{AdamekEA90}). We proceed to recall requisite background on
nominal sets and universal coalgebra.

\subparagraph*{Nominal Sets} serve the principled treatment of
\emph{names} and \emph{renaming} in syntax and
semantics~\cite{Pitts13}; a key insight behind the framework is that many
issues around renaming are best handled by permuting names. Fix a
countably infinite set~$\Names$ of \emph{names}, and let~$\Perm$ be
the group of finite permutations on~$\Names$\lsnote{Clashes with
  functor $G$}, which is generated by the permutations $(a\ b)$ that
swap $a,b\in\Names$. A \emph{$\Perm$-set} $(X,{\cdot})$, or just~$X$,
is a set~$X$ equipped with a left action $(\pi,x)\mapsto \pi\cdot x$
of~$\Perm$. Then,~$X$ decomposes disjointly into the \emph{orbits} of
its elements $x\in X$, i.e.~the sets $\{\pi\cdot x\mid\pi\in
\Perm\}$. If~$X$ has only finitely many orbits, then~$X$ is
\emph{orbit-finite}.

For $x\in X$, $A\subseteq X$,
we write
\begin{equation*}\textstyle
  \fix(x)=\{\pi\in \Perm\mid \pi\cdot x=x\}\quad\text{and}\quad
  \Fix(A)=\bigcap_{x\in A}\fix(x);
\end{equation*}
that is, $\fix(x)$ is the set of permutations fixing the element~$x$,
and~$\Fix(A)$ is the set of permutations fixing the set~$A$ pointwise.
Notice that~$\Names$ itself is a $\Perm$-set with the action given by
$\pi\cdot a=\pi(a)$. A set~$S\subseteq\Names$ is a \emph{support}
of~$x\in X$ if $\Fix(S)\subseteq\fix(x)$, i.e.~if every permutation
that fixes all names in~$S$ fixes also~$x$; in this case, we think
of~$S$ as giving an upper bound on the names occurring in~$x$. We say
that~$x$ is \emph{finitely supported} if~$x$ has some finite support.
A $\Perm$-set~$X$ is a \emph{nominal set} if every $x\in X$ is
finitely supported. One can then show that every $x\in X$ has a
\emph{least} finite support, denoted $\supp(x)$; we think
of~$\supp(x)$ as the set of names occurring in~$x$. A name
$a\in\Names$ is \emph{fresh} for $x\in X$ (in symbols, $a\fresh x$) if
$a\notin\supp(x)$. Every set~$X$ can be made into a \emph{discrete}
nominal set by equipping~$X$ with the trivial $\Perm$-action
$\pi\cdot x=x$. Then, $\supp(x)=\emptyset$ for all $x\in X$. For
every~$n\in\Nat$,~$\Names^n$ is a nominal set under the componentwise
action; then, $\supp(a_1,\dots,a_n)=\{a_1,\dots,a_n\}$.

A map $f\colon X\to Y$ between $\Perm$-sets $X,Y$ is
\emph{equivariant} if $f(\pi\cdot x)=\pi\cdot f(x)$ for all $x\in X$,
$\pi\in \Perm$. (This term extends to subsets $Y\subseteq X$ of a
nominal set~$X$ by viewing them as predicates; explicitly,~$Y$ is
equivariant if $y\in Y$ implies $\pi\cdot y\in Y$ for all
$\pi\in \Perm$.) Nominal sets and equivariant maps form a
category~$\Nom$.

For every $\Perm$-set~$X$, the finitely supported elements of~$X$ form
a nominal set. For instance, the powerset $\pow(X)$ of a nominal
set~$X$ is a $\Perm$-set under the action
$\pi\cdot A=\{\pi\cdot x\mid x\in A\}$. We obtain a nominal set
$\powfs(X)$, the \emph{finitely supported powerset}, consisting of the
finitely supported elements of~$\pow(X)$. Similarly, for nominal
sets~$X$,~$Y$, the set $X\to Y$ of maps from~$X$ to~$Y$ is a
$\Perm$-set under the action
$(\pi\cdot f)(x)=\pi\cdot f(\pi^{-1}\cdot x))$. The set $X\to_\fs Y$
of finitely supported elements of $X\to Y$ forms a nominal set, the
set of \emph{finitely supported maps}~$X\to Y$. A set $A\in\powfs(X)$
is \emph{uniformly finitely supported} if $\supp(x)\subseteq\supp(A)$
for all $x\in A$; the \emph{uniformly finitely supported powerset}
$\powufs(X)$ consists of the uniformly finitely supported subsets
of~$X$. If~$X$ is orbit-finite, then the uniformly finitely supported
subsets of~$X$ are precisely the finite subsets. (Contrastingly, for
instance, a subset of~$\names$ is finitely supported iff it is either
finite or cofinite.)

Given a nominal set~$X$, the \emph{abstraction set} $[\Names]X$ is the
quotient of $\Names\times X$ modulo the equivalence relation~$\sim$
defined by
\begin{equation*}
  (a,x)\sim(b,y)\iff (c\ a)\cdot x=(c\ b)\cdot y\text{ for some fresh~$c\in\Names$}.
\end{equation*}
We write $\bind{a}x$ for the equivalence class of $(a,x)$
under~$\sim$. We think of $\bind{a}x$ as arising from~$x$ by
abstracting the name~$a$ in analogy to $\lambda$-abstraction in
$\lambda$-calculus or name abstraction in the $\pi$-calculus, i.e.~by
allowing~$a$ to be $\alpha$-renamed. Indeed we have
$\supp(\bind{a}x)=\supp(x)\setminus\{a\}$.

\subparagraph*{Universal Coalgebra} Recall that a \emph{functor}
$F\colon \BC\to\BC$ on a category~$\BC$ assigns to each
$\BC$-object~$C$ a $\BC$-object $FC$ and additionally acts on
$\BC$-morphisms, compatibly with domains and codomains of morphisms as
well as identities and composition. It is useful to think of~$FC$ as
consisting of structured objects made up of inhabitants of~$C$. We
will primarily be interested in the case $\BC=\Nom$. Indeed, all
constructions of nominal sets recalled above extend to functors. For
instance, the construction of the abstraction set $[\Names]X$ extends
to an \emph{abstraction functor} $[\Names](-)$ taking an equivariant
map $f\colon X\to Y$ to the equivariant map
$[\Names]f \colon [\Names]X\to [\Names]Y$ given by
$[\Names]f(\bind{a}x)=\bind{a}f(x)$.

Universal coalgebra~\cite{Rutten00} provides a unified framework for
automata models and concurrent systems by abstracting the type of a
system as a functor~$F\colon\BC\to\BC$ on a base category~$\BC$. An
\emph{$F$-coalgebra} $(C,\gamma)$ then consists of a $\BC$-object~$C$,
thought of as an object of \emph{states}, and a $\BC$-morphism
$\gamma\colon C\to FC$ embodying the \emph{transition} structure of
the system. As a first example, we have a functor~$G$ on $\Nom$
defined on objects by $GX=2\times\powfs(\Names\times X)$ where
$2=\{\top,\bot\}$ is a two-element discrete nominal set. A
$G$-coalgebra $\gamma\colon C\to GC$ then assigns to each state
$c\in C$ a binary value, read as determining finality, and a set of
label/successor pairs; it is thus equivalently described as a triple
$(C,F,\to)$ where $F\subseteq C$ is an equivariant set of final states
and ${\to}\subseteq C\times\Names\times C$ is an equivariant
$\Names$-labelled transition relation. Thus, orbit-finite
$G$-coalgebras are precisely non-deterministic orbit-finite automata
(NOFAs) over the alphabet~$\Names$~\cite{BojanczykEA14}.

A $\BC$-morphism $f\colon C\to D$ between $F$-coalgebras $(C,\gamma)$
and $(D,\delta)$ is an \emph{$F$-coalgebra morphism} if
$Ff\cdot\gamma=\delta\cdot f$. Such $F$-coalgebra morphisms are
thought of as behaviour-preserving maps. Correspondingly, universal
coalgebra yields a canonical notion of behavioural equivalence: States
$c\in C$, $d\in D$ in $F$-coalgebras $(C,\gamma)$, $(D,\delta)$ are
\emph{behaviourally equivalent} if there exists an $F$-coalgebra
$(E,\epsilon)$ and $F$-coalgebra morphisms $f\colon C\to E$,
$g\colon D\to E$ such that $f(c)=g(d)$. This notion typically captures
branching-time equivalences, such as bisimilarity on labelled
transition systems. However, one is often interested in more
coarse-grained notions of equivalence, such as trace equivalence on
labelled transition systems or language equivalence of automata. There
have been quite a number of approaches to modelling such equivalences
in universal coalgebra (e.g.~\cite{HasuoEA07,AdamekEA12,MiliusEA15});
the specific framework of graded semantics is recalled in detail in
\cref{sec:graded-monads}.

\section{Nominal Systems  with Name Allocation}
\label{sec:rnna}
\noindent We proceed to recall regular non-deterministic nominal
automata (RNNA)~\cite{SchroderEA17} and the associated subclass of
nominal transition systems~\cite{ParrowEA21}, discussing their global
and local freshness semantics. The main driving principle behind RNNA
is to let automata natively accept words with bound names indicating
that a fresh name is read from the input string, so-called bar
strings, and then generate data words from bar strings by taking all
instantiations of the bound letters. Here, instantiation of bound
names proceeds in two steps: First, form all $\alpha$-equivalent
variants of the bar string, and then forget that bound names are
bound. Bound names are indicated by a preceding `$\newletter$'; that
is, $\newletter a$ indicates that a fresh letter~$a$ is read from the
input. Thus, a \emph{bar string} is just a word over the extended
alphabet $\barNames=\Names\cup\{\newletter a\mid a\in\Names\}$.
Formally, an occurrence of $\newletter a$ in a bar string binds the
name~$a$, with the scope of the binding extending to the end of the
string. Bound names in this sense can be renamed according to the
usual discipline of $\alpha$-equivalence known from $\lambda$-calculus
or $\pi$-calculus; explicitly:
\begin{defn}[$\alpha$-Equivalence of bar strings]
  The relation of \emph{$\alpha$-equivalence} on bar strings is the
  equivalence~\(\alphaeq\) generated by
  \begin{equation}\label{eq:bar-alpha}
    w\newletter{a}v \alphaeq w\newletter{b}u \quad \text{if } \bind av = \bind bu
    \text{ in } [\Names]\,\barNames^*.
  \end{equation}
  The $\alpha$-equivalence class of a bar string \(w\in \barNames^*\)
  is denoted \([w]_{\alpha}\). The set $\FN(w)$ of \emph{free names}
  in~$w$ is given recursively by $\FN(\epsilon)=\emptyset$ (where as
  usual~$\epsilon$ denotes the empty string),
  $\FN(aw)=\{a\}\cup\FN(w)$, and
  $\FN(\newletter aw)=\FN(w)\setminus\{a\}$. A bar string~$w$ is
  \emph{closed} if $\FN(w)=\emptyset$, and \emph{clean} if all bound
  names~$\newletter a$ in~$w$ are mutually distinct and distinct from
  all free names in~$w$.
\end{defn}
\noindent For instance,
$\newletter a\newletter b\alphaeq\newletter a\newletter c\alphaeq
\newletter a\newletter a$, while
$\newletter a\newletter b a\alphaeq\newletter a\newletter
ca\not\alphaeq\newletter a\newletter a a$. Crucially, the renaming
of~$\newletter c$ into $\newletter a$ in $\newletter a\newletter c a$
is \emph{blocked} by the free occurrence of~$a$ in $\newletter ca$;
formally, this is due to the (straightforward) observation that
$\supp([w]_\alpha)=\FN(w)$.

We will be interested in three types of languages: \emph{Data
  languages} are subsets of~$\Names^*$, whose elements we refer to as
\emph{data words}; \emph{literal languages} are subsets
of~$\barNames^*$; and \emph{bar languages} are subsets
of~$\barNames^*/{\alphaeq}$. We convert bar languages into data
languages in two alternative ways respectively corresponding to
\emph{global} and \emph{local freshness} semantics of name binding. We
define the auxiliary operation
$\ub\colon\barNames^*\to\Names^*$ recursively by
$\ub(\epsilon)=\epsilon$, $\ub(aw)=a\ub(w)$, and
$\ub(\newletter aw)=a\ub(w)$; that is, $\ub$ removes all bars
`$\newletter$' from a bar string. Then given a bar language~$L$, we
define data languages~$N(L)$,~$D(L)$ by
\begin{equation*}
  N(L)=\{\ub(w)\mid [w]_\alpha\in L, w\text{ clean}\} \quad\text{and}\quad D(L)=\{\ub(w)\mid[w]_\alpha\in L\}.
\end{equation*}
We briefly write $D(w)=D(\{w\})$, $N(w)=N(\{w\})$. For instance,
$N(\newletter a\newletter b)=\{ab\mid a,b\in\Names, a\neq b\}$,
$D(\newletter a\newletter b)=\{ab\mid a,b\in\Names\}$, and
$D(\newletter a\newletter ba)=\{aba\mid a,b\in\Names,a\neq
b\}$. Roughly speaking, global freshness semantics, embodied by
the~$N$ operator, enforces instantiations of bound letters to be
distinct from all letters seen previously in the input word, while
local freshness, embodied by the~$D$ operator, is more permissive,
enforcing distinctness only from previously seen letters that are
expected to be seen again (in an automaton model, this implies in
particular that the previous letter must still be in memory). In fact
the~$N$ operator is injective on languages of closed bar strings, so
global freshness semantics is essentially equivalent to a semantics
given directly in terms of bar languages.

The automata model of RNNA is designed to accept bar languages,
subsequently equipped with global or local freshness semantics as
indicated above.

\begin{defn}
  A \emph{regular nondeterministic nominal automaton (RNNA)} is a
  tuple \(A=(Q, \to, s, F)\) consisting of an orbit-finite set \(Q\)
  of \emph{states}; an equivariant subset
  \({\to} \subseteq Q \times \barNames \times Q\), the
  \emph{transition relation}; an \emph{initial state} \(s \in Q\); and
  an equivariant subset \(F \subseteq Q\) of \emph{final states} such
  that the following conditions are satisfied:
    \begin{itemize}
    \item The relation \(\to\) is \emph{$\alpha$-invariant}: If
      \(s \xrightarrow{\scriptnew a} q\) and
      \(\bind aq = \bind{a'}q'\), then
      \(s \xrightarrow{\scriptnew a'} q'\).
    \item The relation \(\to\) is \emph{finitely branching up to
        $\alpha$-equivalence}: For every state \(q \in Q\), the sets
      \(\{(a, q') \mid q \xrightarrow{a} q'\}\) and
      \(\{\bind aq' \mid q \xrightarrow{\scriptnew a} q'\}\) are
      finite.
    \end{itemize}
    The relation \(\to\) is extended to bar strings in the usual
    manner; a bar string~$w$ is \emph{accepted} by a state $q\in Q$ if
    there exists a state $q'\in F$ such that $q\xrightarrow{w}
    q'$. The \emph{literal language} $L_0(A,q)$ accepted by~$q$ is the
    set of bar strings accepted by~$q$. Correspondingly, the \emph{bar
      language} $L(A,q)$ accepted by~$q$ is the
    set~$L_0(A,q)/{\alphaeq}$ of equivalence classes of bar strings
    accepted by~$q$. The \emph{data languages} accepted by~$q$ under
    \emph{global} and \emph{local freshness semantics} are the sets
    $N(L(A,q))$ and $D(L(A,q))$, respectively. When mention of~$q$ is
    omitted, we assume $q=s$; e.g.\ $L_0(A)=L_0(A,s)$ is the literal
    language accepted by~$A$, etc. A \emph{regular nominal transition
      system (RNTS)} is an RNNA $A$ as above with $F=Q$, i.e.~with all
    states final; we then refer to elements of $L_0(A,q)$, $L(A,q)$,
    $N(L(A,q))$, or $D(L(A,q))$ as \emph{literal traces}, \emph{bar
      traces}, or \emph{data traces} (under global/local freshness)
    of~$q$, respectively, and similarly to languages as \emph{trace
      sets}. States $q$, $q'$ in regular nominal
    transitions~$A$,~$A'$, respectively, are \emph{trace equivalent
      under global (local) freshness semantics} if
    $N(L(A,q))=N(L(A',q'))$ (resp.~$D(L(A,q))=D(L(A',q'))$).
\end{defn}
\noindent For readability, we will restrict the technical development
to RNTS (see however \cref{rem:automata}), which can be seen as a
slight simplification of nominal transition systems in the sense of
Parrow et al.~\cite{ParrowEA21} given by omitting propositional atoms
(\emph{state predicates}), restricting the number of bound names per
transition to at most one, and imposing finite
branching. % Languages accepted by
% states in regular nominal transition systems are thus largely
% analogous to trace sets of states in standard labelled transition
% systems, and indeed regular nominal transition systems may be seen
% as a subclass of nominal transition systems~\cite{ParrowEA21}.
RNTS can be viewed as orbit-finite coalgebras (cf.~\cref{sec:prelims})
for the functor \(H\colon \Nom \to \Nom\) given on objects by
\begin{equation}\label{eq:rnna-functor}
  HX = \powufs(\Names \times X) \times \powufs([\Names]X).
\end{equation}
The induced coalgebraic notion of behavioural equivalence can be
described concretely as a form of bisimilarity in a fairly
straightforward manner, and as such in particular does not coincide
with trace equivalence under either local or global freshness
semantics. Instead, we discuss in
\cref{sec:graded-monads,sec:local-freshness} how these semantics are
captured using graded semantics.

An important property of RNTS is that names are
obtained only by reading them from the input word; formally, this is
reflected in the properties that
\begin{equation*}
  q\xrightarrow{a}q' \implies \{a\} \cup\supp(q')\subseteq\supp(q)
  \quad\text{and}\quad
  q\xrightarrow{\scriptnew a}q' \implies \supp(q')\subseteq\supp(q)\cup\{a\}.
\end{equation*}
The following diagram depicts an example RNTS~$A$ by picking
representatives of orbits:
\begin{equation*}
  \begin{tikzcd}[row sep=-3.0mm, column sep=6mm,baseline=-1mm]
    & s()\arrow[dr,"\scriptnew a"]\\
    v(b) \arrow[ur,"b"]& & t(a)\arrow[dl,"\scriptnew b"]\\
    & u(a,b)\arrow[ul,"a"]
  \end{tikzcd}
\end{equation*}
The bar trace set $L(A,s())$ of $s()$ consists of the
$\alpha$-equivalence classes of all bar strings of the form
$(\newletter a\newletter bab)^nw$ where
$w\in\{\epsilon,\newletter a,\newletter a \newletter b,\newletter a
\newletter ba\}$. Correspondingly, the data trace set $D(L(A,s()))$
under local freshness consists of the data traces of the form
$a_1b_1a_1b_1\dots a_nb_na_nb_nw$ where
$a_1,\dots,a_n,b_1,\dots,b_n\in\Names$, $a_i\neq b_i$ for
$i=1,\dots,n$, and $w\in\Names^*$ is a data trace of length at
most~$3$ in which the third letter, if any, equals the first and
differs from the second.

\section{Graded Monads and Graded Theories}\label{sec:graded-monads}

\noindent We next briefly summarize the framework of \emph{graded
  semantics}~\cite{MiliusEA15,DorschEA19}, which serves to unify
system semantics of varying granularity, such as found, for example,
on the well-known linear-time / branching-time
spectrum~\cite{Glabbeek90}. The central notion of the framework is the
following.

\begin{defn}\label{def:graded-monad}
  A \emph{graded monad}~\cite{Smirnov08} on a category \(\BC\) is a
  tuple $\monad=((M_n)_{n \in \Nat}, \eta,$
  $(\mu^{nk})_{n, k \in \Nat})$ consisting of
    \begin{itemize}
    \item a family of endofunctors \(M_n\colon \BC \to \BC\), indexed over
      \(n \in \Nat\);
    \item a natural transformation \(\eta: \Id \to M_0\), the
      \emph{unit}; and
    \item a family of natural transformations
      \(\mu^{nk}: M_n M_k \to M_{n + k}\), indexed over
      \(n, k \in \Nat\), the \emph{multiplication}
    \end{itemize}
    satisfying graded variants of the usual \emph{unit} and
    \emph{associative} laws
    ($\mu^{0,n} \cdot \eta M_n = \id_{M_n} = \mu^{n,0} \cdot M_n
    \eta$,
    $\mu^{n,k+m} \cdot M_n \mu^{k,m} = \mu^{n+k,m} \cdot \mu^{n,k}
    M_m$). The \emph{graded Kleisli star} operation $(-)_k^*$
    of~$\monad$, indexed over $k\ge 0$, lifts a morphism
    $f\colon X\to M_nY$ to the morphism
    $f_k^*=\mu^{kn}\cdot M_kf\colon M_kX\to M_{k+n}Y$.
  \end{defn}
  We recall how finitary graded monads on the category $\Set$ of sets
  and maps can be generated from algebraic
  presentations~\cite{Smirnov08,MiliusEA15}; one of our present
  contributions is to generalize this principle to graded monads on
  the category $\Nom$ of nominal sets (\cref{sec:graded-algebra}). A
  \emph{graded signature}~$\Sigma$ is a collection of function
  symbols~$f$ with assigned \emph{arity}~$\ar(f)$ and
  \emph{depth}~$d(f)$. This induces a notion of term of uniform depth:
  Variables are terms of uniform depth~\(0\), and if \(d(f) = k\) and
  $\ar(f)=n$, then \(f(t_1, \ldots, t_n)\) is a term of uniform depth
  \(m + k\) if all~\(t_i\) are terms of uniform depth~\(m\). (Hence,
  if~$\ar(c)=0$ and $d(c)=k$, then~$c$ is a term of uniform depth~$m$
  for every $m\ge k$.) A \emph{graded theory} $(\Sigma,E)$ then
  consists of a graded signature~$\Sigma$ and a set~\(E\) of
  equational axioms of \emph{uniform depth}, i.e.\ for every equation
  $s=t$ in~$E$, there is some~$k$ such that~$s$ and~$t$ have uniform
  depth~$k$. Such a graded theory induces a graded monad
  \(((M_n), \eta, (\mu^{n, k}))\) on $\Set$ where \(M_n X\) consists
  of terms over \(X\) of uniform depth~\(n\) modulo derivable
  equality,~\(\eta\) converts variables into terms, and \(\mu^{n, k}\)
  collapses layered terms~\cite{MiliusEA15}.

  Generally, we think of~$M_nX$ as a set of $n$-step behaviours ending
  in poststates in~$X$, and of~$M_n1$ (where~$1$ is a terminal object)
  as a set of $n$-step behaviours. Correspondingly, one has the
  following notion of \emph{graded semantics}:
  \begin{defn}\label{defn:graded-semantics}
    A \emph{graded semantics} $(\monad,\beta)$ for a functor~$G$ on a
    category~$\BC$ with a terminal object~$1$ consists of a graded
    monad \(\monad=((M_n), \eta, (\mu^{nk}))\) and a natural
    transformation \(\beta: G \to M_1\).  Given a \(G\)-coalgebra
    \(\gamma\colon X \to GX\), the \emph{behaviour maps}
    $(\gamma^{(n)}\colon X\to M_n1)_{n<\omega}$ are given recursively as the composites    \lsnote{! Later uses of $\gamma^{(n)}$ refer to pretraces, adapt}

    \begin{align*}
        \gamma^{(0)} & = (X \xrightarrow{\eta_X} M_0X \xrightarrow{M_0!} M_01), \\
        \gamma^{(n + 1)}&  = (X \xrightarrow{\beta_X \cdot \gamma} M_1X \xrightarrow{M_1\gamma^{(n)}} M_1M_n1 \xrightarrow{\mu^{1n}_X} M_{n+1}1).
    \end{align*}
    States $x\in X$, $y\in Y$ in $G$-coalgebras
    \(\gamma\colon X \to GX\) and $\delta\colon Y\to GY$ are
    \emph{($\beta$)-behaviourally equivalent} if
    \(\gamma^{(n)}(x) = \delta^{(n)}(y)\) for all \(n \in \Nat\).
\end{defn}

\begin{example}[Trace semantics of LTS~\cite{DorschEA19}]
  \label{expl:traces}
  We can view finitely branching $\Act$-labelled transition systems as
  coalgebras for the functor \(G\colon \Set \to \Set\) given by
  $GX = \powf(\Act \times X)$ where $\powf$ denotes finite
  powerset. We have a graded monad~$\monad$ with functors~$M_n$ given
  on objects by \(M_nX = \powf(\Act^n \times X)\) and with evident
  unit and multiplications, and obtain a graded semantics
  $(\monad,\beta)$ for~$G$ where \(\beta\) is just the identity on
  $G=M_1$. This graded semantics captures exactly the trace semantics
  of LTS; in particular, two states are $\beta$-behaviourally
  equivalent iff they are trace equivalent in the usual sense. The
  graded monad~$\monad$ is induced by the graded algebraic theory that
  has a constant~$0$ and a binary function~$+$ of depth~$0$, and for
  every $a\in\Act$ a unary function $a(-)$ of depth~$1$; the equations
  are the usual join-semilattice axioms for~$0$ and~$+$, and depth-$1$
  equations $a(0)=0$ and $a(x+y)=a(x)+a(y)$.
    % Let \(\gamma: X \to G(X)\) describe the the LTS shown in \cref{fig:lts}.
    % \begin{figure}[h!]
    %     \centering
    %     \begin{tikzpicture}[node distance=1cm,initial text={}]
    %         \node[state]                         (s0) {$s_0$};
    %         \node[state,right=of s0]             (s1) {$s_1$};
    %         \node[state,right=of s1]             (s2) {$s_2$};
    %         \node[state,right=of s1,below=of s2] (s3) {$s_3$};
    %         \path[->] (s0) edge node [above] {$a$} (s1)
    %                   (s1) edge node [above] {$b$} (s2)
    %                   (s1) edge node [above] {$c$} (s3);
    %     \end{tikzpicture}
    %     \caption{A depiction of an LTS.}
    %     \label{fig:lts}
    % \end{figure}
    % At depth \(2\), we have
    % \begin{align*}
    %     \gamma^{(2)}(s_0) &= \{ (ab, s_2), (ac, s_3) \}, \\
    %     (M_n ! \cdot \gamma^{(2)})(s_0) &\cong \{ ab, ac \},
    % \end{align*}
    % matching exactly the traces of length \(2\).
\end{example}
\begin{remark}\label{rem:automata}
  The coalgebraic modelling of automata models differs from that of
  transition systems in having an additional output or acceptance
  component. % ; for instance, finitely branching non-deterministic
  % automata are coalgebras for the functor
  % $2\times\powf(\Act\times(-))$
  Graded monads for the language semantics of automata retain a trace
  component capturing words that have a run; the language semantics is
  then obtained by applying a canonical projection $M_n1\to M_n0$ to
  the graded semantics~\cite{KurzEA15,MiliusEA15}. In the algebraic
  view, accepted words are captured by a depth-1 constant for
  acceptance, and the canonical projection substitutes the single
  variable in the set~$1$ by the bottom
  element. % For instance, the graded monad
  % for the language semantics of finitely branching non-deterministic
  % automata has $M_nX=\powf(\Act^{<n}+\Act^n\times X)$ where
  % $\Act^{<n}$ is the set of words over~$\Act$ of length $<n$
  % (capturing accepted words) and $\Act^n\times X$ captures ongoing
  % runs.
  In the present discussion, we elide this complication purely in the
  interest of readability.
\end{remark}

\noindent One has the following graded analogues of monad algebras:
\begin{defn}[Graded algebras]
  Given \(n \in \Nat\) and a graded monad
  \(\monad=((M_n), \eta, (\mu^{n, k}))\) on \(\BC\), an
  \emph{\(M_n\)-algebra} \(A\) consists of
    \begin{itemize}
        \item a family \((A_k)_{0 \le k \le n}\) of $\BC$-objects \(A_k\) called \emph{carriers}, and
        \item a family \((a^{m, k})_{0 \le m + k \le n}\) of morphisms \(a^{m, k}: M_m A_k \to A_{m + k}\)
    \end{itemize}
    satisfying evident graded variants of the usual laws
    % \begin{enumerate}
    %     \item for \(m \le n\),
    (\(a^{0, m} \cdot \eta_{A_m} = \id_{A_m}\),
    % \item for \(m + r + k \le n\),
    \(a^{m, r + k} \cdot M_m a^{r, k} = a^{m + r, k} \cdot \mu^{m,
      r}_{A_k}\)).
%    \end{enumerate}
    A \emph{homomorphism} $A\to B$ of \(M_n\)-algebras \(A, B\) is a
    family \((f_k)_{0 \le k \le n}\) of morphisms
    \(f_k\colon A_k \to B_k\) such that, for \(m + k \le n\),
    \(b^{m, k} \cdot M_m f_k = f_{m + k} \cdot a^{m, k}\).
\end{defn}
\noindent Most results on graded semantics, such as logical and
game-based characterizations of behavioural
equivalence~\cite{MiliusEA15,DorschEA19,FordEA22,ForsterEA25a,ForsterEA25b},
hinge on the monad being \emph{depth-$1$} in the following sense.
\begin{defn}\label{def:depth-1}
  A graded monad \(\monad=((M_n), \eta, (\mu^{nk}))\) is
  \emph{depth-1} if the following diagram (which commutes by the
  associative law of~$\monad$) is a coequalizer in the category of
  \(M_0\)-algebras for all \(X\) and \(n \in \Nat\):
  \begin{equation}
    \label{eq:depth-1}
    \begin{tikzcd}
      M_1 M_0 M_n X    \arrow[r,swap, "\mu^{1,0}_{M_n X }", shift right] \arrow[r, "M_1 \mu^{0,n}_X ", shift left] & M_1 M_n X   \arrow[r, "\mu^{1,n}_X"] & M_{1+n} X .
    \end{tikzcd}
  \end{equation}
\end{defn}
\noindent
On \(\Set\), a graded monad is depth-1 iff it is generated by a
\emph{depth-$1$ graded theory}, i.e.\ one whose operations and axioms
all have depth at most~1 \cite[Proposition~7.3]{MiliusEA15} (where the
`only if' direction depends on size considerations, i.e.~on either
allowing infinitary operations or restricting to finitary graded
monads). For instance, the presentation of the graded monad for trace
semantics (\cref{expl:traces}) implies that the graded monad is
depth-$1$. Similar results (in particular for the `if' direction) have
been obtained for graded monads on posets~\cite{FordEA21}, relational
structures~\cite{FordThesis}, and metric spaces~\cite{ForsterEA25a},
and indeed we prove the `if' direction for our notion of graded
nominal theory, introduced next (\Cref{thm:depth1}). In the
universal-algebraic view, the coequalizer condition in
\cref{def:depth-1} says roughly that all identifications caused by
collapsing depth-$1$ terms over depth-$n$ terms into depth-$(n+1)$
terms come from shifting depth-$0$ operations between the depth-$1$
layer and the depth-$n$ layer, in combination with the depth-$0$
axioms.

\section{Graded Nominal Algebra}\label{sec:graded-algebra}

\noindent We now present our framework of \emph{graded nominal
  algebra}, a universal-algebraic system in which theories define
graded monads on $\Nom$. The system builds on \emph{nominal
  algebra}~\cite{Gabbay09,GabbayMathijssen09}, from which it differs
most prominently by adding the grading.  Further differences are
discussed in \cref{rem:expressiveness}.

%We use operations that carry free or bound names:
\begin{defn}
  A \emph{nominal graded signature}~\(\Sigma\) is a disjoint union of
  sets $\sigp$,~$\sigf$, and~$\sigb$ of \emph{pure}, \emph{free}, and
  \emph{bound} operations, respectively. Each operation~$f$ is
  equipped with an \emph{arity} \(\ar(f) \in \Nat\) and a \emph{depth}
  \(d(f) \in \Nat\).  We write \(f/n \in \sigp\) if \(f \in \sigp\)
  and \(\ar(f) = n\), similarly for~\(\sigf\), \(\sigb\),
  and~$\Sigma$.
\end{defn}

\begin{defn}[Nominal Terms]
  Given a nominal set \(X\) of \emph{variables} and a graded
  signature~$\Sigma$, we define \emph{nominal $\Sigma$-terms $t,\dots$
    (over \(X\))} by the grammar
    \[
      t ::= x \mid f(t_1, \ldots, t_p) \mid a.g(t_1, \ldots, t_p) \mid
      \nu a.\,h(t_1, \ldots, t_p)
    \]
    where \(x \in X\), \(a \in \Names\), \(f/p \in \sigp\),
    \(g/p \in \sigf\), and \(h/p \in \sigb\). For brevity, we
    occasionally unify the three last clauses notationally as
    $\eta.f(t_1,\dots,t_p)$ for $f\in\Sigma$, where $\eta$ is either
    nothing or of the form~$a$ or~$\nu a$, with use of the notation
    implying the assumption of well-formedness.
\end{defn}
\noindent Note in particular that we work with given nominal structures
on sets of variables; cf.~\cref{rem:expressiveness}. In terms
$a.g(t_1,\dots,t_p)$,~$a$ is a \emph{free} name, while in terms
$\nu a.\,h(t_1, \ldots, t_p)$, the name~$a$ is \emph{bound}. This
reading gives rise to an evident notion of \emph{free names} in a
term, where we count both explicit occurrences of names and
occurrences of names in the support of variables; we write $\FN(t)$
for the set of free names in~$t$. Explicitly, $\FN(x)=\supp(x)$,
$\FN(f(t_1,\dots,t_p))=\FN(t_1)\cup\dots\cup\FN(t_p)$,
$\FN(a.g(t_1,\dots,t_p))=\FN(t_1)\cup\dots\cup\FN(t_p)\cup\{a\}$, and
$\FN(\nu
a.\,h(t_1,\dots,t_p))=(\FN(t_1)\cup\dots\cup\FN(t_p))\setminus
\{a\}$.

We have a notion of \emph{uniform depth} of terms defined completely
analogously to the set-based case (\cref{sec:graded-monads}). Then,
a substitution~$\sigma$ (a map assigning nominal terms~$\sigma(x)$ to
variables~$x\in X$) has \emph{uniform depth} if all $\sigma(x)$ are of
the same uniform depth. As usual, we write $t\sigma$ for the
application of~$\sigma$ to a nominal term~$t$, which just replaces all
occurrences of variables~$x$ in~$t$ with $\sigma(x)$. Like in nominal
algebra~\cite{GabbayMathijssen07,Gabbay09,GabbayMathijssen09},
application of substitutions thus explicitly does \emph{not} avoid
capture of names; these issues are instead dealt with at the level of
the derivation system. We write $\Term_{\Sigma,m}(X)$ for the set of
nominal terms of uniform depth~$m$ over~$X$; equipped with the evident
action of~$\Perm$ (in particular,
$\pi\cdot\nu a.\,f(t_1,\dots,t_n)=\nu\pi(a).\,f(\pi\cdot
t_1,\dots,\pi\cdot t_n)$), $\Term_{\Sigma,m}(X)$ is a nominal set.

\begin{defn}[Equations and Theories]
  A \emph{depth-n $\Sigma$-equation}, written \(X \vdash_n t = u\),
  consists of an orbit-finite nominal set~\(X\) of \emph{variables}
  and terms \(t, u \in \Term_{\Sigma, n}(X)\).  We often omit mention
  of~$n$.  A \emph{graded (nominal) theory} \(T = (\Sigma, E)\)
  consists of a graded signature~\(\Sigma\) and a set~\(E\) of
  \(\Sigma\)-equations, referred to as \emph{axioms}.
\end{defn}

\noindent Given a graded theory \(T = (\Sigma, E)\), we define
\emph{derivable $\Sigma$-equations} inductively by the derivation
rules in \cref{fig:derivationrules}.\lsnote{eventually rename
  (perm) into $(\alpha)$}
\begin{figure}
    \begin{framed}
        \begin{mathpar}
            \lrule{refl}{}{X \vdash_0 x = x} \and
            \lrule{symm}{X \vdash_m u = t}{X \vdash_m t = u} \and
            \lrule{trans}{X \vdash_m t = v \qquad X \vdash_m v = u}{X \vdash_m t = u} \and
            \lrule{cong}{X \vdash_m t_i = u_i \qquad (i=1,\dots,p)}{X \vdash_{m + d(f)} \eta.f(t_1, \ldots, t_p) = \eta.f(u_1, \ldots, u_p)}  \;\; (f/p \in \Sigma) \and
            %  \lrule{cong-f}{X \vdash_m t_i = u_i \qquad (i=1,\dots,p)}{X \vdash_{m + d(f)} a.f(t_1, \ldots, t_p) = a.f(u_1, \ldots, u_p)} \quad (f/p \in \sigf, a \in \Names)
            % \\[0.5em]
            % \lrule{cong-b}{X \vdash_m t_i = u_i \qquad (i=1,\dots,p)}{X \vdash_{m + d(f)} \nu a.f(t_1, \ldots, t_p) = \nu a.f(u_1, \ldots, u_p)} \quad (f/p \in \sigb, a \in \Names) \\[0.5em]
            \lrule{ax}{X \vdash_l \pi\cdot \sigma(y) = \sigma(\pi\cdot y) \qquad \forall \pi \in \Perm, y \in Y}{X \vdash_{m + l} (\tau\cdot r)\sigma = (\tau\cdot s)\sigma} \;\; (Y \vdash_m r = s \in E, \tau \in \Perm)  \and
            \lrule{perm}{X \vdash_m t_i = (a\ b)\cdot u_i \qquad (i=1,\dots,p)}{X \vdash_{m + d(f)} \nu a.f(t_1, \ldots, t_p) = \nu b.f(u_1, \ldots, u_p)} \;\; (f/p \in \sigb, a\neq b \in \Names, a \fresh u_i)
        \end{mathpar}
    \end{framed}
    \caption{Equational inference rules over a graded theory \(T = (\Sigma, E)\).}
    \label{fig:derivationrules}
\end{figure}
Note that the (ax) rule allows instantiation of axioms both under a
permutation~$\tau$ of the names and under a substitution~$\sigma$ of
the variables. We refer to the condition imposed on~$\sigma$ in the
premise of the rule as \emph{derivable equivariance}. This condition
ensures \emph{equivariance of derivability} (i.e., if
$X \vdash_m t =u$ is derivable, then so is
$X \vdash_m \pi \cdot t = \pi \cdot u$ for every $\pi \in \Perm$).

\begin{remark}\label{rem:ax}
  Rule (ax) as phrased in \cref{fig:derivationrules} has infinitely
  many premisses. However, it can be equivalently replaced with a
  finitary version where $\sigma$ is defined only on the finitely many
  variables occurring in~$r$ and~$s$ (and on one representative of
  every orbit of~$Y$, in the maybe pathological case that~$Y$ has
  orbits not mentioned in $r,s$) and then extended to all of~$Y$
  ensuring derivable equivariance. This is made possible by imposing
  premisses ensuring well-definedness up-to-derivability of such an
  extension. The set of these premisses is orbit-finite, so that in a
  simultaneous inductive proof with derivable equivariance,
  % \cref{lem:equivariance}.\eqref{item:derive-equivariance}
  we may restrict to finitely many representative premisses.
\end{remark}
% \begin{remark}
%   Alternatively to including operations with bound names in the
%   signature, one could impose $\alpha$-equivalence by means of
%   dedicated axioms. For instance, in place of a unary operation
%   $f\in\sigb$ we could use $f\in\sigf$ and impose axioms
%   $X\vdash a.f(x)=b.f((a\ b)\cdot x)$ for $b\fresh x$ (the latter just being
%   a side condition). Similar shifts have occurred between earlier
%   versions of nominal algebra\lsnote{Add citation}. We opt for the
%   present variant for added harmony with standard notions of nominal
%   signature, which often include binding (such as binding
%   signatures~\cite{FioreEA99}).
% \end{remark}
\noindent Like in standard universal algebra, the notion of graded
nominal theory comes equipped with a notion of model for which we will
establish soundness and completeness:
\begin{defn}[Graded nominal algebras]
  Given a graded signature \(\Sigma\) and a depth \(n < \omega\), a
  \emph{nominal \((\Sigma, n)\)-algebra} \(A\) consists of a family
  \((A_i)_{0 \le i \le n}\) of nominal sets, called the
  \emph{carriers}, and for each \(f/p \in \Sigma\) and
  $m \le n- d(f)$, an equivariant function $f_{A, m}$ of type
  \(A_m^p \to A_{m + d(f)}\) if $f\in\sigp$, of type
  \(\Names \times A_m^p \to A_{m + d(f)}\) if $f\in\sigf$, and of type
  \( [\Names]A_m^p \to A_{m + d(f)}\) if $f\in\sigb$. A
  \emph{homomorphism} $A\to B$ of \((\Sigma, n)\)-algebras \(A, B\) is
  a family \((h_i)_{0 \le i \le n}\) of equivariant functions
  \(h_i\colon A_i \to B_i\) satisfying the following equalities for
  \(f/p \in \Sigma\), $a\in\Names$, and \(x = (x_1, \ldots, x_p) \in A_m^p\):
  \begin{align*}
    h_{m + d(f)}(f_{A, m}(x)) & = f_{B, m}(h_m(x_1), \ldots, h_m(x_p)) && \text{if $f\in\sigp$}\\
    h_{m + d(f)}(f_{A, m}(a,x)) & = f_{B, m}(a,h_m(x_1), \ldots, h_m(x_p))&& \text{if $f\in\sigf$}\\
    h_{m + d(f)}(f_{A, m}(\bind{a}x)) & = f_{B, m}(\bind{a}(h_m(x_1), \ldots, h_m(x_p))) && \text{if $f\in\sigb$}.
  \end{align*}
  We write \(\Alg(\Sigma, n)\) for the category of nominal
  \((\Sigma, n)\)-algebras and their homomorphisms. We also define
  $(\Sigma,\omega)$-algebras and their homomorphisms; definitions are
  the same except that the range of indices is $i,m<\omega$ (rather
  than $\le n$)
\end{defn}
\noindent Algebras actually satisfying the axioms of the theory are
singled out as follows:
\begin{defn}[Term evaluation, satisfaction, models]
  Given a nominal \((\Sigma, n)\)-algebra \(A\) (where $n\le\omega$)
  and a base depth \(k \in \Nat_0, k \le n\), the \emph{value}
  \(\llbracket t \rrbracket^\iota_m\in A_{k + m}\) of a depth-$m$
  term~$t$ under an equivariant \emph{valuation}
  \(\iota\colon X \to A_k\) is defined by
  \begin{align*}
    \llbracket x \rrbracket^\iota_0  = \iota(x) \qquad
    \llbracket f(t_1, \ldots, t_p) \rrbracket^\iota_{m + d(f)}  &= f_{A,
      k + m}(\llbracket t_1 \rrbracket^\iota_m, \ldots, \llbracket t_p
    \rrbracket^\iota_m)\\
    \llbracket a.f(t_1, \ldots, t_p) \rrbracket^\iota_{m + d(f)} &=
    f_{A, k + m}(a, \llbracket t_1 \rrbracket^\iota_m, \ldots,
    \llbracket t_p \rrbracket^\iota_m) \\
    \llbracket \nu a.f(t_1,
    \ldots, t_p) \rrbracket^\iota_{m + d(f)}  &= f_{A, k + m}(\langle a
    \rangle (\llbracket t_1 \rrbracket^\iota_m, \ldots, \llbracket t_p
    \rrbracket^\iota_m))
  \end{align*}
  We say that \(A\) \emph{satisfies} a $\Sigma$-equation
  \(X \vdash_m t = u\) if
  \(\llbracket t \rrbracket^\iota_m = \llbracket u
  \rrbracket^\iota_m\) for every valuation \(\iota\colon X \to A_k\)
  with \(k + m \le n\) (a vacuous condition if $m>n$). Finally, given
  a graded theory \(T = (\Sigma, E)\), a \emph{\((T, n)\)-model} is a
  nominal \((\Sigma, n)\)-algebra satisfying every axiom in \(E\).  We
  write \(\Alg(T, n)\) for the full subcategory of \((T, n)\)-models
  in \(\Alg(\Sigma, n)\).
\end{defn}
\noindent % It is easily checked that term evaluation is equivariant in
% both the term and the valuation.
% Furthermore, term evaluation is compatible with (semantically equivariant) uniform substitution:

\noindent We obtain soundness w.r.t.~this notion of model:

\begin{theorem}[Soundness]\label{thm:soundness}
  The derivation system in \cref{fig:derivationrules} is sound: If
  an equation is derivable in a graded theory~$T$, then it is
  satisfied by every \((T, n)\)-model.
\end{theorem}

% \begin{proof}[Proof (Sketch)]
%   Induction on derivations. The case for rule (ax) uses the substitution
%   lemma, where the semantic equivariance of the substitution is guaranteed by derivable equivariance.
% \end{proof}

\noindent We prove completeness of the system by constructing free
graded nominal algebras, which at the same time yields our
construction of a graded monad on $\Nom$ from a graded nominal
theory. We write~$\sim$ for the equivalence relation on terms defined
by derivable equality, which by equivariance of derivability
is equivariant, and denote
the equivalence class of a depth-$n$ term~$t$ by $[t]_n$. Given a
nominal set~$X$ and $n\le\omega$, we define a $(\Sigma,n)$-algebra
$FX$ by carriers $(FX)_i = \Term_{\sig, i}(X)/{\sim}$ (for $i \le n$
if $n<\omega$, or for $i<\omega$ if $n=\omega$), and for
$f/p \in \Sigma$:
\begin{align*}
  f_{FX, m}([t_1]_m, \ldots, [t_p]_m) &= [f(t_1, \ldots, t_p)]_{m + d(f)} && \text{if $f \in \sigp$,} \\
  f_{FX, m}(a, [t_1]_m, \ldots, [t_p]_m) &= [a.f(t_1, \ldots, t_p)]_{m + d(f)} && \text{if $f \in \sigf$,} \\
  f_{FX, m}(\bind{a} ([t_1]_m, \ldots, [t_p]_m)) &= [\nu a.f(t_1, \ldots, t_p)]_{m + d(f)} && \text{if $f \in \sigb$.}
\end{align*}
This algebra is equipped with a \emph{canonical valuation}
$\eta\colon X\to (FX)_0$ given by $\eta(x)=[x]_0$. The rules of the
system, specifically (cong) and (perm), guarantee that this structure
is well-defined.
\begin{theorem}[Free $(T,n)$-models]\label{thm:free}
  For a graded nominal theory~$T$ and $n\le\omega$, the algebra $FX$
  is a free $(T,n)$-model over~$X$ w.r.t.\ the forgetful functor that
  sends a $(T,n)$-model~$A$ to the nominal set~$A_0$, with~$\eta$
  being the universal arrow.
\end{theorem}
% \begin{proof}[Proof (Sketch)]
%   The main step is to show that $FX$ is indeed a $(T,n)$-model. The
%   key point in the proof of this property is that a given equivariant
%   valuation in $FX$ induces a derivably equivariant substitution, so
%   that (ax) can be applied to establish that the corresponding
%   instance of a given axiom is satisfied in~$FX$.
% \end{proof}
\noindent Since~$FX$ identifies terms precisely when they are derivably
equal, we obtain
\begin{corollary}[Completeness]
  The derivation system in \cref{fig:derivationrules} is complete:
  If a depth-$m$ equation is satisfied by every \((T, n)\)-model of a
  graded nominal theory~$T$ where $m\le n\le\omega$, then it is
  derivable in~$T$.
\end{corollary}
\noindent From the adjunction of \cref{thm:free} for the case
$n=\omega$, we obtain a graded monad by standard
results~\cite[Section~3]{FujiiEA16}. \lsnote{Display this more
  prominently} Specifically, we have a (strict) \emph{action}~$\star$
of the discrete monoidal category $(\Nat,+,0)$ on $\Alg(T,\omega)$
given on objects by taking $n\star A$, for a $(T,\omega)$-model~$A$,
to be the $(T,\omega)$-model whose $i$-th carrier is $A_{i+n}$ and whose
interpretation $f_{n\star A,i}$ of $f\in\sig$ is $f_{A,i+n}$. We then
obtain a graded monad $\monad^T$ by sandwiching this action between
the two adjoint functors as per \cref{thm:free}. In particular, the
functor parts $M^T_n$ of $\monad^T$ are given as
$M^T_nX=(n\star FX)_0=(FX)_n$; that is, we have $M^T_nX=\Term_{\Sigma,n}(X)/\sim$
as expected. The unit of $\monad$ is just the canonical
valuation~$\eta$, and the multiplication is obtained from the counit
of the adjunction; intuitively,~$\eta$ thus converts variables into
terms, and the multiplication collapses layered terms.

\begin{example}[Traces in nominal systems]\label{expl:nominal-ts}
  As a first example of a graded nominal theory, take~$\sigp$ to
  consist of depth-$0$ operations~$0$,~$+$, with~$0$ a constant
  and~$+$ being binary and written in infix notation, and~$\sigf$ to
  consist of a single unary operation $\pre$ of depth~$1$. We impose
  depth-$0$ equations ensuring that $0$,~$+$ form a \emph{nominal join-semilattice} (i.e.~a nominal set carrying an equivariant
  join-semilattice structure), explicitly $X\vdash_0 x+0=x$,
  $X\vdash_0 x+x=x$, $X\vdash_0 x+y=y+x$, and
  $X\vdash_0 (x+y)+z=x+(y+z)$ for all orbit-finite~$X$ (more
  precisely, representatives of all such~$X$ modulo isomorphism to
  ensure that the axioms form a set, a point we will elide henceforth)
  and $x,y,z\in X$. Additionally, we have depth-$1$ axioms
  \begin{gather}
    X\vdash_1a.\pre(x+y)=a.\pre(x)+a.\pre(y)
    \quad\text{and}\quad
    \emptyset\vdash_1 a.\pre(0)=0\label{ax:distributivity-pre}
  \end{gather}
  for all~$a\in\Names$, all orbit-finite~$X$, and all $x,y\in X$.
  Notice here that postulating the equalities for \emph{all} $X,x,y$
  ensures that they apply to all elements of all nominal sets (see
  also \cref{rem:expressiveness}); on the other hand, postulating the
  equalities for all rather than just some~$a$ is in fact inessential,
  since rule (ax) allows applying permutations to axioms. Similarly as
  in \cref{expl:traces}, the induced graded monad~$\monad$ has functor
  parts~$M_n$ given on objects by $M_nX=\powf(\Names^n\times
  X)$. Coalgebras for the functor $G=M_1=\powf(\Names\times(-))$ are
  finitely branching \emph{non-deterministic nominal transition
    systems}, that is, NOFAs (\cref{sec:prelims}) in which all states
  are accepting. The graded semantics $(\monad,\id)$ for~$G$ captures
  precisely the trace semantics of such systems (i.e.\ the language
  semantics of the associated NOFA). We see an example with
  name-binding operations in \cref{sec:local-freshness}.
\end{example}

\begin{remark}\label{rem:expressiveness}
  The necessity of combining operations with free or bound names
  instead of having names and name binding as separate syntactic
  constructs is owed to the grading. For instance, in
  \cref{expl:nominal-ts}, it is crucial that $a.\pre(-)$ has
  depth~$1$ as a whole, rather than being composed of two components,
  one of which would necessarily have depth~$0$.

  In some previous systems of nominal
  algebra~\cite{Gabbay09,GabbayMathijssen09}, contexts of equations
  are presented in terms of variables and freshness conditions, which
  in our terms essentially amounts to restricting to contexts~$X$ that
  are \emph{strong} nominal sets~\cite{TzevelekosThesis}, i.e.~for
  every $\pi\in \Perm$ and $x\in X$, one has $\pi\cdot x = x$ if and
  only if $\pi(a)=a$ for all $a\in \supp(x)$. % However, we do not
  % currently claim to have examples where the added generality of our
  % system is needed.
  Roughly speaking, strong nominal sets thus do not have symmetries;
  more formally, every orbit in a strong nominal set is isomorphic to
  some nominal set of the form $\Names^{\fresh n}$ for $n\in\Nat$,
  i.e.~the set of $n$-tuples over~$\Names$ with pairwise distinct
  entries~\cite[Lemma~2.4.2]{Petrisan11}. For instance, the set
  $\{\{a,b\}\subseteq\Names\mid a\neq b\}$ fails to be strong, due to
  the symmetry embodied in the equality
  $(a\ b)\cdot\{a,b\}=\{a,b\}$. Our system is thus more expressive in
  that it allows equations to be conditioned on symmetries in the
  context; e.g.~we could impose an equation $f(x)=x$ conditioned on
  $(a\ b)\cdot x=x$. We do not currently claim to have real-life
  examples where this is needed; for instance, in
  \cref{expl:nominal-ts} one could equivalently restrict all equations
  to strong nominal sets of variables.

  % On \(\Set\), it can be shown that every (finitary) graded monad is
  % induced by a graded theory~\cite{Smirnov08,MiliusEA15}. To capture
  % all (finitary) graded monads or even just monads on $\Nom$, it will
  % presumably be necessary to extend the framework of (graded) nominal
  % algebra: In the known algebraic description of finitary monads on
  % locally finitely presentable (lfp) categories~\cite{KellyPower93},
  % arities are finitely presentable (fp) objects. The category $\Nom$
  % is lfp, with the fp objects being precisely the orbit-finite
  % sets. This entails that algebraic descriptions of monads on $\Nom$
  % may contain partial operations with domains determined by symmetry
  % conditions in the sense discussed above, a feature not currently
  % supported by any system of nominal algebra.
\end{remark}

\noindent Finally, we establish the announced criterion for the graded
monad induced by a graded nominal theory to be depth-$1$
(\cref{def:depth-1}).
\begin{defn}
  A graded theory \(T\) is \emph{depth-1} if all its operations and
  axioms are of depth at most $1$.
\end{defn}
\begin{theorem}\label{thm:depth1}
  Graded monads induced by depth-$1$ graded nominal theories are
  depth-$1$.
\end{theorem}
% \begin{proof}[Proof (Sketch)]
%   In the main part of the proof that~\eqref{eq:depth-1} is a
%   coequalizer, one shows well-definedness of a factoring morphism
%   $\bar q\colon M_{1+n}X\to Q$, where $q\colon M_1M_nX\to Q$ is a
%   morphism of $M_0$-algebras satisfying
%   $q\cdot M_1\mu^{0,n}_X=q\cdot\mu^{1,0}_{M_nX}$, by induction on
%   derivations at depth $1+n$. In the induction, one exploits that in
%   depth-$1$ graded nominal theories, one can split off a top-most
%   depth-$1$ part of terms and equalities.
% \end{proof}
\noindent For instance, the graded monad of \cref{expl:nominal-ts} is
depth-$1$.

\section{Case Study: Graded Monads for Global and  Local Freshness}
\label{sec:local-freshness}

\noindent We now present our main examples, in which we cast the
global and local freshness semantics of regular nominal transition
systems (RNTS; \cref{sec:rnna}) as graded semantics. We begin with
global freshness, exploiting that the induced notion of state
equivalence coincides with bar language semantics. The graded
theory~$T^{\Br}$ for bar language semantics extends the theory for
traces of non-deterministic nominal transition systems
(\cref{expl:nominal-ts}) by an additional depth-$1$ operation
$\abs\in\sigb$, defining a signature~$\Sigma^{\Br}$, and additional
depth-$1$ equations
\begin{gather}
  X\vdash_1\nu a.\,\abs(x+y)=(\nu a.\abs(x))+(\nu a.\abs(y))
  \quad\text{and}\quad
  \emptyset\vdash_1\nu a.\,\abs(0)=0 \label{ax:distributivity-abs}
\end{gather}
for all orbit-finite~$X$ and $x,y\in X$. We write
$at=a.\pre(t)$ and $\newletter at=\nu a.\,\abs(t)$ for all
terms~$t$. A \emph{pretrace} over a nominal set~$X$ is an element of
$(w,x)\in\barNames^*\times X$, with~$x$ thought of as a poststate
reached after executing~$w$. We can literally extend the notion of
\emph{$\alpha$-equivalence}~$\alphaeq$ of bar strings to pretraces.
% (allowing~$v$ to be a pretrace in~\eqref{eq:bar-alpha}).
We now define a graded monad $\monad^\Br=((M^\Br_n),\eta,(\mu^{nm}))$
of pretraces modulo $\alpha$-equivalence by
\begin{equation*}
  M^\Br_nX=\powf((\barNames^n\times X)/\alphaeq),
\end{equation*}
(hence $M^\Br_0X=\powf X$) with $\eta(x)=[x]_\alpha$ and
$\mu^{nm}(S)=\{[uv]_\alpha\mid [uV]_\alpha\in S,[v]_\alpha\in V\}$.
\begin{theorem}\label{thm:global-freshness-monad}
  The graded monad $\monad^\Br$ is induced by the graded nominal
  theory~$T^\Br$, and in particular is depth-$1$ by \cref{thm:depth1}.
\end{theorem}
% \begin{proof}[Proof (Sketch)]
%   We can surjectively map terms $t\in\Term_{\Sigma,n}(X)$ to elements
%   of $M_n^\Br X$ by distributing the join-semilattice operations to
%   the top and then taking $\alpha$-equivalence classes of
%   pretraces. This map~$\iota$ is homomorphically compatible (speaking
%   informally at the moment because terms do not form an algebra) with
%   a $(\Sigma^\Br,\omega)$-algebra structure~$A$ on the carriers
%   $A_n=M_n^\Br X$ extending the usual join-semilattice structure of
%   $\powf$ with $\pre_{A,n}(a,S)=\{[aw]_\alpha\mid [w]_\alpha\in S\}$
%   and
%   $\abs_{A,n}(\bind{a}S)=\{[\newletter aw]_\alpha\mid[w]_\alpha\in
%   S\}$. One shows that~$A$ is a $(T^\Br,\omega)$-model, so
%   that~$\iota$ induces a surjective homomorphism
%   $\iota^\sharp\colon FX\to A$; this homomorphism commutes with the
%   respective units of the adjunction with~$F$ and of~$\monad^\Br$. By
%   construction of~$\iota$ and \cref{lem:pretrace-alpha}, terms
%   identified by~$\iota$ are derivably equal, so that~$\iota^\sharp$ is
%   also injective.
% \end{proof}
Recall that regular nominal transition systems are coalgebras for the
functor~$H$ as per~\eqref{eq:rnna-functor}. To avoid discussion of
infinitary theories, we restrict~$H$ to
$H_\omega=\powf(\Names\times(-))\times\powf([\Names](-))$ (and hence
restrict to finitely branching RNTS). We
have an obvious isomorphism $\beta\colon H_\omega\to M_1$, giving
rise to a graded semantics $(\monad^\Br,\beta)$ for~$H$. This graded
semantics captures global freshness semantics:
\begin{theorem}\label{thm:global-freshness}
  Two states in finitely branching RNTS are trace equivalent under
  global freshness semantics iff they are behaviourally equivalent
  w.r.t.~the graded semantics $(\monad^\Br,\beta)$.
\end{theorem}

The \emph{local freshness} semantics of an RNTS (\cref{sec:rnna}) is
computed from its bar language semantics, and imposes additional
identifications. We define a graded theory $T^\loc$ for local
freshness semantics by extending $T^\Br$ with depth-$1$ axioms
\begin{equation}\label{ax:loc}
  X\vdash_1 a.\pre(x)+\nu a.\abs(x)=\nu a.\abs(x)
\end{equation}
for all orbit-finite~$X$ and $x\in X$, saying that every bound name
can be instantiated by itself. % (It may seem surprising that
% the axiom refers to coincidence of a free name with a bound name;
% however, recall that~$X$ is a nominal set, so renaming~$a$ into,
% say,~$b$ in $\nu a.\,x$ will change $x$ to $(a\ b)\cdot x$.)
We will see that this is already sufficient to describe local
freshness.  However, we can simplify some computations by introducing
\emph{name restrictions}: Similar to the computationally useful idea
of closing RNNA under name dropping and thus closing the literal
language under $\alpha$-equivalence~\cite{SchroderEA17}, we will
introduce an algebraic operation that can arbitrarily restrict the
support of a term.  This construction will be essential for the
description of equivalence games for local freshness semantics in
\cref{sec:games}. We define the graded theory~$T^\drop$ by
extending~$T^\loc$ with a bound operation $\res/1$ at depth~$0$, which
hides a name in its argument from the outside.  Correspondingly, we
add the \emph{name restriction
  axioms}~\cite[Chapter~9]{Pitts13}: % , used previously, for instance, to describe extensions of the $\lambda$-calculus with locally scoped names~\cite{Odersky94,PittsStark98}:
\begin{gather}
  X \vdash_0 \nu a.\res(x) = x \quad \text{for } a\fresh x
  \quad\text{and}\quad
  X \vdash_0 \nu a.\res(\nu b.\res(x)) = \nu b.\res(\nu a.\res(x)). \label{ax:namerestriction}
\end{gather}
We require this operation to be compatible with our join-semilattice
structure, in that
\begin{gather}
  X \vdash_0 \nu a.\res(x + y) = \nu a.\res(x) + \nu a.\res(y)
  \quad\text{and}\quad
  X \vdash_0 x = x + \nu a.\res(x), \label{ax:namerestrictionsemilattice}
\end{gather}
with $X \vdash_0 \nu a.\res(0) = 0$ already implied by \eqref{ax:namerestriction}.
For the depth-$1$ operations, we need the axioms
\begin{gather}
  X \vdash_1 \nu a.\res(b.\pre(x)) = b.\pre(\nu a.\res(x)) \quad\text{and}\quad X \vdash_1 \nu a.\res(a.\pre(x)) = 0, \label{ax:respre} \\
  X \vdash_1 \nu a.\res(\nu b.\abs(x)) = \nu b.\abs(\nu a.\res(x)) \label{ax:resabs}
\end{gather}
for $a \neq b$, where $X \vdash_1 \nu a.\res(\nu a.\abs(x)) = \nu a.\abs(x)$ is again implied by \eqref{ax:namerestriction}.

Let $\monad^\loc=((M^\loc),\eta,(\mu^{nm}))$ denote the graded monad
induced by~$T^\loc$ (we refrain from decorating~$\eta$ and~$\mu$ with
more indices) and similarly for~$\monad^\drop$ and $T^\drop$. Again,
we have by \cref{thm:depth1} that $\monad^\loc$ and $\monad^\drop$ are
depth-$1$. From theory inclusion, we then have canonical natural
transformations
\begin{equation*}
  \theta_n\colon M_n^\Br \to M_n^\loc\quad \text{and}\quad\kappa_n\colon M_n^\loc \to M_n^\drop
\end{equation*}
for every~$n$, and obtain graded semantics
$(\monad^\loc,\theta_1\cdot\beta)$ and
$(\monad^\drop,\kappa_1\cdot\theta_1\cdot\beta)$
for~$H$. Both of these capture local freshness:
\begin{theorem}\label{thm:local-freshness}
  Two states in finitely branching RNTS are trace equivalent under
  local freshness semantics iff they are behaviourally equivalent
  w.r.t.~the graded semantics \mbox{
    $(\monad^\loc,\theta_1\cdot\beta)$} iff they are behaviourally
  equivalent w.r.t.~the graded semantics \mbox{
    $(\monad^\drop,\kappa_1\cdot\theta_1\cdot\beta)$}.
\end{theorem}
\begin{remark}
  \label{rem:namerestrictionmodel}
  Given a nominal set~$X$, $x\in X$, and $N \subseteq \supp(x)$, write
  $x|_N = \Fix(N) \cdot x = \{\pi \cdot x \mid \pi \in \Fix(N)\}$ for
  the orbit of $x$ under $\Fix(N)$ (similar to the states in the
  construction of name-dropping RNNA~\cite{SchroderEA17}). We can
  then describe elements of $M^\drop_0X$ as finite sets~$S$ of such
  orbits such that $x|_{N'} \in S$ whenever $x|_N \in S$ and
  $N' \subseteq N$.
\end{remark}

\section{Graded Behavioural Equivalence Games}
\label{sec:games}

For illustration of the benefits of having a depth-$1$ graded
semantics, we discuss how graded behavioural equivalence games,
previously considered for set-based coalgebras~\cite{FordEA22},
transfer to the nominal setting. We then instantiate the generic game
to trace semantics of RNTS (\cref{sec:local-freshness}). The game has
a categorical formulation that is analogous to the set-based case. \hsnote{Reference} %, and deferred to the appendix.
The main novelty in the present work is the
formulation of the game in nominal-algebraic terms using our system of
graded nominal algebra from \cref{sec:graded-algebra}, presented
next. To this end, we fix a depth-$1$ graded
semantics~$(\monad,\beta)$ for a functor~$F$ on~$\Nom$, with~$\monad$
presented by a graded theory~$T$, and an $F$-coalgebra
$(C,\gamma)$. Throughout, we exploit that in join semilattices,
inequalities $s\le t$ are expressible as equations $t=s+t$.

\subparagraph*{Predeterminization} Graded behavioural equivalence
games are morally played on a \emph{predetermization} of $(C,\gamma)$,
that is, on a coalgebra $M_0C\to\mbar M_0C$ for a functor $\mbar$ on
$M_0$-algebras induced from a depth-$1$ graded semantics. (The name of
the construction is owed to the fact that in case $M_01=1$, it yields
an actual determinization% , i.e.~an $\mbar$-coalgebra in which
% the graded semantics is turned into coalgebraic behavioural
% equivalence (cf.~\cref{sec:rnna})
~\cite{FordEA22}). For our present purposes, we only need the
construction of~$\mbar$. Given an $M_0$-algebra~$A$, the
$M_0$-algebra~$\mbar(A,a)$ may be described categorically via a
coequalizer construction.\hsnote{Add here?} % found in the appendix
Algebraically,~$\mbar A$ is obtained by freely extending~$A$ to an
$M_1$-algebra and then forgetting~$A$. That is, $\mbar A$ consists of
depth-$1$ terms over~$A$, quotiented by equalities derivable in the
given graded theory from depth-$0$ equalities that hold over~$A$. It can be shown that
$\mbar(M_nX,\mu^{0n})\cong(M_{n+1}X,\mu^{0,n+1})$~\cite{FordEA22}; we thus
briefly write $\mbar M_n=M_{n+1}$.

\begin{example}
  \label{exp:m1-bar}
  \begin{enumerate}[wide]
  \item \label{item:m1-bar-global-freshness} \emph{Global freshness
      semantics (graded monad\/ $\monad^\Br$):} We have
    \(\mbar A = A^{\Names}_\omega \times [\Names]A\), where the
    endofunctor \((-)^{\Names}_\omega\) on $M_0$-algebras sends \(A\)
    to the set of functions \(f \colon \Names \to A\) with
    \(f(a) = 0\) for almost all \(a\).
  \item \label{item:m1-bar-local-freshness} \emph{Local freshness
      semantics with name-dropping (graded monad\/ $\monad^\drop$):} An
    \(M_0\)-algebra~\(A\) is a join semilattice $(A,+,0)$ with a name
    restriction operation
    \([\Names]A \to A, \bind{a}x \mapsto a \setminus x\).  As such, we
    can describe \(\mbar A\) as the set of finitely supported
    functions \(f\colon \Names \to_\fs A\) such that
    \((a\ c) \cdot (a \setminus f(c))\le f(a) \) for all
    \(a \in \Names\) and \(c \fresh (f, a)\). % (similar to
    %\eqref{eq:restriction}).  % It can be shown that this construction
    % is functorial for \(M_0\)-algebra morphisms using pointwise
    % application; in particular, this makes~$\mbar$ a subfunctor of the
    % functor $\Names \to_\fs (-)$, implying preservation of monos.
    % Intuitively speaking, this is a variation of
    % \eqref{item:m1-bar-global-freshness}, removing the distinction
    % between traces from bound prefixes and free prefixes.

    \end{enumerate}
    In both cases, the given description
    implies immediately that~$\mbar$ preserves monomorphisms, as
    needed for the correctness of the game (\cref{thm:games}).
\end{example}

\subparagraph*{Equivalence games} Graded behavioural equivalence games
come in a finite-depth and an infinite-depth
variant~\cite{FordEA22}; % The latter yields sensible results
% essentially only in case $M_01=1$. Although the latter condition can
% be ensured in our running examples by restricting to deadlock-free
% systems, i.e.~by replacing the finite powerset functor~$\powf$ with
% non-empty finite powerset throughout,
we focus on finite depth. Generally, we blur the distinction between
terms and elements of $M_nC$, i.e.~equivalence classes of terms.  The
game is played by \emph{Spoiler~(S)} and \emph{Duplicator~(D)}, with~D
trying to prove and~S trying to disprove behavioural
equivalence. Configurations of the game are pairs
$(s,t)\in M_0C\times M_0C$, read as equalities $C\vdash_0 s=t$ claimed
by~D. Given a set $Z\subseteq M_0C\times M_0C$ of depth-$0$
equalities, we write $Z\centails C\vdash_n s=t$ for terms
$s,t\in\Term_{\Sigma, n}(C)$ if the depth-$n$ equation $C\vdash_n s=t$
can be derived from the closure of~$Z$ under equivariance in the
system of \cref{sec:graded-algebra} (that is, equations in~$Z$ may be
renamed but not substituted into). We view the map
$\beta\cdot\gamma\colon C\to M_1C$ as a uniform depth-$1$
substitution~$\sigma$. The game is parametrized by a number~$n$ of
rounds: % ;
% more strictly speaking, we have a uniform depth-$1$ substitution
% such that for each $c\in C$, $(\beta\cdot\gamma)(c)$ is the
% equivalence class of $\sigma(c)$ in $M_1C$.

\begin{defn}
  The \emph{$n$-round $(\monad,\beta)$-behavioural equivalence game}
  $\game_n(\gamma)$ on~$(C,\gamma)$ is played in~$n$ rounds. A round
  starting in configuration $(s,t)$ is played as follows:
  \begin{enumerate}
  \item Duplicator plays a finite set $Z\subseteq M_0C\times M_0C$ such
    that $Z\centails C\vdash_1 s\sigma=t\sigma$.
  \item Spoiler picks an element $(s',t')\in Z$, which becomes the next
    configuration.
  \end{enumerate}
  Any player who cannot move on their turn, loses. After~$n$ rounds
  have been played, ending in configuration $(s,t)$,~D wins if
  $1\vdash_0 s\sigma_0=t\sigma_0$ is derivable, where $\sigma_0$ is the
  depth-$0$ substitution mapping all variables to $\star\in 1$; we
  refer to this step as \emph{calling the bluff}. Otherwise,~S wins.
\end{defn}
The game is intuitively understood as follows. We may regard
$\beta\cdot\gamma\colon C\to M_1C$ as defining the behaviour of states
$c\in C$ by algebraic equations $c=\sigma(c)$.  The behaviour map
$\gamma^{(n)}\colon C\to M_n1$ as per \cref{defn:graded-semantics}
unfolds definitions of states~$n$ times and finally applies the
substitution~$\sigma_0$ mapping all variables to $\star$ as above;
this defines a depth-$n$ substitution~$\sigma_n$ such that
$\gamma^{(n)}(c)$ is the equivalence class of $\sigma_n(c)$ in
$M_n1$. We can thus understand the $n$-round game from configuration
$(s,t)$ as D trying to construct a nominal-algebraic proof of the
depth-$n$ equation $1\vdash_n s\sigma_n=t\sigma_n$, with S challenging
selected claims made by~D.

Correctness of the game is formulated as follows.
\begin{theorem}[Game correctness] \label{thm:games} Let
  $(\monad,\beta)$ be a graded semantics for a functor~$F$ on~$\Nom$,
  with~$\monad$ presented by a graded nominal theory, and let
  $\gamma\colon C\to FC$ be a $F$-coalgebra. Suppose that~$\mbar$
  preserves monomorphisms (i.e.~injective $M_0$-homomorphisms). For
  $n\ge 0$,~D wins the configuration $(s,t)$ in $\game_n(\gamma)$ iff
  $(\gamma^{(n)})^*_0(s)=(\gamma^{(n)})^*_0(t)$.
% \end{theorem}
% \begin{corollary}
%   Let $(\monad,\beta)$ be a graded semantics for a functor~$F$
%   on~$\Nom$, and let $\gamma\colon C\to FC$ be a
%   $F$-coalgebra. Suppose that~$\mbar$ preserves monomorphisms and
%   that~$M_0$ is regularly algebraic.
  In particular, states $c,d\in C$ are $\beta$-behaviourally
  equivalent iff~D wins $(c,d)$ in $\game_n(\gamma)$ for
  every~$n\ge 0$.
  % \end{corollary}
\end{theorem}
% \begin{remark}
%   The equivariant set~$Z$ of equalities played by Duplicator will
%   typically be infinite; however, in many cases, in particular in our
%   main example, Duplicator can restrict to playing
%   orbit-finite~$Z$. In such cases,~$Z$ is still infinite but can be
%   finitely represented by picking one element per
%   orbit. Syntactically, closure of the representative elements under
%   equivariance will be usefully captured by introducing a dedicated
%   equational deduction rule for introducing assumptions from~$Z$: If
%   $C\vdash_0 u=v$ is a representative element of~$Z$, one may
%   introduce $C\vdash_0\tau\cdot u=\tau\cdot v$ for $\tau\in \Perm$.
% \end{remark}

\begin{example}
  \label{exp:explicit-games}
  For our main examples (\cref{sec:local-freshness}), we give explicit
  descriptions of the game, which we can rearrange to let S move
  first. \hsnote{Details?} In both cases, we use normal
  forms of~$M_1C$ implicit in the equation $M_1=\mbar M_0C$ and the
  respective description of~$\mbar$ (\autoref{exp:m1-bar}), and the
  substitution~$\sigma$ representing the coalgebra structure.
  \begin{enumerate}[wide]
  \item \emph{Global freshness semantics (graded monad\/
      $\monad^\Br$):} Given a term $s \in M_0C$, i.e.~a finite
    set of states, we can describe
    $s\sigma \in M_1 C = \mbar M_0 C=(M_0 C)^{\Names}_\omega \times
    [\Names]M_0 C$ as
    $s\sigma =((s_a)_{a\in\names}, \bind{c}s_{\scriptnew{c}})$ where
      \begin{equation*}
        s_a = \{y \mid x \in s, x \xrightarrow{a} y\} \quad\text{and}\quad s_{\scriptnew{c}} = \{y \mid x \in s, x \xrightarrow{\scriptnew{c}} y\}) \quad\text{for } c\fresh s.
      \end{equation*}
      In a configuration $(s,t)$ with $c\fresh(s,t)$, S either picks an $a \in \Names$ and $x \in s_a$ or an $x \in s_{\scriptnew{c}}$ (or analogously for~$t$).
      Duplicator then responds with a subset $t' \subseteq t_a$ or $t' \subseteq t_{\scriptnew{c}}$ correspondingly, yielding the new configuration $(t' \cup \{x\}, t')$ corresponding to the claim $x\le t'$ (and analogously for~$s$).
      This also means that S cannot switch sides after the initial move.

    \item \emph{Local freshness semantics with name-dropping (graded
        monad\/ $\monad^\drop$):} Viewing terms $s\in M_0C$ as sets of
      name restrictions closed under further restriction as detailed
      in \cref{rem:namerestrictionmodel}, we can describe
      $s\sigma \in M_1C = \mbar M_0 C$ as a function
      $\Names \to M_0 C, a \mapsto s_a$ where
      \begin{align*}
        s_a
          =~&\{ y|_{N'} \mid x|_N \in s, x \xrightarrow{a} y, a \in N, N' \subseteq \supp(y) \cap N \} \\
          \cup~&\{ (a\ c) \cdot (y|_{N' \setminus \{a\}}) \mid x|_N \in s, x \xrightarrow{\scriptnew{c}} y, N' \subseteq \supp(y) \cap (N \cup \{c\}) \}.
      \end{align*}
      In a configuration $(s, t)$, S then picks an $a \in \Names$ and
      $x|_N \in s_a$ (or $y|_N\in t_a$), to which D responds
      with a subset $t' \subseteq t_a$ closed under further
      restriction, yielding the new configuration
      $(t' \cup \{x|_{N'} \mid N' \subseteq N\}, t')$ (or
      symmetrically), effectively corresponding to the claim
      $x|_N\le t'$. Again, this means that S cannot switch sides
      after the initial move. As a rough summary, S picks a side
      where they play restricted successors, and D answers
      with restriction-closed sets of restricted successors under
      matching labels.
  \end{enumerate}
\end{example}

\section{Conclusions and Future Work}

\noindent We have introduced a universal-algebraic system for the
presentation of graded monads on the category of nominal sets. In a
subsequent case study, we have applied this system to model the global
and local freshness semantics of regular non-deterministic transition
systems (a slightly simplified form of \emph{regular non-deterministic
  nominal automata (RNNA)}~\cite{SchroderEA17}) in the framework of
\emph{graded semantics}~\cite{MiliusEA15,DorschEA19}. Notably, the
axiom distinguishing local from global freshness semantics just states
that every bound name subsumes the corresponding free name. % , a
% consequence of the fact that every bound name is $\alpha$-equivalent
% to itself, and a key factor in the comparative algorithmic
% tractability of RNNA (e.g.~in the elementary complexity of inclusion
% checking~\cite{SchroderEA17}, which contrasts with undecidability of
% inclusion for non-deterministic register automata with more than two
% registers~\cite{KaminskiFrancez94})
We have proved soundness and completeness of our system of
\emph{graded nominal algebra}, and we have shown that the graded monad
induced by graded nominal theory is \emph{depth-$1$} if the graded
theory itself is depth-$1$. In particular, this shows that the graded
monads featuring in the local and global freshness semantics of
regular non-deterministic nominal automata are depth-$1$.

In work on graded semantics over
sets~\cite{MiliusEA15,DorschEA19,FordEA22}, partial
orders~\cite{FordEA21}, metric spaces~\cite{ForsterEA25a}, and
relational structures~\cite{ForsterEA25b}, the somewhat
technical-looking depth-$1$ property has played a key role in enabling
characterization results for graded semantics in terms of modal logics
and Spoiler-Duplicator games. We have illustrated this point by
introducing a nominal variant of \emph{graded equivalence
  games}~\cite{FordEA22}, whose correctness depends on the depth-$1$
property, and instantiating the generic game to obtain game
characterizations of trace equivalence in nominal transition systems
under both local and global freshness semantics. Our system of graded
nominal algebra plays a key role in graded equivalence games, which
like in the set-based case can be viewed as playing out a graded
algebraic proof. An important direction for future work is to extend
further results in graded semantics to the nominal setting, notably
results on characteristic modal
logics~\cite{MiliusEA15,DorschEA19}. We expect that such results will,
like the game characterization that we have already materialized in
the present work, depend on the involved graded monads being
\mbox{depth-$1$}. An additional point of future research is coverage
of additional examples, such as fresh labelled transition
systems~\cite{BandukaraTzevelekos25}.

\bibliography{coalgml}

\iffullversion

% \end{document}
\clearpage
\appendix

\section{Omitted Details and Proofs}

In several places in the technical development, we need equivariance
of derivability:
\begin{lemma}\label{lem:equivariance}
  Let $\pi\in \Perm$, let $t,s\in\Term_{\Sigma,m}(Y)$, and let
  \(\sigma\colon Y \to \Term_{\Sigma, l}(X)\) be derivably
  equivariant. Then
  \begin{enumerate}
  \item\label{item:subst-equivariance}
    $X\vdash_{m+l}(\pi\cdot t)\sigma=\pi\cdot(t\sigma)$ is derivable.

  \item\label{item:derive-equivariance} If $Y\vdash_mt=s$ is
    derivable, then
    \(Y \vdash_m \pi\cdot t = \pi\cdot s\) is
    derivable.
  \end{enumerate}
\end{lemma}

\begin{proof}
  Claim~\ref{item:subst-equivariance} is shown by straightforward
  induction on~$t$; the base case $t=y\in Y$ is precisely derivable
  equivariance of~$\sigma$.  Claim~\ref{item:derive-equivariance} is
  shown by straightforward induction on derivations. In the case for
  rule (ax), one uses Claim~\ref{item:subst-equivariance}.
\end{proof}
\subsection*{Details for  \cref{rem:ax}}

Explicitly, the orbit-finite version (ax-f) of rule (ax) is the following:
\begin{equation*}
  \frac{X \vdash_l \pi\cdot \sigma(y) = \pi'\cdot\sigma(y') \quad
    \forall y,y' \in Y_0, \pi,\pi' \text{ s.t.~$\pi\cdot y=\pi'\cdot y'$}}{X \vdash_{m + l} (\tau\cdot r)\sigma =
      (\tau\cdot s)\sigma} \;(*)
\end{equation*}
where $Y_0$ is the set of variables occurring in~$r$ or~$s$ plus (if
necessary) one representative of every orbit of~$Y$ not mentioned
in~$r,s$, and the side condition~$(*)$ requires that
$Y \vdash_m r = s \in E, \tau \in \Perm$ and that $\sigma$ is a map
$Y_0\to\Term_{\Sigma,l}(X)$. Since~$Y$ is orbit-finite,~$Y_0$ is
finite. To show that (ax-f) is derivable, first extend~$\sigma$ to a
substitution $\overline\sigma$ defined on all of~$Y$, as follows.
Since~$Y_0$ contains at least one representative of every orbit
of~$Y$, we can pick, for every~$y\in Y\setminus Y_0$, elements
$\pi_y\in \Perm$ and $z_y\in Y_0$ such that $y=\pi_y\cdot z_y$ in such a
way that for $y\in Y_0$, we have $\pi_y=\id$, $z_y=y$. Then, define
\begin{equation*}
  \bar\sigma(y)=\pi_y\cdot\sigma(z_y).
\end{equation*}
We are done once we show that $\bar\sigma$ is derivably equivariant,
as this will allow deriving the conclusion by (ax). So let $\pi\in \Perm$,
$y\in Y$; we have to show that
$\pi\cdot\bar\sigma(y)=\bar\sigma(\pi\cdot y)$ is derivable. The
left-hand term is $\pi\cdot\pi_y\cdot\sigma( z_y)$, and the right-hand
term is $\pi_{\pi\cdot y}\cdot\sigma(z_{\pi\cdot y})$. So the required
equality is actually one of the premisses of (ax-f), since
$\pi\cdot\pi_y\cdot z_y=\pi\cdot y=\pi_{\pi\cdot y}\cdot z_{\pi\cdot
  y}$. For the other direction, given a derivably equivariant substitution
$\bar\sigma$, the premisses of (ax-f) are derivable for the
restriction~$\sigma$ of~$\bar\sigma$ to~$Y_0$: Given $\pi,\pi'\in \Perm$
and $y,y'\in Y_0$ such that $\pi\cdot y=\pi'\cdot y'$, we have the
following chain of derivable or literal equalities
over~$X$:
\begin{align*}
  & \pi\cdot\sigma(y) \\
  & = \pi\cdot\bar\sigma(y) \\
  & = \bar\sigma(\pi\cdot y) \\
  & = \bar\sigma(\pi'\cdot y') \\
  & = \pi'\cdot\bar\sigma(y') \\
  & = \pi'\cdot\sigma(y').
\end{align*}
Thus, we can alternatively apply (ax-f) to~$\sigma$ to obtain the
conclusion.

We can thus, altogether, replace (ax) with (ax-f). The set of
premisses of (ax-f) is clearly orbit-finite. We prove that we can
replace (ax-f) with an actual finitary version (ax-ff) where we only
require one representative premise for each orbit, by simultaneous
induction with
\cref{lem:equivariance}.\eqref{item:derive-equivariance} as
indicated in \cref{rem:ax}.

\subsection*{Proof of \cref{thm:soundness}}
The proof relies on the following property:
\begin{lemma}[Substitution Lemma]\label{thm:substitutionlemma}
    Let \(A\) be a nominal \((\Sigma, n)\)-algebra, \(\iota: Y \to A_k\) a valuation, and \(\sigma: X \to \Term_{\Sigma, l}(X)\) a substitution.
    If \(\kappa: X \to A_{k + l}\) with \(\kappa(x) = \llbracket \sigma(x) \rrbracket^\iota_l\) is equivariant, then \(\llbracket t\sigma \rrbracket^\iota_{m + l} = \llbracket t \rrbracket^\kappa_m\) for every term \(t \in \Term_{\Sigma, m}(X)\) with \(k + m + l \le n\).
\end{lemma}
\begin{proof}
  By routine induction on $t$, unfolding the definition of term
  evaluation at each step.
\end{proof}

The proof of \Cref{thm:soundness} then proceeds as follows:

Let \(A\) be a \((T, n)\)-model, and suppose that \(X \vdash_m t = u\) is derivable.

    We will show that for every equivariant valuation \(\iota: X \to A_k\) with \(k + m \le n\), it follows that \(\llbracket t \rrbracket^\iota_m = \llbracket u \rrbracket^\iota_m\), by induction on the derivation of \(X \vdash_m t = u\).
    \begin{itemize}
        \item
            Cases (refl), (symm), (trans), (cong) are straightforward.
        \item
            \textit{For (ax):}
            We know that \(A\) satisfies \(Y \vdash_{m'} r = s\) with \(r, s \in \Term_{\Sigma, m'}(Y)\) (because it is a \((T, n)\)-model), and that \(t = (\tau r)\sigma, u = (\tau s)\sigma\) for a derivably equivariant substitution \(\sigma: Y \to \Term_{\Sigma, l}(X)\) and a permutation \(\tau \in \Perm\).

            Let \(\kappa: Y \to A_{k + l}\) be defined as \(\kappa(x) = \llbracket \sigma(x) \rrbracket^\iota_l\).
            We will first show that \(\kappa\) is equivariant:
            For \(\pi \in \Perm\) and \(x \in Y\), we know that \(X \vdash_l \pi \sigma(x) = \sigma(\pi x)\) is derivable.
            It follows that
%            \begin{align*}
%                \kappa(\pi x)
%                &= \llbracket \sigma(\pi x) \rrbracket^\iota_l && \text{definition} \\
%                &= \llbracket \pi \sigma(x) \rrbracket^\iota_l && \text{inductive hypothesis} \\
%                &= \pi\llbracket \sigma(x) \rrbracket^\iota_l && \text{equivariance} \\
%                &= \pi\kappa(x) && \text{definition}.
%            \end{align*}
            \[ \kappa(\pi x)
            = \llbracket \sigma(\pi x) \rrbracket^\iota_l
            = \llbracket \pi \sigma(x) \rrbracket^\iota_l
            = \pi\llbracket \sigma(x) \rrbracket^\iota_l
            = \pi\kappa(x). \]
            due to the induction hypothesis and equivariance. Thus, we conclude that
           \begin{align*}
               \llbracket t \rrbracket^\iota_m
               &= \llbracket (\tau r)\sigma) \rrbracket^\iota_{m' + l} \\
               &= \llbracket \tau r \rrbracket^\kappa_{m'} && \text{\cref{thm:substitutionlemma}} \\
               &= \tau\llbracket r \rrbracket^\kappa_{m'} && \text{equivariance} \\
               &= \tau\llbracket s \rrbracket^\kappa_{m'} && \text{model} \\
               &= \llbracket \tau s \rrbracket^\kappa_{m'} && \text{equivariance} \\
               &= \llbracket (\tau s)\sigma \rrbracket^\iota_{m' + l} && \text{\cref{thm:substitutionlemma}} \\
               &= \llbracket u \rrbracket^\iota_m.
           \end{align*}
            where the fourth equation is due to $A$ satisfying $Y \vdash_{m'} r = s$.
          \item \textit{For (perm):} We know that
            \(t = \nu a.f(t_1, \ldots, t_p)\) and
            \(u = \nu b.f(u_1, \ldots, u_p)\), where \(a \fresh u_i\) and
            \(X \vdash_{m'} t_i = (a\ b)u_i\) is derivable for each
            \(i \in \{1, \ldots, p\}\) with \(m = m' + d(f)\).  Since
            \(a \neq b\) and, by equivariance of
            \(\llbracket \cdot \rrbracket^\iota_{m'}\),
            \(a \fresh \llbracket u_i \rrbracket^\iota_{m'}\) for every
            \(i\), it follows by standard characterizations of
            $\alpha$-equivalence~\cite{Pitts13} that
            \begin{gather}
                \langle a \rangle (a\ b)(\llbracket u_1 \rrbracket^\iota_{m'}, \ldots, \llbracket u_p \rrbracket^\iota_{m'}) = \langle b \rangle (\llbracket u_1 \rrbracket^\iota_{m'}, \ldots, \llbracket u_p \rrbracket^\iota_{m'}). \label{eq:alphaequivalenceofparameters}
            \end{gather}
            It then follows by computation that
            \begin{align*}
              \llbracket t \rrbracket^\iota_m
              &= \llbracket \nu a.f(t_1, \ldots, t_p) \rrbracket^\iota_{m' + d(f)} \\
              &= f_{A, m'}(\langle a \rangle (\llbracket t_1 \rrbracket^\iota_{m'}, \ldots, \llbracket t_p \rrbracket^\iota_{m'})) && \text{semantics} \\
              &= f_{A, m'}(\langle a \rangle (\llbracket (a\ b)u_1 \rrbracket^\iota_{m'}, \ldots, \llbracket (a\ b)u_p \rrbracket^\iota_{m'})) && \text{induction} \\
              &= f_{A, m'}(\langle a \rangle ((a\ b)\llbracket u_1 \rrbracket^\iota_{m'}, \ldots, (a\ b)\llbracket u_p \rrbracket^\iota_{m'})) && \text{equivariance} \\
              &= f_{A, m'}(\langle b \rangle (\llbracket u_1 \rrbracket^\iota_{m'}, \ldots, \llbracket u_p \rrbracket^\iota_{m'})) && \text{\cref{eq:alphaequivalenceofparameters}} \\
              &= \llbracket \nu b.f(u_1, \ldots, u_p) \rrbracket^\iota_{m' + d(f)} && \text{semantics} \\
              &= \llbracket u \rrbracket^\iota_m.
            \end{align*}
          \end{itemize}
\qed

\subsection*{Details for \cref{thm:free}}
We state and prove well-definedness and equivariance of the structure of $FX$, claimed inline in the main body:
\begin{lemma}\label{thm:freemodelisalgebra}
  The definition of \(FX\) indeed yields a well-defined nominal
  \((\sig, n)\)-algebra, in that \(f_{FX, m}\) is well-defined and equivariant for all \(f \in \sig\).
%    \begin{enumerate}
%        \item \(f_{FX, m}\) is well-defined and equivariant for all \(f \in \sigp\).
%        \item \(f_{FX, m}\) is well-defined and equivariant for all \(f \in \sigf\).
%        \item \(f_{FX, m}\) is well-defined and equivariant for all \(f \in \sigb\).
%    \end{enumerate}
\end{lemma}

\begin{proof}
    To show well-definedness for \(f \in \sigp\), let \([t_i]_m = [u_i]_m\) for all \(i \in \{1, \ldots, p\}\).
    It follows from (cong) that \(f(t_1, \ldots, t_p) \sim f(u_1, \ldots, u_p)\) and thus
    \begin{align*}
        f_{FX, m}([t_1]_m, \ldots, [t_p]_m)
        &= [f(t_1, \ldots, t_p)]_{m + d(f)} \\
        &= [f(u_1, \ldots u_p)]_{m + d(f)} \\
        &= f_{FX, m}([u_1]_m, \ldots, [u_p]_m).
    \end{align*}

    The case \(f \in \sigf\) is analogous. For \(f \in \sigb\), let
    \(\langle a \rangle ([t_1]_m, \ldots, [t_p]_m) = \langle b
    \rangle ([u_1]_m, \ldots, [u_p]_m)\).  By standard
    properties of $\alpha$-equivalence~\cite{Pitts13}, we only
    need to consider two cases:

    If \(a = b\), then \([t_i]_m = [u_i]_m\) for every \(i \in \{1, \ldots, p\}\) and the statement follows similar to the above case.

    Otherwise, \(a \fresh [u_i]_m\) and \([t_i]_m = (a\ b)[u_i]_m\) for every \(i \in \{1, \ldots, p\}\).
    By standard properties of equivariant equivalences~\cite{Pitts13}, we know that there is some \(\tilde{u}_i \in \Term_{\sig, m}(X)\) with \(u_i \sim \tilde{u}_i\) and \(a \fresh \tilde{u}_i\) for every \(i\).
    Thus, \([t_i]_m = (a\ b)[\tilde{u}_i]_m = [(a\ b)\tilde{u}_i]_m\) for every \(i\).
    We then have
    \begin{align*}
        &f_{FX, m}(\langle a \rangle ([t_1]_m, \ldots, [t_p]_m)) \\
        =~&[\nu a.f(t_1, \ldots, t_p)]_{m + d(f)} && \text{definition} \\
        =~&[\nu b.f(\tilde{u}_1, \ldots, \tilde{u}_p)]_{m + d(f)} && \text{(perm)} \\
        =~&[\nu b.f(u_1, \ldots, u_p)]_{m + d(f)} && \text{(cong)} \\
        =~&f_{FX, m}(\langle b \rangle ([u_1]_m, \ldots, [u_p]_m)) && \text{definition}.
    \end{align*}

    For equivariance of \(f \in \sigp\), let \(\pi \in \Perm\).
    Then we have
    \begin{align*}
        &f_{FX, m}(\pi[t_1]_m, \ldots, \pi[t_p]_m) \\
        =~&f_{FX, m}([\pi t_1]_m, \ldots, [\pi t_p]_m) && \text{equivariance} \\
        =~&[f(\pi t_1, \ldots, \pi t_p)]_{m + d(f)} && \text{definition} \\
        =~&[\pi(f(t_1, \ldots, t_p))]_{m + d(f)} && \text{definition} \\
        =~&\pi[f(t_1, \ldots, t_p)]_{m + d(f)} && \text{equivariance} \\
        =~&\pi(f_{FX, m}([t_1]_m, \ldots, [t_p]_m)) && \text{definition}.
    \end{align*}
    The two remaining cases are again analogous.

\end{proof}

\begin{lemma}\label{thm:canonicalenvironmentevaluation}
    If \(t \in \Term_{\sig, m}(X)\), then \(\llbracket t \rrbracket^{\eta_X}_m = [t]_m\).
\end{lemma}

\begin{proof}
    By induction on \(t\).
    \begin{itemize}
        \item
            \textit{For \(t = x\) with \(x \in X\):}
            By definition, we have \(\llbracket t \rrbracket^{\eta_X}_m = \llbracket x \rrbracket^{\eta_X}_0 = \eta_X(x) = [x]_0 = [t]_m\).
        \item
            \textit{For \(t = f(t_1, \ldots, t_p)\) with \(f/p \in \sigp\), \(t_1, \ldots, t_p \in \Term_{\sig, m'}(X)\), and \(m = m' + d(f)\):}
            \begin{align*}
                \llbracket t \rrbracket^{\eta_X}_m
                &= \llbracket f(t_1, \ldots, t_p) \rrbracket^{\eta_X}_{m' + d(f)} \\
                &= f_{FX, m'}(\llbracket t_1 \rrbracket^{\eta_X}_{m'}, \ldots, \llbracket t_p \rrbracket^{\eta_X}_{m'}) && \text{definition} \\
                &= f_{FX, m'}([t_1]_{m'}, \ldots, [t_p]^{\eta_X}_{m'}) && \text{induction} \\
                &= [f(t_1, \ldots, t_p)]_{m' + d(f)} && \text{definition} \\
                &= [t]_m.
            \end{align*}
        \item
            \textit{For \(t = a.f(t_1, \ldots, t_p)\) and \(t = \nu a.f(t_1, \ldots, t_p)\):}
            Analogous to the above case.
          \end{itemize}

\end{proof}

\begin{lemma}
    The graded nominal algebra \(FX\) is a \((T, n)\)-model.
\end{lemma}

\begin{proof}
    Since we know that \(FX\) is a nominal \((\sig, n)\)-algebra (by \cref{thm:freemodelisalgebra}), we only need to prove that \(FX\) satisfies every axiom in \(E\).

    Let \(Y \vdash_m t = u \in E\) be an axiom and \(\iota: Y \to (FX)_k\) an equivariant valuation.
    We will show that \(\llbracket t \rrbracket^\iota_m = \llbracket u \rrbracket^\iota_m\).

    Fix a splitting \(u_k: (FX)_k \to \Term_{\sig, k}(X)\) of the canonical projection (in that \([-]_k \circ u_k = \id\)) and let \(\sigma = u_k \circ \iota\).
    This definition of \(\sigma\) yields a derivably equivariant substitution:
    Let \(x \in Y\) and \(\pi \in \Perm\).
    Then
    \begin{align*}
        [\sigma(\pi x)]_k
        &= [u_k(\iota(\pi x))]_k && \text{by definition of } \sigma \\
        &= \iota(\pi x) && \text{because } [-]_k \circ u_k = \id \\
        &= \pi \iota(x) && \text{because } \iota \text{ is equivariant} \\
        &= \pi[u_k(\iota(x))]_k && \text{because } [-]_k \circ u_k = \id \\
        &= \pi[\sigma(x)]_k && \text{by definition of } \sigma \\
        &= [\pi \sigma(x)]_k, &&
    \end{align*}
    and thus \(X \vdash_k \pi \sigma(x) = \sigma(\pi x)\) is derivable.
    By applying (ax$_{t = u}$) with \(\tau = \id\), we then have \(X \vdash_{m + k} t\sigma = u\sigma\).

    Since we have
    \begin{align*}
        \iota(x)
        &= [u_k(\iota(x))]_k && \text{because } [-]_k \circ u_k = \id \\
        &= [\sigma(x)]_k && \text{by definition of } \sigma \\
        &= \llbracket \sigma(x) \rrbracket^{\eta_X}_k && \text{by \cref{thm:canonicalenvironmentevaluation}},
    \end{align*}
    we can conclude that
    \begin{align*}
        \llbracket t \rrbracket^\iota_m
        &= \llbracket t\sigma \rrbracket^{\eta_X}_{m + k} && \text{by \cref{thm:substitutionlemma}} \\
        &= [t\sigma]_{m + k} && \text{by \cref{thm:canonicalenvironmentevaluation}} \\
        &= [u\sigma]_{m + k} && \text{because } X \vdash_{m + k} t\sigma = u\sigma \\
        &= \llbracket u\sigma \rrbracket^{\eta_X}_{m + k} && \text{by \cref{thm:canonicalenvironmentevaluation}} \\
        &= \llbracket u \rrbracket^\iota_m && \text{by \cref{thm:substitutionlemma}}.
    \end{align*}
\end{proof}
\noindent In the proof of freeness of $FX$ (\cref{thm:free}), we
will in fact establish a full-blown adjoint situation with free
functor, unit, and counit. We begin by explicitly defining the
forgetful functor featuring in the statement of \cref{thm:free}:

\begin{defn}\label{def:forgetfulfunctor}
    We define the forgetful functor \(G: \Alg(T, n) \to \Nom\) by
    \begin{gather*}
        GA = A_0 \\
        Gh = h_0.
    \end{gather*}
\end{defn}
\noindent We next extend~$F$ to a functor:
\begin{defn}\label{def:freefunctor}
    We define the free functor \(F\colon \Nom \to \Alg(T, n)\) with \(FX\) as given above and for every \(f \in \Nom(X, Y)\),
    \[
        (Ff)_i([t]_i) = [t\sigma_f]_i,
    \]
    with the substitution \(\sigma_f = (X \xrightarrow{f} Y \hookrightarrow \Term_{\sig, 0}(Y))\).
\end{defn}

\begin{lemma}
    The above description of \(F\) is indeed a well-defined functor, in that
    \begin{enumerate}
        \item \(Ff\) is a well-defined morphism between \((T, n)\)-algebras for every \(f \in \Nom(X, Y)\),
        \item \(F\id_X = \id_{FX}\) for every nominal set~\(X\),
        \item \(F(g \circ f) = Fg \circ Ff\) for every \(f, g \in \Nom(X, Y)\).
    \end{enumerate}
\end{lemma}

\begin{proof}
    We will prove the statements individually:
    \begin{enumerate}[wide]
        \item \label{itm:freemodelfunctorequivariance}
            Let \(f: X \to Y\) be an equivariant function.

            We define a valuation \(\kappa: X \to (FY)_0\) with \(\kappa(x) = \llbracket \sigma_f(x) \rrbracket^{\eta_X}_0\).
            Note that \(\kappa\) is indeed equivariant:
            For every \(\pi \in \Perm\), we have \(\kappa(\pi x) = \llbracket f(\pi x) \rrbracket^{\eta_X}_0 = \llbracket \pi f(x) \rrbracket^{\eta_X}_0 = \pi\llbracket f(x) \rrbracket^{\eta_X}_0 = \pi\kappa(x)\) by equivariance of term evaluation and equivariance of \(f\).
            It then follows that, for every \([t]_m \in (FX)_m\), we have
            \begin{align*}
                (Ff)_m([t]_m)
                &= [t\sigma_f]_m && \text{definition} \\
                &= \llbracket t\sigma_f \rrbracket^{\eta_X}_m && \text{\cref{thm:canonicalenvironmentevaluation}} \\
                &= \llbracket t \rrbracket^\kappa_m && \text{\cref{thm:substitutionlemma}}.
            \end{align*}

            Thus, we can conclude well-definedness and equivariance from \cref{thm:soundness} and equivariance of term evaluation.

            To show homomorphy w.r.t.~\(g/p \in \sigp\), let \([t_1]_m, \ldots, [t_p]_m \in (FX)_m\).
            Then
            \begin{align*}
                &(Ff)_{m + d(g)}(g_{FX, m}([t_1]_m, \ldots, [t_p]_m)) \\
                =~&(Ff)_{m + d(g)}([g(t_1, \ldots, t_p)]_{m + d(g)}) && \text{definition} \\
                =~&[(g(t_1, \ldots, t_p))\sigma_f] && \text{definition} \\
                =~&[g(t_1\sigma_f, \ldots, t_p\sigma_f)] && \text{definition} \\
                =~&g_{FX, m}([t_1\sigma_f]_m, \ldots, [t_p\sigma_f]_m) && \text{definition} \\
                =~&g_{FX, m}((Ff)_m([t_1]_m), \ldots, (Ff)_m([t_p]_m)) && \text{definition}.
            \end{align*}
            Homomorphy w.r.t.\ operations in $\sigf$ and $\sigb$ is shown similarly.
        \item
            Let \([t] \in (FX)_m\).

            As seen in \eqref{itm:freemodelfunctorequivariance}, we have \((F\id_X)_i([t]_m) = \llbracket t \rrbracket^\kappa_m\) for \(\kappa: X \to (FX)\) with \(\kappa(x) = \llbracket \id_X(x) \rrbracket^{\eta_X}_0 = \llbracket x \rrbracket^{\eta_X}_0 = \eta_X(x)\).
            Thus, \(\kappa = \eta_X\) and it follows from \cref{thm:canonicalenvironmentevaluation} that \((F\id_X)_i([t]_m) = [t]_m = \id([t]_m)\).
        \item
            Let \(t \in \Term_{\sig, m}(X)\).
            We will show that \(t\sigma_{g \circ f} = (t\sigma_f)\sigma_g\) by induction on \(t\).
            \begin{itemize}
                \item
                    \textit{For \(t = x\) with \(x \in X\):}
                    By definition, we have
                    \begin{align*}
                        t\sigma_{g \circ f}
                        &= \sigma_{g \circ f}(x) \\
                        &= g(f(x)) \\
                        &= \sigma_g(f(x)) \\
                        &= (f(x))\sigma_g \\
                        &= (\sigma_f(x))\sigma_g \\
                        &= (t\sigma_f)\sigma_g.
                    \end{align*}
                \item
                    \textit{For \(t = f(t_1, \ldots, t_p)\) with \(f/p \in \sigp\), \(t_1, \ldots, t_p \in \Term_{\sig, m'}(X)\), and \(m = m' + d(f)\):}
                    \begin{align*}
                        t\sigma_{g \circ f}
                        &= f(t_1\sigma_{g \circ f}, \ldots, t_p\sigma_{g \circ f}) && \text{assumption} \\
                        &= f((t_1\sigma_f)\sigma_g, \ldots, (t_p\sigma_f)\sigma_g) && \text{induction} \\
                        &= (f(t_1\sigma_f, \ldots, t_p\sigma_f))\sigma_g && \text{definition} \\
                        &= ((f(t_1, \ldots, t_p))\sigma_f)\sigma_g && \text{definition} \\
                        &= (t\sigma_f)\sigma_g && \text{assumption}.
                    \end{align*}
                \item
                    \textit{For \(t = a.f(t_1, \ldots, t_p)\) and \(t = \nu a.f(t_1, \ldots, t_p)\):}
                    Analogous to the previous case.
            \end{itemize}

            With this, we can conclude that
            \begin{align*}
                (F(g \circ f))_m([t]_m)
                &= [t\sigma_{g \circ f}]_m \\
                &= [(t\sigma_f)\sigma_g]_m \\
                &= (Fg)_m([t\sigma_f]_m) \\
                &= (Fg)_m((Ff)_m([t]_m) \\
                &= (Fg \circ Ff)_m([t]_m).
            \end{align*}
    \end{enumerate}
\end{proof}
\noindent With these data, we show, as announced, the following equivalent version of \cref{thm:free}:
\begin{theorem}\label{thm:adjunction}
    The functor \(F\) is a left adjoint to the forgetful functor \(G\) with the unit \(\eta\) and the counit \(\varepsilon\) given by
    \begin{gather*}
        \varepsilon_A\colon FA_0 \to A, \\
        (\varepsilon_A)_i([t]_i) = \llbracket t \rrbracket^{\id}_i.
    \end{gather*}
\end{theorem}

\begin{proof}
    We will first show that \(\varepsilon_A\) is indeed a well-defined morphism between algebras for every $(T,n)$-model \(A\):
    \begin{itemize}
        \item
            Well-definedness and equivariance follow directly from \cref{thm:soundness} and equivariance of term evaluation.
        \item
            To show homomorphy w.r.t.~\(f/p \in \sigp\), let \([t_1]_m, \ldots, [t_p]_m \in (FA_0)_m\).
            By definition, we have
            \begin{align*}
                &(\varepsilon_A)_{m + d(f)}(f_{FA_0, m}([t_1]_m, \ldots, [t_p]_m)) \\
                =~&(\varepsilon_A)_{m + d(f)}([f(t_1, \ldots, t_p)]_{m + d(f)}) \\
                =~&\llbracket f(t_1, \ldots, t_p) \rrbracket^{\id}_{m + d(f)} \\
                =~&f_{A, m}(\llbracket t_1 \rrbracket^{\id}_m, \ldots, \llbracket t_p \rrbracket^{\id}_m) \\
                =~&f_{A, m}((\varepsilon_A)_m([t_1]_m), \ldots, (\varepsilon_A)_m([t_p]_m)).
            \end{align*}
            Homomorphy w.r.t.~operations in $\sigf$ and $\sigb$ is shown similarly.
    \end{itemize}

    Next, we will show that \(\eta\) is indeed natural.  So let
    \(X, Y\) be nominal sets, let $f\colon X\to Y$ be equivariant, and
    let \(x \in X\).  Let \(\sigma_f\) be defined as in
    \cref{def:freefunctor}.  By definition, we have
    \begin{align*}
        \eta_Y(f(x))
        &= [f(x)]_m \\
        &= [\sigma_f(x)]_m \\
        &= [x\sigma_f]_m \\
        &= (Ff)_m([x]) \\
        &= (Ff)_m(\eta_X(x)).
    \end{align*}

    To show that \(\varepsilon\) is also natural, let \(A, B\) be
    $(T,n)$-models, and let \(h\colon A\to B\) be a homomorphism.  We
    will show
    \((\varepsilon_B)_m((Fh_0)_m([t]_m)) =
    h_m((\varepsilon_A)_m([t]_m))\) for every
    \(t \in \Term_{\sig, m}(A_0)\) by induction on \(t\).
    \begin{itemize}
        \item
            \textit{For \(t = x\) with \(x \in A_0\):} By definition, we have
            \begin{align*}
                (\varepsilon_B)_0((Fh_0)_0([x]_0))
                &= (\varepsilon_B)_0([x\sigma_{h_0}]_0) \\
                &= (\varepsilon_B)_0([h_0(x)]_0) \\
                &= \llbracket h_0(x) \rrbracket^{\id}_0 \\
                &= h_0(x) \\
                &= h_0(\llbracket x \rrbracket^{\id}_0) \\
                &= h_0((\varepsilon_A)_0([x])).
            \end{align*}
        \item
            \textit{For \(t = f(t_1, \ldots, t_p)\) with \(f/p \in \sigp\), \(t_1, \ldots, t_p \in \Term_{\sig, m'}(X)\), and \(m = m' + d(f)\):}
            \begin{align*}
                &(\varepsilon_B)_m((Fh_0)_m([f(t_1, \ldots, t_p)]_m)) \\
                =~&\quad\text{(definition)} \\
                &(\varepsilon_B)_m((Fh_0)_m(f_{FA_0, m'}([t_1]_{m'}, \ldots, [t_p]_{m'}))) \\
                =~&\quad\text{(homomorphy)} \\
                &(\varepsilon_B)_m(f_{FB_0, m'}((Fh_0)_{m'}([t_1]_{m'}), \ldots, (Fh_0)_{m'}([t_p]_{m'}))) \\
                =~&\quad\text{(homomorphy)} \\
                &f_{B, m'}((\varepsilon_B)_{m'}((Fh_0)_{m'}([t_1]_{m'})), \ldots, (\varepsilon_B)_{m'}((Fh_0)_{m'}([t_p]_{m'}))) \\
                =~&\quad\text{(inductive hypothesis)} \\
                &f_{B, m'}(h_{m'}((\varepsilon_A)_{m'}([t_1]_{m'})), \ldots, h_{m'}((\varepsilon_A)_{m'}([t_p]_{m'}))) \\
                =~&\quad\text{(homomorphy)} \\
                &h_m(f_{A, m'}((\varepsilon_A)_{m'}([t_1]_{m'}), \ldots, (\varepsilon_A)_{m'}([t_p]_{m'}))) \\
                =~&\quad\text{(homomorphy)} \\
                &h_m((\varepsilon_A)_m(f_{FA_0, m'}([t_1]_{m'}, \ldots, [t_p]_{m'}))) \\
                =~&\quad\text{(definition)} \\
                &h_m((\varepsilon_A)_m([f(t_1, \ldots, t_p)]_{m})).
            \end{align*}
        \item
            \textit{For \(t = a.f(t_1, \ldots, t_p)\) and \(t = \nu a.f(t_1, \ldots, t_p)\):}
            Analogous to the previous case.
    \end{itemize}

    Finally, we will prove the adjunction using the counit-unit equations
    \begin{gather}
        \Id_F = \varepsilon F \circ F \eta, \label{eq:counitunit1} \\
        \Id_G = G \varepsilon \circ \eta G. \label{eq:counitunit2}
    \end{gather}

    Let \(A\) be a $(T,n)$-model, and let \(X\) be a nominal set.

    For \eqref{eq:counitunit1}, we will show \((\varepsilon_{FX})_m((F\eta_X)_m([t]_m)) = [t]_m\) for every \(t \in \Term_{\sig, m}(X)\) by induction on \(t\).
    \begin{itemize}
        \item
            \textit{For \(t = x\) with \(x \in X\):}
            \begin{align*}
                (\varepsilon_{FX})_0((F\eta_X)_0([x]_0))
                &= (\varepsilon_{FX})_0([x\sigma_{\eta_X}]_0) && \text{definition} \\
                &= (\varepsilon_{FX})_0([[x]_0]_0) && \text{definition} \\
                &= \llbracket [x]_0 \rrbracket^{\id}_0 && \text{definition of } \varepsilon \\
                &= [x]_0 && \text{definition}.
            \end{align*}
        \item
            \textit{For \(t = f(t_1, \ldots, t_p)\) with \(f/p \in \sigp\), \(t_1, \ldots, t_p \in \Term_{\sig, m'}(X)\), and \(m = m' + d(f)\):}
            \begin{align*}
                &(\varepsilon_{FX})_m((F\eta_X)_m([f(t_1, \ldots, t_p)]_m)) \\
                =~&\quad\text{(definition)} \\
                &(\varepsilon_{FX})_m((F\eta_X)_m(f_{FX, m'}([t_1]_{m'}, \ldots, [t_p]_{m'}))) \\
                =~&\quad\text{(homomorphy)} \\
                &(\varepsilon_{FX})_m(f_{F((FX)_0), m'}((F\eta_X)_{m'}([t_1]_{m'}), \ldots, (F\eta_X)_{m'}([t_p]_{m'}))) \\
                =~&\quad\text{(homomorphy)} \\
                &f_{FX, m'}((\varepsilon_{FX} \cdot F\eta_X)_{m'}([t_1]_{m'}), \ldots, (\varepsilon_{FX} \cdot F\eta_X)_{m'}([t_p]_{m'})) \\
                =~&\quad\text{(inductive hypothesis)} \\
                &f_{FX, m'}([t_1]_{m'}, \ldots [t_p]_{m'}) \\
                =~&\quad\text{(definition)} \\
                &[f(t_1, \ldots, t_p)]_m.
            \end{align*}
        \item
            \textit{For \(t = a.f(t_1, \ldots, t_p)\) and \(t = \nu a.f(t_1, \ldots, t_p)\):}
            Analogous to the previous case.
    \end{itemize}

    For \eqref{eq:counitunit2}, let \(x \in A_0\).
    Then we have
    \begin{align*}
        (\varepsilon_A)_0(\eta_{A_0}(x))
        &= (\varepsilon_A)_0([x]_0) && \text{definition} \\
        &= \llbracket x \rrbracket^{\id}_0 && \text{definition of } \varepsilon \\
        &= x && \text{definition}.
    \end{align*}

    With this, it follows that \((\varepsilon, \eta)\colon F \dashv G\).
\end{proof}

\subsection{Details for the Construction of $\monad^T$}\label{def:inducedmonad}

We use the adjunction of \cref{thm:adjunction} for
$(T,\omega)$-models and the strict action
\(\star\colon \Nat_0 \times \Alg(T, \omega) \to \Alg(T, \omega)\) of
the discrete monoidal category \((\Nat_0, +, 0)\) on
\(\Alg(T, \omega)\) defined in full as
\begin{gather*}
  n \star ((A_i), (f_{A, i})) = ((A_{i + n}), (f_{A, i + n})), \\
  n \star (f_i) = (f_{i + n}).
\end{gather*}
Using the sandwiching construction {\cite[Section 3]{FujiiEA16}}, we
obtain a graded monad \(((M^T_n), \eta, (\mu^{nk}))\) where
\begin{gather*}
  M^T_nX = G(n \star FX) = (FX)_n = \Term_{\sig, n}(X)/{\sim}, \\
  M_n^Tf([t]_n) = G(n \star Ff)([t]_n) = (Ff)_n([t]_n) = [t\sigma_f]_n,
\end{gather*}
\(\eta\) being the unit of the adjunction and
\begin{gather*}
  \mu^{nk}_X\colon M_n^TM_k^TX = G(n \star FG(k \star FX)) \to G(n \star (k \star FX)) = M^T_{n + k}X, \\
  \mu^{nk}_X([t]_n) = G(n \star \varepsilon_{k \star FX})([t]_n) = (\varepsilon_{k \star FX})_n([t]_n) = \llbracket t \rrbracket^{\id}_n.
\end{gather*}

\subsection*{Details for \cref{thm:depth1}}

We start with two lemmas clarifying in general that substitution and the monad multiplications relate in the expected manner:
\begin{lemma}\label{thm:multiplicationsubstitution}
  Let \(T = (\sig, E)\) be a graded theory and
  \(((M_n), \eta, (\mu^{nk}))\) be the graded monad induced by it.  If
  \(t \in \Term_{\sig, n}(Y)\) and
  \(\sigma: Y \to \Term_{\sig, k}(X)\), then
  \([t\sigma]_{n + k} = \mu^{n, k}_X([t\bar{\sigma}]_n)\), where
  \(\bar{\sigma}: Y \to \Term_{\sig, 0}(M_k X)\) is defined as
  \(\bar{\sigma}(x) = [\sigma(x)]_k\).
\end{lemma}

\begin{proof}
    By induction on \(t\).
    \begin{itemize}
        \item
            \textit{For \(t = x\) with \(x \in Y\):}
            \[
                [\sigma(x)]_{k} = \bar{\sigma}(x) = \llbracket \bar{\sigma}(x) \rrbracket^\id_0 = \mu^{0, k}_X([\bar{\sigma}(x)]_0).
            \]
        \item
            \textit{For \(t = f(t_1, \ldots, t_p)\) with \(f/p \in \sigp\), \(t_1, \ldots, t_p \in \Term_{\sig, m}(Y)\), and \(1 = m + d(f)\):}
            By inductive hypothesis, we know that \([t_i\sigma]_{m + k} = \mu^{m, k}_X([t_i\bar{\sigma}]_m)\) for every \(i\).
            It then follows that
            \begin{align*}
                &[t\sigma]_{n + k} \\
                =~&[f(t_1\sigma, \ldots, t_p\sigma)]_{n + k} && \text{definition} \\
                =~&f_{F(X), m + k}([t_1\sigma]_{m + k}, \ldots, [t_p\sigma]_{m + k}) && \text{definition} \\
                =~&f_{F(X), m + k}(\mu^{m, k}_X([t_1\bar{\sigma}]_m), \ldots, \mu^{m, k}_X([t_p\bar{\sigma}]_m)) && \text{induction} \\
                =~&f_{F(X), m + k}(\llbracket t_1\bar{\sigma} \rrbracket^\id_m, \ldots, \llbracket t_p\bar{\sigma} \rrbracket^\id_m) && \text{\cref{def:inducedmonad}} \\
                =~&\llbracket f(t_1\bar{\sigma}, \ldots, t_p\bar{\sigma}) \rrbracket^\id_{n} && \text{definition} \\
                =~&\mu^{n, k}_X([t\bar{\sigma}]_n) && \text{\cref{def:inducedmonad}}.
            \end{align*}
        \item
            \textit{For \(t = a.f(t_1, \ldots, t_p)\) and \(t = \nu a.f(t_1, \ldots, t_p)\):}
            Analogous to the previous case.
    \end{itemize}

\end{proof}
\begin{lemma}\label{thm:multiplicationsplitting}
    Let \(T = (\sig, E)\) be a graded theory and \(((M_n), \eta, (\mu^{nk}))\) be the graded monad induced by it.
    If \(\sigma: M_k X \to \Term_{\sig, k}(X)\) is a splitting (in that \([-]_k \circ \sigma = \id\)) and \(t \in \Term_{\sig, n}(M_k X)\), then \(\mu^{nk}_X([t]_n) = [t\sigma]_{n + k}\).
\end{lemma}

\begin{proof}
    First note that, for \([u]_k \in M_k X\), we have
    \begin{align*}
        \llbracket \sigma([u]_k) \rrbracket^{\eta_X}_{k}
        &= [\sigma([u]_k)]_k && \text{\cref{thm:canonicalenvironmentevaluation}} \\
        &= \id([u]_k) && \text{because } [-]_k \circ \sigma = \id.
    \end{align*}

    It then follows that, for \(t \in \Term_{\sig, n}(M_k X)\),
    \begin{align*}
        \mu^{nk}_X([t]_n)
        &= \llbracket t \rrbracket^{\id}_n && \text{\cref{def:inducedmonad}} \\
        &= \llbracket t\sigma \rrbracket^{\eta_X}_{n + k} && \text{\cref{thm:substitutionlemma} with } \kappa = \id \\
        &= [t\sigma]_{n + k} && \text{\cref{thm:canonicalenvironmentevaluation}}.
    \end{align*}

\end{proof}

Moreover, we need the following fact on the interpretation of operations in
induced graded monads:

\begin{lemma}\label{thm:multiplicationwithzero}
    Let \(T = (\sig, E)\) be a graded theory and \(((M_n), \eta, (\mu^{nk}))\) be the graded monad induced by it.
    For \(f/p \in \sig\) and \(t_1, \ldots, t_p \in \Term_{\sig, n}(X)\):
    \begin{equation*}
        [\eta.f(t_1, \ldots, t_p)]_{n + d(f)} = \mu^{d(f), n}_X([\eta.f([t_1]_n, \ldots, [t_p]_n)]_{d(f)}).
    \end{equation*}
\end{lemma}

\begin{proof}
    We will show the statement for $f/p\in\sigp$ by simple computation:
    \begin{align*}
        &\mu^{d(f), n}_X([f([t_1]_n, \ldots, [t_p]_n)]_{d(f)} \\
        =~&\llbracket f([t_1]_n, \ldots, [t_p]_n) \rrbracket^\id_{d(f)} && \text{\cref{def:inducedmonad}} \\
        =~&f_{F(X), n}(\llbracket [t_1]_n \rrbracket^\id_0, \ldots, \llbracket [t_p]_n \rrbracket^\id_0) && \text{definition} \\
        =~&f_{F(X), n}([t_1]_n, \ldots, [t_p]_n) && \text{definition} \\
        =~&[f(t_1, \ldots, t_p)]_{n + d(f)} && \text{definition}.
    \end{align*}

\end{proof}

Fix a depth-1 graded theory \(T = (\sig, E)\) and let
\(((M_n), \eta, (\mu^{nk}))\) be the graded monad induced by it as
described in \cref{def:inducedmonad}.

The monad multiplication \(\mu^{1, n}_X\) collapses depth-1 terms with embedded depth-n terms into terms of depth \(1 + n\).
Conversely, we can split a depth-\(1 + n\) term into such a layered term, given that all operations are at most depth-1.

To do this, let \(s^n_X: \Term_{\sig, 1 + n}(X) \to \Term_{\sig, 1}(M_n X)\) be defined as
\begin{align*}
    s^n_X(f(t_1, \ldots, t_p)) &= f(s^n_X(t_1), \ldots, s^n_X(t_p)) && \text{if } f/p \in \sigp \text{ and } d(f) = 0, \\
    s^n_X(f(t_1, \ldots, t_p)) &= f([t_1]_n, \ldots, [t_p]_n) && \text{if } f/p \in \sigp \text{ and } d(f) = 1,
\end{align*}
and similarly for \(\sigf\) and \(\sigb\).

\begin{lemma}\label{thm:splitequivariance}
    The function \(s^n_X\) is equivariant.
\end{lemma}

\begin{proof}
    Let \(\pi \in \Perm\). We will show \(s^n_X(\pi t) = \pi s^n_X(t)\) for all \(t \in \Term_{\sig, 1 + n}\) by induction on \(t\), only considering cases where \(t\) has at least uniform depth 1.
    \begin{itemize}
        \item
            \textit{For \(t = f(t_1, \ldots, t_p)\) with \(f/p \in \sigp\), \(t_1, \ldots, t_p \in \Term_{\sig, m}(X)\), and \(1 + n = m + d(f)\):}
            Since all operations are at most depth-1, we only have to consider two cases:

            If \(d(f) = 0\), then
            \begin{align*}
                s^n_X(\pi t)
                &= s^n_X(f(\pi t_1, \ldots, \pi t_p)) && \text{definition} \\
                &= f(s^n_X(\pi t_1), \ldots, s^n_X(\pi t_p)) && \text{definition of } s^n_X \\
                &= f(\pi s^n_X(t_1), \ldots, \pi s^n_X(t_p)) && \text{induction} \\
                &= \pi f(s^n_X(t_1), \ldots, s^n_X(t_p)) && \text{definition} \\
                &= \pi s^n_X(t) && \text{definition of } s^n_X.
            \end{align*}

            If \(d(f) = 1\), then
            \begin{align*}
                s^n_X(\pi t)
                &= s^n_X(f(\pi t_1, \ldots, \pi t_p)) && \text{definition} \\
                &= f([\pi t_1]_n, \ldots, [\pi t_p]_n) && \text{definition of } s^n_X \\
                &= f(\pi[t_1]_n, \ldots, \pi[t_p]_n) && \text{equivariance} \\
                &= \pi f([t_1]_n, \ldots, [t_p]_n) && \text{definition} \\
                &= \pi s^n_X(t) && \text{definition of } s^n_X.
            \end{align*}
        \item
            \textit{For \(t = a.f(t_1, \ldots, t_p)\) and \(t = \nu a.f(t_1, \ldots, t_p)\):}
            Analogous to the previous case.
    \end{itemize}

\end{proof}

Of course, where we split exactly is chosen arbitrarily:
Here, we use the outermost term inside a depth-1 operation, however this term could have more depth-0 operations we could include.
We will see that this doesn't matter in the context of morphisms satisfying the universal property of a coequalizer.
To do this, we will first show some general properties of such morphisms:
\begin{lemma}\label{thm:coequalizerproperties}
    Let \((Q, h)\) be an \(M_0\)-algebra and \(q: M_1 M_n X \to Q\) a morphism between \(M_0\)-algebras such that \(q \circ M_1 \mu^{0, n}_X = q \circ \mu^{1, 0}_{M_n X}\).
    \begin{enumerate}
        \item \label{itm:coequalizerpropertiesdepth0}
            If \(f/p \in \sig\) with \(d(f) = 0\) and \(t_1, \ldots, t_p \in \Term_{\sig, 1}(M_n X)\), then
            \begin{equation*}
                q([\eta.f(t_1, \ldots, t_p)]_1) = h([\eta.f(q([t_1]_1), \ldots, q([t_p]_1))]_0).
            \end{equation*}
        \item \label{itm:coequalizerpropertiesdepth1}
            If \(f/p \in \sig\) with \(d(f) = 1\) and \(t_1, \ldots, t_p \in \Term_{\sig, 0}(M_n X)\), then
            \begin{equation*}
                q([\eta.f(t_1, \ldots, t_p)]_1) = q([\eta.f(\mu^{0, n}_X([t_1]_0), \ldots, \mu^{0, n}_X([t_p]_0))]_1).
            \end{equation*}
        \item \label{itm:substitutionlemmadepth0}
            If \(t \in \Term_{\sig, 0}(Y)\) and \(\sigma: Y \to \Term_{\sig, 1 + n}(X)\), then \(q([s^n_X(t\sigma)]_1) = h([t\bar{\sigma}]_0)\) for \(\bar{\sigma}: Y \to \Term_{\sig, 0}(Q)\) with \(\bar{\sigma}(x) = q([s^n_X(\sigma(x))]_1)\).
        \item \label{itm:substitutionlemmadepth1}
            If \(t \in \Term_{\sig, 1}(Y)\) and \(\sigma: Y \to \Term_{\sig, n}(X)\), then \(q([s^n_X(t\sigma)]_1) = q([t\bar{\sigma}]_1)\) for \(\bar{\sigma}: Y \to \Term_{\sig, 0}(M_n X)\) with \(\bar{\sigma}(x) = [\sigma(x)]_n\).
    \end{enumerate}
\end{lemma}

\begin{proof}
    Since \(q\) is a morphism between the \(M_0\)-algebras \((M_1 M_n X, \mu^{0, 1}_{M_n X})\) and \((Q, h)\), we know that
    \begin{align}
        h \circ M_0 q = q \circ \mu^{0, 1}_{M_n X}. \label{eq:m0algebramorphism}
    \end{align}

    We will proceed to show each statement individually:
    \begin{enumerate}[wide]
        \item
            We will only show the statement for \(\sigp\), the arguments for the others are completely analogous.
            By computation, we have
            \begin{align*}
                &q([f(t_1, \ldots, t_p)]_1) \\
                =~&q(\mu^{0, 1}_{M_n X}([f([t_1]_1, \ldots, [t_p]_1)]_0)) && \text{\cref{thm:multiplicationwithzero}} \\
                =~&h(M_0 q([f([t_1]_1, \ldots, [t_p]_1)]_0)) && \text{\cref{eq:m0algebramorphism}} \\
                =~&h([f([t_1]_1, \ldots, [t_p]_1)\sigma_q]_0) && \text{\cref{def:inducedmonad}} \\
                =~&h([f(q([t_1]_1), \ldots, q([t_p]_1))]_0) && \text{definition}.
            \end{align*}
        \item
            Once again, we will only show the statement for \(\sigp\).
            By computation, we have
            \begin{align*}
                &q([f(t_1, \ldots, t_p)]_1) \\
                =~&q(\mu^{1, 0}_{M_n X}([f([t_1]_0, \ldots, [t_p]_0)]_1)) && \text{\cref{thm:multiplicationwithzero}} \\
                =~&q(M_1 \mu^{0, n}_X([f([t_1]_0, \ldots, [t_p]_0)]_1)) && \text{assumption} \\
                =~&q([f([t_1]_0, \ldots, [t_p]_0) \sigma_{\mu^{0, n}_X}]_1) && \text{\cref{def:inducedmonad}} \\
                =~&q([f(\mu^{0, n}_X([t_1]_0), \ldots, \mu^{0, n}_X([t_p]_0))]_1) && \text{definition}.
            \end{align*}
        \item
            By induction on \(t\).
            \begin{itemize}
                \item
                    \textit{For \(t = x\) with \(x \in Y\):}
                    \begin{align*}
                        &q([s^n_X(t\sigma)]_1) \\
                        =~&q(\llbracket [s^n_X(t\sigma)]_1 \rrbracket^\id_0) && \text{definition} \\
                        =~&q(\mu^{0, 1}_{M_n X}([[s^n_X(t\sigma)]_1]_0)) && \text{\cref{def:inducedmonad}} \\
                        =~&h(M_0 q([[s^n_X(t\sigma)]_1]_0)) && \text{\cref{eq:m0algebramorphism}} \\
                        =~&h([q([s^n_X(\sigma(x))]_1)]_0) && \text{\cref{def:inducedmonad}} \\
                        =~&h([\bar{\sigma}(x)]_0) && \text{definition of } \bar{\sigma} \\
                        =~&h([t\bar{\sigma}]_0) && \text{definition}.
                    \end{align*}
                \item
                    \textit{For \(t = f(t_1, \ldots, t_p)\) with \(f/p \in \sigp\):}
                    We know that \(d(f) = 0\) (because \(t\) is depth-0). It then follows that
                    \begin{align*}
                        &q([s^n_X(t\sigma)]_1) \\
                        =~&q([s^n_X(f(t_1\sigma, \ldots, t_p\sigma))]_1) && \text{definition} \\
                        =~&q([f(s^n_X(t_1\sigma), \ldots, s^n_X(t_p\sigma))]_1) && \text{definition of } s^n_X \\
                        =~&h([f(q([s^n_X(t_1\sigma)]_1), \ldots, q([s^n_X(t_p\sigma)]_1))]_0) && \text{\eqref{itm:coequalizerpropertiesdepth0}} \\
                        =~&h([f(h([t_1\bar{\sigma}]_0), \ldots, h([t_p\bar{\sigma}]_0))]_0) && \text{induction} \\
                        =~&h(M_0 h([f([t_1\bar{\sigma}]_0, \ldots, [t_p\bar{\sigma}]_0)]_0)) && \text{\cref{def:inducedmonad}} \\
                        =~&h(\mu^{0, 0}_{Q}([f([t_1\bar{\sigma}]_0, \ldots, [t_p\bar{\sigma}]_0)]_0)) && (\star) \\
                        =~&h([f(t_1\bar{\sigma}, \ldots, t_p\bar{\sigma})]_0) && \text{\cref{thm:multiplicationwithzero}} \\
                        =~&h([t\bar{\sigma}]_0) && \text{definition},
                    \end{align*}
                    where $(\star)$ follows from $(Q, h)$ being an $M_0$-algebra.
                \item
                    \textit{For \(t = a.f(t_1, \ldots, t_p)\) and \(t = \nu a.f(t_1, \ldots, t_p)\):}
                    Analogous to the previous case.
            \end{itemize}
        \item
            By induction on \(t\).
            \begin{itemize}
                \item
                    \textit{For \(t = x\) with \(x \in Y\):}
                    This would imply that \(t\) is depth-0, but we know that \(t\) has uniform depth 1.
                \item
                    \textit{For \(t = f(t_1, \ldots, t_p)\) with \(f/p \in \sigp\):}
                    Since \(t\) has uniform depth 1, we have to consider two cases:

                    For \(d(f) = 0\), we have
                    \begin{align*}
                        &q([s^n_X(t\sigma)]_1) \\
                        =~&q([s^n_X(f(t_1\sigma, \ldots, t_p\sigma))]_1) && \text{definition} \\
                        =~&q([f(s^n_X(t_1\sigma), \ldots, s^n_X(t_p\sigma))]_1) && \text{definition of } s^n_X \\
                        =~&h([f(q([s^n_X(t_1\sigma)]_1), \ldots, q([s^n_X(t_p\sigma)]_1)]_0) && \text{\eqref{itm:coequalizerpropertiesdepth0}} \\
                        =~&h([f(q([t_1\bar{\sigma}]_1), \ldots, q([t_p\bar{\sigma}]_1)]_0) && \text{induction} \\
                        =~&q([f(t_1\bar{\sigma}, \ldots, t_p\bar{\sigma})]_1) && \text{\eqref{itm:coequalizerpropertiesdepth0}} \\
                        =~&q([t\bar{\sigma}]_1) && \text{definition}.
                    \end{align*}

                    For \(d(f) = 1\), we have
                    \begin{align*}
                        &q([s^n_X(t\sigma)]_1) \\
                        =~&q([s^n_X(f(t_1\sigma, \ldots, t_p\sigma))]_1) && \text{definition} \\
                        =~&q([f([t_1\sigma]_n, \ldots, [t_p\sigma]_n)]_1) && \text{definition of } s^n_X \\
                        =~&q([f(\mu^{0, n}_X([t_1 \bar{\sigma}]_0), \ldots, \mu^{0, n}_X([t_p \bar{\sigma}]_0)]))]_1) && \text{\cref{thm:multiplicationsubstitution}} \\
                        =~&q([f(t_1 \bar{\sigma}, \ldots, t_p \bar{\sigma})]_1) && \text{\eqref{itm:coequalizerpropertiesdepth1}} \\
                        =~&q([t\bar{\sigma}]_1) && \text{definition}.
                    \end{align*}
                \item
                    \textit{For \(t = a.f(t_1, \ldots, t_p)\) and \(t = \nu a.f(t_1, \ldots, t_p)\):}
                    Analogous to the previous case.
            \end{itemize}
    \end{enumerate}

\end{proof}

With this, we can finally show the coequalizer property for \(\mu^{1, n}_X\) and thus the desired result.

\begin{theorem}
    The graded monad induced by a depth-1 graded theory \(T = (\sig, E)\) is depth-1.
\end{theorem}

\begin{proof}
    We show that \eqref{eq:depth-1} is a coequalizer.
    % By \cref{def:gradedmonad}, we already know that \(\mu^{1, n}_X \circ M_1 \mu^{0, n}_X = \mu^{1, n}_X \circ \mu^{1, 0}_{M_n X}\).
    So let \((Q, h)\) be an \(M_0\)-algebra and
    \(q: M_1 M_n X \to Q\) a morphism between \(M_0\)-algebras such
    that \(q \circ M_1 \mu^{0, n}_X = q \circ \mu^{1, 0}_{M_n X}\).
    We now have to show that there exists a unique morphism
    \(\bar{q}: M_{1 + n} X \to Q\) with
    \(\bar{q} \circ \mu^{1, n}_X = q\).

    So let \(\bar{q}\) be defined as \(\bar{q}([t]_{1 + n}) = q([s^n_X(t)]_1)\).
    We will show that this definition satisfies our requirements:
    \begin{enumerate}[wide]
        \item \label{itm:coequalizerequivariance}
            If \(\pi \in \Perm\) and \(t \in \Term_{\sig, 1 + n}(X)\), then
            \begin{align*}
                \bar{q}([\pi t]_{1 + n})
                &= q([s^n_X(\pi t)]_1) && \text{definition of } \bar{q} \\
                &= q([\pi s^n_X(t)]_1) && \text{\cref{thm:splitequivariance}} \\
                &= q(\pi[s^n_X(t)]_1) && \text{\cref{lem:equivariance}} \\
                &= \pi q([s^n_X(t)]_1) && \text{equivariance of } q \\
                &= \pi \bar{q}([t]_{1 + n}) && \text{definition of } \bar{q}.
            \end{align*}

            Furthermore, it follows by equivariance of \(s^n_X\), the canonical projection, and \(q\), that
            \[
                \supp(\bar{q}([t]_{1 + n})) = \supp(q([s^n_X(t)]_1)) \subseteq \supp(t).
            \]
        \item \label{itm:coequalizerwelldefinedness}
            \(\bar{q}\) is well-defined; i.e., if \(X \vdash_{1 + n} t = u\) is derivable, then \(\bar{q}([t]_{1 + n}) = \bar{q}([u]_{1 + n})\).
            We will show this by induction on the derivation.
            \begin{itemize}
                \item
                    \textit{For (refl):}
                    This would imply that \(t = x\) for some \(x \in X\), however \(t\) has uniform depth \(1 + n\), whereas \(x\) has uniform depth \(0\).
                \item
                    \textit{For (symm):}
                    We know that \(X \vdash_{1 + n} u = t\) is derivable and, by inductive hypothesis, we have \(\bar{q}([u]_{1 + n}) = \bar{q}([t]_{1 + n})\).
                \item
                    \textit{For (trans):}
                    We know that \(X \vdash_{1 + n} t = v\) and \(X \vdash_{1 + n} v = u\) are derivable for some \(v\).
                    Thus, by inductive hypothesis, we have \(\bar{q}([t]_{1 + n}) = \bar{q}([v]_{1 + n}) = \bar{q}([u]_{1 + n})\).
                \item
                    \textit{For (cong):} First, let $f\in\sigp$.
                    We know that \(t = f(t_1, \ldots, t_p)\), \(u = f(u_1, \ldots, u_p)\), and \(X \vdash_m t_i = u_i\) is derivable for all \(i \in \{1, \ldots, p\}\) with \(1 + n = m + d(f)\).
                    Since the operations are at most depth-1, we only have to consider two cases:

                    If \(d(f) = 0\), we have
                    \begin{align*}
                        &\bar{q}([t]_{1 + n}) \\
                        =~&q([s^n_X(f(t_1, \ldots, t_p))]_1) && \text{definition of } \bar{q} \\
                        =~&q([f(s^n_X(t_1), \ldots, s^n_X(t_p))]_1) && \text{definition of } s^n_X \\
                        =~&h([f(q([s^n_X(t_1)]_1), \ldots, q([s^n_X(t_p)]_1))]_0) && \text{\cref{thm:coequalizerproperties} \eqref{itm:coequalizerpropertiesdepth0}} \\
                        =~&h([f(\bar{q}([t_1]_{1 + n}), \ldots, \bar{q}([t_p]_{1 + n}))]_0) && \text{definition of } \bar{q},
                    \end{align*}
                    and similarly
                    \[
                        \bar{q}([u]_{1 + n}) = h([f(\bar{q}([u_1]_{1 + n}), \ldots, \bar{q}([u_p]_{1 + n})]_0).
                    \]
                    The equality then follows by inductive hypothesis.

                    If \(d(f) = 1\), then we have
                    \begin{align*}
                        &\bar{q}([t]_{1 + n}) \\
                        =~&q([s^n_X(f(t_1, \ldots, t_p))]_1) && \text{definition of } \bar{q} \\
                        =~&q([f([t_1]_n, \ldots, [t_p]_n)]_1) && \text{definition of } s^n_X \\
                        =~&q([f([u_1]_n, \ldots, [u_p]_n)]_1) && \text{assumption} \\
                        =~&q([s^n_X(f(u_1, \ldots, u_p))]_1) && \text{definition of } s^n_X \\
                        =~&\bar{q}([u]_{1 + n}) && \text{definition of } \bar{q}.
                    \end{align*}
                    The cases for $f\in\sigf$ and $f\in\sigb$ are analogous.
                \item
                    \textit{For (ax):}
                    We know that \(Y \vdash_m r = s \in E\) with \(t = (\tau r)\sigma\) and \(u = (\tau s)\sigma\) for a derivably equivariant substitution \(\sigma: Y \to \Term_{\sig, l}(X)\) and a permutation \(\tau \in \Perm\).
                    Since we have assumed \(T\) to be depth-1, we have to consider two cases:

                    If \(r\) and \(s\) are depth-0, then \(l = 1 + n\).
                    In this case, let \(\bar{\sigma}: Y \to \Term_{\sig, 0}(Q)\) be defined as
                    \[
                        \bar{\sigma}(x) = \bar{q}([\sigma(x)]_{1 + n}) = q([s^n_X(\sigma(x))]_1).
                    \]
                    It then follows by \cref{thm:coequalizerproperties} \eqref{itm:substitutionlemmadepth1} that
                    \begin{align*}
                        \bar{q}([t]_{1 + n}) &= q([s^n_X((\tau r)\sigma)]_1) = h([(\tau r)\bar{\sigma}]_0) \quad \text{and} \\
                        \bar{q}([u]_{1 + n}) &= q([s^n_X((\tau s)\sigma)]_1) = h([(\tau s)\bar{\sigma}]_0).
                    \end{align*}
                    Finally, we can show \(Q \vdash_0 (\tau r)\bar{\sigma} = (\tau s)\bar{\sigma}\) using (ax$_{r = s}$) by showing that \(\bar{\sigma}\) is derivably equivariant:
                    Let \(\pi \in \Perm\) and \(x \in Y\).
                    It then follows by inductive hypothesis that
                    \begin{align*}
                        \pi\bar{\sigma}(x)
                        &= \pi\bar{q}([\sigma(x)]_{1 + n}) && \text{definition of } \bar{\sigma} \\
                        &= \bar{q}([\pi \sigma(x)]_{1 + n}) && \text{\eqref{itm:coequalizerequivariance}} \\
                        &= \bar{q}([\sigma(\pi x)]_{1 + n}) && \text{inductive hypothesis} \\
                        &= \bar{\sigma}(\pi x) && \text{definition of } \bar{\sigma}.
                    \end{align*}

                    If \(r\) and \(s\) are depth-1, then \(l = n\).
                    In this case, let \(\bar{\sigma}: Y \to \Term_{\sig, 0}(M_n X)\) be defined as \(\bar{\sigma}(x) = [\sigma(x)]_n\).
                    It then follows by \cref{thm:coequalizerproperties} \eqref{itm:substitutionlemmadepth1} that
                    \begin{align*}
                        \bar{q}([t]_{1 + n}) &= q([s^n_X((\tau r)\sigma)]_1) = q([(\tau r)\bar{\sigma}]_1) \quad \text{and} \\
                        \bar{q}([u]_{1 + n}) &= q([s^n_X((\tau s)\sigma)]_1) = q([(\tau s)\bar{\sigma}]_1).
                    \end{align*}
                    We can then derive \(Q \vdash_0 (\tau r)\bar{\sigma} = (\tau s)\bar{\sigma}\) by once again showing that \(\bar{\sigma}\) is derivably equivariant:
                    Let \(\pi \in \Perm\) and \(x \in Y\).
                    Then \(\pi\bar{\sigma}(x) = \pi[\sigma(x)]_n = [\pi \sigma(x)]_n = [\sigma(\pi x)]_n\) by assumption.
                \item
                    \textit{For (perm):}
                    We know that \(t = \nu a.f(t_1, \ldots, t_p)\) and \(u = \nu b.f(u_1, \ldots, u_p)\) where \(a \neq b\), \(a \fresh u_i\), and \(X \vdash_m t_i = (a\ b)u_i\) is derivable for all \(i \in \{1, \ldots, p\}\).
                    Since the operations are at most depth-1, we only have to consider two cases:

                    If \(d(f) = 0\), we have
                    \begin{align*}
                        \bar{q}([t]_{1 + n}) &= h([\nu a.f(\bar{q}([t_1]_{1 + n}), \ldots, \bar{q}([t_p]_{1 + n}))]_0) \quad \text{and} \\
                        \bar{q}([u]_{1 + n}) &= h([\nu b.f(\bar{q}([u_1]_{1 + n}), \ldots, \bar{q}([u_p]_{1 + n}))]_0)
                    \end{align*}
                    analogously to (cong$_f$).
                    Furthermore, for \(i \in \{1, \ldots, p\}\), we have
                    \begin{align*}
                        \bar{q}([t_i]_{1 + n})
                        &= \bar{q}([(a\ b)u_i]_{1 + n} && \text{inductive hypothesis} \\
                        &= (a\ b) \bar{q}([u_i]_{1 + n}) && \text{\eqref{itm:coequalizerequivariance}},
                    \end{align*}
                    and also \(a \notin \supp(\bar{q}([u_i]_{1 + n})) \subseteq \supp(u_i)\) by \eqref{itm:coequalizerequivariance}.
                    Thus, the equality follows by (perm).

                    If \(d(f) = 1\), we have
                    \begin{align*}
                        \bar{q}([t]_{1 + n}) &= \bar{q}([\nu a.f([t_1]_{n}, \ldots, [t_p]_{n})]_1) \quad \text{and} \\
                        \bar{q}([u]_{1 + n}) &= \bar{q}([\nu b.f([u_1]_{n}, \ldots, [u_p]_{n})]_1)
                    \end{align*}
                    analogously to (cong$_f$).
                    Furthermore, for \(i \in \{1, \ldots, p\}\), we have \([t_i]_n = [(a\ b)u_i]_n = (a\ b)[u_i]_n\) and \(a \notin \supp([u_i]_n) \subseteq \supp(u_i)\) by assumption.
                    The result then follows by (perm).
            \end{itemize}
        \item \label{itm:coequalizermorphism}
            \(\bar{q}: (M_{1 + n} X, \mu^{0, 1 + n}_X) \to (Q, h)\) is indeed a morphism between \(M_0\)-algebras:
            First, note that equivariance follows from \eqref{itm:coequalizerequivariance}.
            We will also have to show that \(h \circ M_0 \bar{q} = \bar{q} \circ \mu^{0, 1 + n}_X\).
            So let \([t]_0 \in M_0 M_{1 + n} X\).
            Fix a splitting \(\sigma: M_{1 + n} X \to \Term_{\sig, 1 + n}(X)\), in that \([-]_{1 + n} \circ \sigma = \id\).
            Let \(\bar{\sigma}: M_{1 + n} X \to \Term_{\sig, 0}(Q)\) be defined as \(\bar{\sigma}(x) = \bar{q}(x) = \bar{q}([\sigma(x)]_{1 + n}) = q([s^n_X(\sigma(x))]_1)\).
            It then follows that
            \begin{align*}
                h(M_0 \bar{q}([t]_0))
                &= h([t\bar{\sigma}]_0) && \text{\cref{def:inducedmonad} because } \sigma_{\bar{q}} = \bar{\sigma} \\
                &= q([s^n_X(t\sigma)]_1) && \text{\cref{thm:coequalizerproperties} \eqref{itm:substitutionlemmadepth0}} \\
                &= \bar{q}([t\sigma]_{1 + n}) && \text{definition of } \bar{q} \\
                &= \bar{q}(\mu^{0, 1 + n}_X([t]_0)) && \text{\cref{thm:multiplicationsplitting}}.
            \end{align*}
        \item \label{itm:coequalizercorrect}
            \(\bar{q}\) satisfies the given property; i.e., \(\bar{q} \circ \mu^{1, n}_X = q\).
            So let \([t]_1 \in M_1 M_n X\).

            We fix a splitting \(\sigma: M_n X \to \Term_{\sig, n}(X)\), in that \([-]_n \circ \sigma = \id\).
            Let \(\bar{\sigma}: M_n X \to \Term_{\sig, 0}(M_n X)\) be defined as \(\bar{\sigma}(x) = [\sigma(x)]_1 = x\).

            It then follows that
            \begin{align*}
                \bar{q}(\mu^{1, n}_X([t]_1))
                &= \bar{q}([t\sigma]_{1 + n}) && \text{\cref{thm:multiplicationsplitting}} \\
                &= q([s^n_X(t\sigma)]_1) && \text{definition of } \bar{q} \\
                &= q([t\bar{\sigma}]_1) && \text{\cref{thm:coequalizerproperties} \eqref{itm:substitutionlemmadepth1}} \\
                &= q(M_1 \id([t]_1)) && \text{\cref{def:inducedmonad} because } \sigma_{\id} = \bar{\sigma} \\
                &= q([t]_1) && \text{functoriality of } M_1.
            \end{align*}
        \item \label{itm:coequalizerunique}
            If there is another such morphism \(q'\), then \(q'([t]_1) = \bar{q}([t]_1)\) for every \(t \in \Term_{\sig, 1}(M_n X)\).

            So assume that \(q': M_{1 + n} X \to Q\) is a morphism between \(M_0\)-algebras such that
            \begin{gather}
                q' \circ \mu^{1, n}_X = q. \label{eq:qprimehasproperty}
            \end{gather}

            We will proceed by induction on \(t\), only considering cases where \(t\) has at least uniform depth 1.
            \begin{itemize}
                \item
                    \textit{For \(t = f(t_1, \ldots, t_p)\) with \(f/p \in \sigp\), \(t_1, \ldots, t_p \in \Term_{\sig, m}(X)\), and \(1 + n = m + d(f)\):}
                    Since all operations are at most depth-1, we only have to consider two cases:

                    If \(d(f) = 0\), then
                    \begin{align*}
                        &q'([f(t_1, \ldots, t_p)]_{1 + n}) \\
                        =~&q'(\mu^{0, 1 + n}([f([t_1]_1, \ldots, [t_p]_1)]_0)) && \text{\cref{thm:multiplicationwithzero}} \\
                        =~&h(M_0 q'([f([t_1]_1, \ldots, [t_p]_1)]_0)) && \text{homomorphy} \\
                        =~&h([f(q'([t_1]_1), \ldots, q'([t_p]_1)]_0) && \text{\cref{def:inducedmonad}}.
                    \end{align*}
                    With a similar argument, we have
                    \[
                        \bar{q}([f(t_1, \ldots, t_p)]_{1 + n}) = h([f(\bar{q}([t_1]_1), \ldots, \bar{q}([t_p]_1))]_0)
                    \]
                    The equality then follows by inductive hypothesis.

                    If \(d(f) = 1\), then
                    \begin{align*}
                        &q'([f(t_1, \ldots, t_p)]_{1 + n}) \\
                        =~&q'(\mu^{1, n}_X([f([t_1]_n, \ldots, [t_p]_n)]_1)) && \text{\cref{thm:multiplicationwithzero}} \\
                        =~&q([f([t_1]_n, \ldots, [t_p]_n)]_1) && \text{\cref{eq:qprimehasproperty}} \\
                        =~&q([s^n_X(f(t_1, \ldots, t_p)]_1) && \text{definition of } s^n_X \\
                        =~&\bar{q}([f(t_1, \ldots, t_p)]_{1 + n}) && \text{definition of } \bar{q}.
                    \end{align*}
                \item
                    \textit{For \(t = a.f(t_1, \ldots, t_p)\) and \(t = \nu a.f(t_1, \ldots, t_p)\):}
                    Analogous to the previous case.
            \end{itemize}
    \end{enumerate}

\end{proof}

\subsection*{Details for \cref{sec:local-freshness}}
Write $\Sigma^{\trc}$ for the signature obtained from $\Sigma^\Br$ by
removing~$0$ and~$+$, and $T^\trc$ for the empty theory
over~$\Sigma^\trc$. Then, pretraces correspond to (uniform-depth)
terms in $\Sigma^\trc$. In this view, we have the following
characterization of $\alpha$-equivalence of pretraces:

\begin{lemma}\label{lem:pretrace-alpha}
  Pretraces $w,v\in\barNames^n\times X$ are $\alpha$-equivalent iff
  $X\vdash_n w=v$ is derivable in~$T^\trc$.
\end{lemma}

\begin{proof}
  \emph{'If':}
  We will proceed by induction on the derivation of \(X \vdash_n w = v\) to show a stronger statement:
  For all \(u\in \barNames^*\), \(uw \alphaeq uv\), especially for \(u = \epsilon\).
  \begin{itemize}
  \item \emph{For (refl):}
    We know that \(w = v = x \in X\).
    By reflexivity, \(uw = ux \alphaeq ux = uv\).
  \item \emph{For (symm):}
    We know that \(X \vdash_n v = w\) is derivable and, by inductive hypothesis, \(uv \alphaeq uw\).
    The statement then follows by symmetry.
  \item \emph{For (trans):}
    We know that \(X \vdash_n w = x\) and \(X \vdash_n x = v\) are derivable for some \(x\).
    By inductive hypothesis and transitivity, we have \(uw \alphaeq ux \alphaeq uv\).
  \item \emph{For (cong):}
    We know that \(w = aw'\) and \(v = av'\) for some \(a\in \barNames\).
    By applying the inductive hypothesis with \(\tilde{u} = ua\), it follows that \(uw = \tilde{u}w' \alphaeq \tilde{u}v' = uv\).
  \item \emph{For (perm):}
    We know that \(w = \newletter{a}w'\) and \(v = \newletter{b}v'\), where \(a \fresh v'\) and \(X \vdash_{m} w' = (a\ b)v'\) is derivable.
    By applying the inductive hypothesis with \(\tilde{u} = u\newletter{a}\), we have \(uw = u\newletter{a}w' \alphaeq u\newletter{a}((a\ b)v')\).
    Furthermore, we have \(\bind{a}((a\ b)v') = \bind{b}v'\) and, by definition, \(u\newletter{a}((a\ b)v') \alphaeq u\newletter{b}v'\).
    The statement then follows by transitivity of \(\alphaeq\).
  \end{itemize}

  \noindent\emph{'Only if':}
  Assume that \(w \alphaeq w'\). Since derivable equality is an equivalence, we only have to consider the base case \(w = u\newletter{a}v\) and \(w' = u\newletter{a'}v'\) with \(\bind{a}v = \bind{a'}v'\) in \([\Names]\,\barNames^*\), where \(u\in \barNames^m\) and \(v, v' \in \barNames^k\).

  If \(a = a'\), then \(v = v'\) and we are done because derivable equality is reflexive.
  Consider the case \(a \neq a'\) with \(a\fresh v'\) and \(v = (a\ a')v'\).
  By reflexivity, we have \(X \vdash_k v = (a\ a')v'\).
  By applying (perm), we also obtain \(X \vdash_{k + 1} \newletter{a}v = \newletter{a'}v'\).
  Finally, by applying (cong) $m$~times, we have \(X \vdash_{m + 1 + k} u\newletter{a}v = u\newletter{a'}v'\).
\end{proof}
We will also sometimes use the following standard characterization of \(\alphaeq\):

\begin{lemma}\label{lem:inductive-alphaeq}
  Let \(w, v \in \barNames^* \times X\) be traces. Then \(t \alphaeq v\) is derivable iff either
  \begin{enumerate}
    \item\label{itm:inductive-alphaeq-variable} \(w = v = x\) for some \(x \in X\),
    \item\label{itm:inductive-alphaeq-free} \(w = aw'\) and \(v = av'\) with \([w']_\alpha = [v']_\alpha\),
    \item\label{itm:inductive-alphaeq-bound} \(w = \newletter{a}w'\) and \(v = \newletter{b}v'\) with \(\bind{a}[w']_\alpha = \bind{b}[v']_\alpha\).
  \end{enumerate}
\end{lemma}

\begin{proof}
  By \cref{lem:pretrace-alpha}, we can use derivability in \(T^\trc\) instead of the definition of \(\alphaeq\).

  \noindent\emph{'If':}
  We consider all of the given cases.
  \begin{enumerate}
    \item
      We can derive \(x \alphaeq x\) by (refl).
    \item
      We know that \(w' \alphaeq v'\) and can thus derive \(aw' \alphaeq av'\) by (cong).
    \item
      If \(a = b\), then we have \(w' \alphaeq v'\) and \(aw' \alphaeq av'\) follows by (cong).

      If \(a \neq b\), we have \(a \fresh [v']_\alpha\) and \(w \alphaeq (a\ b) \cdot v'\).
      Since \(a \fresh [v']_\alpha\), we can pick some \(u'\) with \(u' \alphaeq v'\) and \(a \fresh u'\) (and, since $\alphaeq$ is equivariant, also \((a\ b) \cdot v' \alphaeq (a\ b) \cdot u'\)).
      It then follows by (perm) and (cong) that $\newletter{a}w' \alphaeq \newletter{b}u' \alphaeq \newletter{b}v'$.
  \end{enumerate}

  \noindent\emph{'Only if':}
  By induction on the derivation of \(X \vdash_n w = v\).
  \begin{itemize}
    \item \emph{For (refl):}
      We know that \(w = v = x \in X\).
      Thus, \eqref{itm:inductive-alphaeq-variable} follows immediately.
    \item \emph{For (symm):}
      We know that \(X \vdash_n v = w\) is derivable.
      Since the conditions for all cases are symmetric, we know that one of them must hold by inductive hypothesis.
    \item \emph{For (trans):}
      We know that, for some \(u \in \barNames^n \times X\), \(X \vdash_n w = u\) and \(X \vdash_n u = v\) are derivable.
      By inductive hypothesis for \(w \alphaeq u\), we only need to consider the following cases:
      \begin{enumerate}
        \item
          \emph{If \(w = u = x\) with \(x \in X\):}
          By inversion on the inductive hypothesis for \(u \alphaeq v\), we know \(u = v = x\).
          Thus, \eqref{itm:inductive-alphaeq-variable} follows immediately.
        \item
          \emph{If \(w = aw'\) and \(u = au'\) with \([w']_\alpha = [u']_\alpha\):}
          By inversion, we know that \(v = av'\) with \([u']_\alpha = [v']_\alpha\).
          Thus, \eqref{itm:inductive-alphaeq-free} holds.
        \item
          \emph{If \(w = \newletter{a}w'\) and \(u = \newletter{b}u'\) with \(\bind{a}[w']_\alpha = \bind{b}[u']_\alpha\).}
          By inversion, we know that \(v = \newletter{c}v'\) with \(\bind{b}[u']_\alpha = \bind{c}[v']_\alpha\).
          Thus, we have \(\bind{a}[w']_\alpha = \bind{c}[v']_\alpha\) and \eqref{itm:inductive-alphaeq-bound} holds.
      \end{enumerate}
    \item
      \emph{For (cong) with \(w = aw'\) and \(v = av'\):}
      We know that \(w' \alphaeq v'\) and thus \eqref{itm:inductive-alphaeq-free} holds.
    \item
      \emph{For (cong) with \(w = \newletter{a}w'\) and \(v = \newletter{a}v'\):}
      We know that \(w' \alphaeq v'\).
      It follows that \(\bind{a}[w']_\alpha = \bind{a}[v']_\alpha\) and thus \eqref{itm:inductive-alphaeq-bound} holds.
    \item
      \emph{For (perm):}
      We know that \(w = \newletter{a}w'\) and \(v = \newletter{b}v'\), where \(a \fresh v'\) and \(w' \alphaeq (a\ b)v'\).
      It then follows that \(a \fresh [v']_\alpha\) and thus \(\bind{a}[w']_\alpha = \bind{b}[v']_\alpha\).
      Thus, \eqref{itm:inductive-alphaeq-bound} holds.
  \end{itemize}
\end{proof}

\subsection*{Proof of \cref{thm:global-freshness-monad}}

We define the following $(\Sigma^\Br, \omega)$-model $A$.
\begin{align*}
A_n = M^\Br_n X, \qquad & 0_{A,n} = \emptyset, \\
& {+}_{A,n}(S, V) = S \cup V, \\
& \pre_{A,n}(a,S)=\{[aw]_\alpha\mid [w]_\alpha\in S\}, \\
& \abs_{A,n}(\bind{a}S)=\{[\newletter aw]_\alpha\mid[w]_\alpha\in S\}
\end{align*}

Note that \(\abs\) is indeed well-defined:
Assume \(\bind{a}S = \bind{a'}S'\) with \(a \neq a'\) and thus \(a \fresh S'\) and \(S' = (a\ a')S\).
We will only show \(\abs_{A, n}(\bind{a}S) \subseteq \abs_{A, n}(\bind{a'}S')\) as the other inclusion is completely analogous.
So let \([\newletter{a}w]_\alpha \in \abs_{A, n}(\bind{a}S)\) with \([w]_\alpha \in S\).
By equivariance of \(\alphaeq\), we have \([(a\ a')w]_\alpha \in S'\) and, by definition, \([\newletter{a'}((a\ a')w)]_\alpha \in \abs_{A, n}(\bind{a'}S')\).
Furthermore, since \(S'\) is finite (and thus ufs), we have \(a \notin \supp([w]_\alpha) \subseteq \supp(S')\).
It then follows from \cref{lem:inductive-alphaeq} that \(\newletter{a'}((a\ a')w) \alphaeq \newletter{a}w\).

Equivariance of $0_{A,n}$ and $+_{A,n}$ are trivial; for $\pre_{A,n}$ (and analogously for $\abs_{A,n}$) observe that
\begin{align*}
    \pi \pre_{A,n}(a,S) &= \pi \{[aw]_\alpha\mid [w]_\alpha\in S\} \\
    &= \{[\pi(a)\pi w]_\alpha\mid [w]_\alpha\in S\} \\
    &= \{[\pi(a) w]_\alpha\mid [w]_\alpha\in \pi S\} \\
    &= \pre_{A,n}(\pi a, \pi S)
\end{align*}

It is straightforward to see that A is a $(T^\Br,\omega)$-model. The join-semilattice axioms already hold by the structure of \(\pow_\omega\). Let \(\kappa: X \to A_k\) be an equivariant valuation and $x,y \in X$. For \eqref{ax:distributivity-pre} (and analogously for \eqref{ax:distributivity-abs}) we have
    \begin{align*}
      \sem{a(x + y)}^\kappa_1
      &= \{ [aw]_\alpha \mid [w]_\alpha \in \kappa(x) \cup \kappa(y) \} \\
      &= \{ [aw]_\alpha \mid [w]_\alpha \in \kappa(x) \} \cup \{ [aw]_\alpha \mid [w]_\alpha \in \kappa(y) \} \\
      &= \sem{ax + ay}^\kappa_1,
    \end{align*}
    and
    \begin{equation*}
      \sem{a 0}^\kappa_n = \{ [aw]_\alpha \mid [w]_\alpha \in \emptyset \} = \emptyset = \sem{0}^\kappa_n.
    \end{equation*}

We define a map $\mathsf{nf}_n \colon \Term_{\Sigma,n}(X) \to \Term_{\Sigma,n}(X)$ to normalize terms by pushing the join-semilattice operations up (and disposing of $0$) via the distributive equations in a derivable way, e.g. $\mathsf{nf}_n(a(x + y)) = \mathsf{nf}_n(ax) + \mathsf{nf}_n(ay)$. Normalized terms have the form
\[\sum_{i \in I} v_i, \qquad v_i \in T^\trc \] (where the empty sum is equal to $0$).
We also define the map $\sem{\cdot}^\iota_n \colon \Term_{\Sigma,n}(X) \to M^\Br_n X$ via the valuation $\iota(x) = \{[x]_\alpha\}$.
%\begin{align*}
%    \iota_n(x) &= \{[x]_\alpha\} \\
%    \iota_n(0) &= \emptyset = 0_{A,n} \\
%    \iota_n(t + s) &= \{\iota_n(t), \iota_n(s)\} = {+}_{A,n} (\iota_n(t), \iota_n(s)) \\
%    \iota_n(at) &= \{[av]_\alpha \mid [v]_\alpha \in \iota_n(t)\} = \pre_{A,n}(a,\iota_n(t)) \\
%    \iota_n(\newletter at) &= \{[\newletter av]_\alpha \mid [v]_\alpha \in \iota_n(t)\} = \abs_{A,n}(a,\iota_n(t))
%\end{align*}

Combining $\sem{\cdot}^\iota_n$ and $\mathsf{nf}_n$ for $n \in \Nat$ induces a homomorphism $\iota^\sharp\colon FX \to A$, where $\iota^\sharp_n([t]_n) = \sem{\cdot}^\iota_n(\mathsf{nf}_n(t))$.
Well-definedness is given by a canonical choice of input for $\sem{\cdot}^\iota_n$. Homomorphy is also preserved, as the equations to check are analogous to those of $(\Sigma^\Br, \omega)$-model $A$. The map $\iota^\sharp$ is surjective as for any set of (depth-n) pretraces there is a straightforward representation as a normalized term. It is also injective, as due to \cref{lem:pretrace-alpha} we have a correspondence between normalized terms modulo derivable equivalence and sets of $\alpha$-equivalent pretraces. As such, $\iota^\sharp([t]_n) = \iota^\sharp([s]_n)$ results in the same set of pretraces, which implies they share the same derivable normalized term, so $[t]_n = [s]_n$.
\qed

\subsection*{Proof of \cref{thm:global-freshness}}

First we prove the following statement
\begin{equation*}
    \gamma^{(n)}(x) = \{[w]_\alpha \mid [wy]_\alpha \in (\barNames^n\times X) / {\alphaeq}, \, x \xrightarrow{w} y\} \qquad (n > 0)
\end{equation*}

\noindent
Through induction on $n$, for $a \in \barNames$ in the base case
\begin{align*}
    &[a]_\alpha \in \gamma^{(1)}(x) \\
    \Leftrightarrow~&[a]_\alpha \in (\mu^{1,0}_X \cdot M_1\gamma^{(0)} \cdot \gamma)(x) && \text{definition} \\
    \Leftrightarrow~&[a]_\alpha \in (\mu^{1,0}_X \cdot M_1M_0! \cdot M_1\eta_X \cdot \gamma)(x) && \text{definition} \\
    \Leftrightarrow~&[a]_\alpha \in (M_1! \cdot \gamma)(x) && \text{naturality, monad laws} \\
    \Leftrightarrow~&\exists y.\, [ay]_\alpha \in \gamma(x) && \text{definition} \\
    \Leftrightarrow~&\exists y,a',y'.\, ay \alphaeq a'y' \text{ and } x \xrightarrow{a'} y' && \text{definition}\\
\end{align*}
and in the induction step
\begin{align*}
    &[aw]_\alpha \in \gamma^{(n)}(x) \\
    \Leftrightarrow~&[aw]_\alpha \in (\mu^{1,n-1}_X \cdot M_1\gamma^{(n-1)} \cdot \gamma)(x) && \text{definition} \\
    \Leftrightarrow~&[aw]_\alpha \in \{[bv]_\alpha \mid [by]_\alpha \in \gamma(x),\, [v]_\alpha \in \gamma^{(n-1)}(y)\}&& \text{definition} \\
    \Leftrightarrow~&\exists y.\, [ay]_\alpha \in \gamma(x),\, [w]_\alpha \in \gamma^{(n-1)}(y)&& \text{definition} \\
    \Leftrightarrow~&\exists y,z.\, x \xrightarrow{a} y,\, y \xrightarrow{w} z && \text{induction} \\
    \Leftrightarrow~&\exists z.\, x \xrightarrow{aw} z && \text{definition}
\end{align*}
where the last two equivalences hold up to $\alpha$-equivalence (c.f. the last line of the base case, omitted in favour of readability).

To prove the theorem, it suffices to show the following as a consequence of injectivity of the global freshness operator $N$.
\begin{equation*}
  L(A,x) = L(A,y) \Leftrightarrow \forall n.\,\gamma^{(n)}(x) = \gamma^{(n)}(y)
\end{equation*}
This follows from straightforward induction on $n$ via the previously shown equality.
\qed

\subsection*{Proof of \cref{thm:local-freshness}}

We give the outline of the proof, relying on several key lemmas
(\cref{thm:semanticsmodelismodel,thm:languageinterpretationofpretraces,thm:localfreshnessinjectivity})
for which we give proofs, in turn involving dedicated lemmas, in the
subsequent sections.

The local freshness semantics is computed from the bar language
semantics by applying the~$D$ operator, which takes all
$\alpha$-equivalent representatives of bar strings and then drops the
bars (\cref{sec:rnna}). We can extend~$D$ to pretraces and sets of
pretraces in the evident manner. In order to show coincidence of the
graded semantics $(\monad^\loc,\theta_1\cdot\beta)$ with local
freshness semantics, it then suffices to show that $D$ and~$\theta$
make the same identifications on~$M_n^\Br X$.

To show this, we define a $(T^\drop, \omega)$-model $F'X$ (despite the notation, $F'$ will not be functorial) into which the free model~$FX$ embeds along a subalgebra inclusion~$e$.
We will then show that, for the canonical transformation
\begin{equation*}
  \kappa_n\colon M_n^\loc \to M_n^\drop,
\end{equation*}
$e_n \cdot \kappa_n \cdot \theta_n$ makes the same identifications as~$D$. % ; in fact, it will coincide with~$D$ on length-$n$ pretraces\lsnote{Hard to see how this types}.
Since we will show that $e_n \cdot \kappa_n$ is already injective, this also means that $\theta$ makes the same identifications as~$D$.

As the first step in this programme, we take $F'X$ to have carriers
\begin{equation*}
  (F' X)_0 = \powf(X), \qquad (F' X)_{n+1} = \{f \colon \Names \to_\fs (F' X)_n \mid (\star)\},
\end{equation*}
(recall that $\Names \to_\fs (F' X)_n$ is the set of finitely
supported maps from~$\Names$ to $(F'X)_n$) where $(\star)$ requires
that, for all $a \in \Names$ and $c \fresh (f,a)$, we have
\begin{equation}
  f(a) = f(a) + (a\ c) \cdot \res_{F' X, n}(\bind{a} f(c)). \label{eq:restriction}
\end{equation}
This ensures a form of closure under restriction that allows unblocking blocked $\alpha$-renamings in the same spirit as the method of name dropping~\cite{SchroderEA17}.
We use the usual (nominal) join-semilattice structure at depth~$0$ (e.g.~$+$ is union) and let $0_{F'X, m+1}$, $+_{F'X, m+1}$ be given by
\begin{equation*}
  0_{F'X, m+1}(b) = 0 \quad\text{and}\quad +_{F'X, m+1}(f_1, f_2)(b) = f_1(b) + f_2(b)
\end{equation*}
Furthermore, we add name restriction by
\begin{align*}
  \res_{F' X, 0}(\bind{a} S) &= \{x \in S \mid a \fresh x\}, \\
  \res_{F' X, m+1}(\bind{a} f)(b) &= \res_{F' X, m}(\bind{c} ((a\ c) \cdot f)(b)) \text{ for } c\fresh(f,a,b),
\end{align*}
and interpret the depth-$1$ operations by
\begin{align*}
  \pre_{F' X, m}(a, x)(b) & =
                            \begin{cases}
                              x & a = b\\
                              0 & \text{otherwise, and}
                            \end{cases}
  \\
  \abs_{F' X, m}(\bind{a} x)(b) & = \res_{F' X, m}(\bind{c} (b\ c) \cdot (a\ c) \cdot x) \text{ for } c\fresh(x,a,b).
\end{align*}
The above definition of the restriction operation matches that of the name restriction operation defined on finitely supported functions given in Pitts \cite{Pitts13}, while the abstraction operation resembles the definition of $\alpha$-equivalence.
The requirement $(\star)$ is necessary to ensure that the traces for an atom in the support include all of the relevant traces for fresh atoms, which is used to prove axiom \eqref{ax:namerestrictionsemilattice} (2).
\begin{lemma}\label{thm:semanticsmodelismodel}
  The $(\Sigma^\drop,\omega)$-algebra $F'X$ is a
  $(T^\drop,\omega)$-model.
\end{lemma}

\noindent By freeness, we thus obtain a morphism $e\colon FX\to F'X$.
Furthermore, by the obvious isomorphism $(\powfs(Y)^X)_\fs \cong \powfs(X \times Y)$ for nominal sets $X, Y$, we can define an injective operator $\overline{(-)} \colon (F'X)_n \to \powfs(\Names^n \times X)$ that "flattens" our carrier set into a set of pretraces.
\begin{lemma}\label{thm:languageinterpretationofpretraces}
  For $S\in M_n^\Br X$, we have $\overline{e_n(\kappa_n(\theta_n(S)))}=D(S)$.
\end{lemma}
% \begin{proof}[Proof (Sketch)]
%   Since $e_n\cdot\kappa_n\cdot\theta_n$ is homomorphic w.r.t.~the join-semilattice
%   structure of $M_n^\Br X$, it suffices to show the claim for the case
%   where~$S$ is a singleton $S=\{[t]_\alpha\}$ for a pretrace~$t$. This
%   case is proved by induction on the length of~$t$.
% \end{proof}
\noindent The remaining last step in our programme is injectivity
of~$e_n \cdot \kappa_n$:
\begin{lemma}\label{thm:localfreshnessinjectivity}
  For all components $e_n : (FX)_n \to (F'X)_n$ of the $(\Sigma^\drop,\omega)$-homomorphism $e\colon FX \to F'X$, $e_n \cdot \kappa_n$ is injective.
\end{lemma}
% \begin{proof}[Proof (Sketch)]
%   Writing $X\vdash_n s\le t$ to abbreviate $X\vdash_n t+s=t$ for terms
%   $s,t\in\Term_{\Sigma^\Br,n}(X)$, we show more generally that
%   whenever $D(s)\subseteq D(t)$, then $X\vdash_n s\le t$ in $T^\loc$,
%   which implies the claim by
%   \cref{thm:languageinterpretationofpretraces}. Since the~$e_n$ are
%   join-semilattice homomorphisms, this reduces immediately to the case
%   that $s$ is a pretrace. This case is then proved by induction
%   on~$s$, crucially using the axioms~\eqref{ax:loc} in the inductive
%   step for $s$ of the form $s=aw$.
% \end{proof}

\subsection*{Details for \cref{thm:semanticsmodelismodel}}

First, note that all operations preserve the condition stated in \eqref{eq:restriction} by the observation that the condition does not depend on the choice of \(c\):

\begin{lemma}
  Let \(X\) be a nominal set with a name restriction operation \([\Names]X \to X, \bind{a} x \mapsto a \setminus x\), \(f\colon \Names \to_\fs X\), and \(a,b \in \Names\).
  For \(c,c'\fresh(f,a,b)\), we have \(c \setminus ((a\ c) \cdot f)(b) = c' \setminus ((a\ c') \cdot f)(b)\).
\end{lemma}

\begin{proof}
  If \(b = a\), it follows by computation that
  \begin{align*}
    &c \setminus ((a\ c) \cdot f)(a) \\
    =~&c \setminus ((a\ c) \cdot (c\ c') \cdot f)(a) && \text{because } c,c'\fresh f \\
    =~&c \setminus (a\ c) \cdot (c\ c') \cdot f(c') && \text{definition} \\
    =~&(a\ c) \cdot (c\ c') \cdot a \setminus f(c') && \text{equivariance} \\
    =~&(a\ c') \cdot a \setminus f(c') && \text{by } (\star) \\
    =~&c' \setminus (a\ c') \cdot f(c') && \text{equivariance} \\
    =~&c' \setminus ((a\ c') \cdot f)(a) && \text{definition},
  \end{align*}
  where \((\star)\) follows by the fact that \((a\ c) \cdot (c\ c') \cdot d = (a\ c') \cdot d\) for all \(d \in \supp(a \setminus f(c'))\).

  If \(b \neq a\), we have
  \begin{align*}
    &c \setminus ((a\ c) \cdot f)(b) \\
    =~&c \setminus (a\ c) \cdot f(b) && \text{definition} \\
    =~&(a\ c) \cdot a \setminus f(b) && \text{equivariance} \\
    =~&a \setminus f(b) && \text{because } a,c\fresh(a \setminus f(b)),
  \end{align*}
  and similarly, \(d \setminus ((a\ d) \cdot f)(b) = a \setminus f(b)\).
\end{proof}

One can then verify that the definitions of \(\abs\) and \(\res\) also do not depend on the choice \(c\).

Most of the depth-$0$ axioms are easy to verify, we will restrict ourselves to the most intersting cases for $(F'X)_{n + 1}$:

\begin{itemize}
  \item For \eqref{ax:namerestrictionsemilattice}:
    First, we have to show that \(\res_{F'X, n+1}(\bind{a} f + g) = \res_{F'X, n+1}(\bind{a} f) + \res_{F'X, n+1}(\bind{a} g)\) for all \(a,f,g\).

    Let \(b \in \Names\) and pick a \(c \in \Names \setminus \supp(f,g,a,b)\).
    We then have
    \begin{align*}
      &\res_{F'X,n+1}(\bind{a} f + g)(b) \\
      =~&\res_{F'X,n}(\bind{c} ((a\ c) \cdot (f + g))(b)) && \text{definition} \\
      =~&\res_{F'X,n}(\bind{c} (a\ c) \cdot (f + g)((a\ c) \cdot b)) && \text{definition} \\
      =~&(a\ c) \cdot \res_{F'X,n}(\bind{c} f((a\ c) \cdot b) + g((a\ c) \cdot b)) && \text{equivariance} \\
      =~&(a\ c) \cdot \res_{F'X,n}(\bind{c} f((a\ c) \cdot b)) + \\&(a\ c) \cdot \res_{F'X,n}(\bind{c} g((a\ c) \cdot b)) && \text{\eqref{ax:namerestrictionsemilattice} for } (F'X)_n \\
      =~&\res_{F'X,n+1}(\bind{a} f)(b) + \res_{F'X,n+1}(\bind{a} g)(b) && \text{definition}.
    \end{align*}

    We also have to show that \(f = f + \res_{F'X, n+1}(\bind{a} f)\) for all \(a,f\).
    So let \(b \in \Names\).

    If \(a = b\), this follows directly from the definition of \((F'X)_{n + 1}\).

    If \(a \neq b\), pick \(c \in \Names \setminus \supp(f,a,b)\).
    \begin{align*}
      f(b)
      &= f(b) + \res_{F'X,n}(\bind{a} f(b)) && \text{\eqref{ax:namerestrictionsemilattice} for } (F'X)_n \\
      &= f(b) + (a\ c) \cdot \res_{F'X, n}(\bind{a} f(b)) && \text{because } a,c\fresh\bind{a}f(b) \\
      &= f(b) + \res_{F'X, n}(\bind{c} (a\ c) \cdot f(b)) && \text{equivariance} \\
      &= f(b) + \res_{F'X, n}(\bind{c} ((a\ c) \cdot f)(b)) && \text{because } (a\ c) \cdot b = b \\
      &= f(b) + \res_{F'X,n+1}(\bind{a}f)(b) && \text{definition}.
    \end{align*}
\end{itemize}

For the depth-$1$ axioms, we will use the following lemma:

\begin{lemma}\label{lem:abs-eq}
  For \(a \in \Names\) and \(x \in (F'X)_n\), we have \(\abs_{F'X, n}(\bind{a} x)(a) = x\).
\end{lemma}

\begin{proof}
  For sufficiently fresh \(c\), we have
  \begin{align*}
    &\abs_{F'X, n}(\bind{a} x)(a) \\
    =~&\res_{F'X, n'}(\bind{c}(a\ c) \cdot (a\ c) \cdot x) && \text{definition} \\
    =~&\res_{F'X, n'}(\bind{c} x) && \text{because } (a\ c) \cdot (a\ c) = \id \\
    =~&x && \text{axioms, because } c\fresh x.
  \end{align*}
\end{proof}

Again, we will restrict ourselves to the most interesting axioms:

\begin{itemize}
  \item For \eqref{ax:respre} (2):
    We have to show that \(\res_{F'X,n+1}(\bind{a} \pre_{F'X,n}(a,x)) = 0_{F'X,n+1}\) for all \(a,x\).
    So let \(b \in \Names\).

    Pick \(c \in \Names \setminus \supp(x, a, b)\).
    We then have \((a\ c) \cdot b \neq a\) by case distinction on whether \(b = a\).

    It then follows immediately that
    \begin{align*}
      &\res_{F'X,n+1}(\bind{a} \pre_{F'X,n}(a,x))(b) \\
      =~&\res_{F'X,n}(\bind{c} (a\ c) \cdot \pre_{F'X,n}(a,x))((a\ c) \cdot b) && \text{definition} \\
      =~&\res_{F'X,n}(\bind{c} (a\ c) \cdot 0_{F'X,n}) && \text{because } (a\ c) \cdot b \neq a \\
      =~&0_{F'X,n} = 0_{F'X,n+1}(b) && \text{definition}.
    \end{align*}
  \item For \eqref{ax:resabs}:
    We have to show that \(\abs_{F'X,n}(\bind{a} \res_{F'X, n}(\bind{b} x)) = \res_{F'X, n+1}(\bind{b} \abs_{F'X,n}(\bind{a} x))\) for all \(a,b,x\) with \(a \neq b\).
    So let \(e \in \Names\).

    If \(b = e\), pick \(c \in \Names \setminus \supp(x,a,b)\) and \(d \in \Names \setminus \supp(x,a,b,c)\).
    It then follows that
    \begin{align*}
      &\abs_{F'X,n}(\bind{a} \res_{F'X, n}(\bind{b} x))(b) \\
      =~&\res_{F'X, n}(\bind{c} (b\ c) \cdot (a\ c) \cdot \res_{F'X,n}(\bind{b} x)) && \text{definition} \\
      =~&(b\ c) \cdot (a\ c) \cdot \res_{F'X,n}(\bind{b} \res_{F'X,n}(\bind{b} x)) && \text{equivariance} \\
      =~&(b\ c) \cdot (a\ c) \cdot \res_{F'X,n}(\bind{b} x) && \text{axioms} \\
      =~&(b\ c) \cdot (a\ c) \cdot (c\ d) \cdot \res_{F'X,n}(\bind{b} x) && \text{because } c,d\fresh x \\
      =~&(b\ c) \cdot (c\ d) \cdot (a\ d) \cdot \res_{F'X,n}(\bind{b} x) && \text{computation} \\
      =~&(b\ c) \cdot \res_{F'X,n}(\bind{b} (c\ d) \cdot (a\ d) \cdot x) && \text{equivariance} \\
      =~&(b\ c) \cdot \res_{F'X,n}(\bind{b} (c\ d) \cdot \res_{F'X,n}(\bind{c} (a\ d) \cdot x)) && \text{axioms} \\
      =~&(b\ c) \cdot \res_{F'X,n}(\bind{b} (c\ d) \cdot \abs_{F'X,n}(\bind{a} x)(c)) && \text{definition} \\
      =~&\res_{F'X, n}(\bind{c} (b\ c) \cdot \abs_{F'X,n}(\bind{a} x)(c)) && \text{equivariance} \\
      =~&\res_{F'X, n}(\bind{c} ((b\ c) \cdot \abs_{F'X,n}(\bind{a} x))(b)) && \text{definition} \\
      =~&\res_{F'X,n+1}(\abs_{F'X,n}(\bind{a} x))(b) && \text{definition}.
    \end{align*}

    If \(b \neq e\), pick \(c \in \Names \setminus \supp(x,a,b,e)\).
    We then have
    \begin{align*}
      &\abs_{F'X,n}(\bind{a} \res_{F'X, n}(\bind{b} x))(e) \\
      =~&\res_{F'X,n}(\bind{c} (e\ c) \cdot (a\ c) \cdot \res_{F'X, n}(\bind{b} x)) && \text{definition} \\
      =~&(e\ c) \cdot \res_{F'X,n}(\bind{e} \res_{F'X,n}(\bind{b} (a\ c) \cdot x)) && \text{equivariance} \\
      =~&(e\ c) \cdot \res_{F'X,n}(\bind{b} \res_{F'X,n}(\bind{e} (a\ c) \cdot x)) && \text{axioms} \\
      =~&\res_{F'X,n}(\bind{b} (e\ c) \cdot \res_{F'X,n}(\bind{e} (a\ c) \cdot x)) && \text{equivariance} \\
      =~&\res_{F'X,n}(\bind{b} \abs_{F'X,n}(\bind{a} x)(e)) && \text{definition} \\
      =~&\res_{F'X,n+1}(\bind{b} \abs_{F'X,n}(\bind{a} x))(e) && \text{because } b \neq e.
    \end{align*}
  \item For \eqref{ax:loc}:
    We have to show that \(\abs_{F'X,n}(\bind{a}x) = \abs_{F'X,n}(\bind{a}x) + \pre_{F'X,n}(a,x).\)

    This follows directly from the definition of \(\pre\) and \cref{lem:abs-eq}.
\end{itemize}

\subsection*{Proof of \cref{thm:languageinterpretationofpretraces}}

Using the characterization of \(\alphaeq\) given in \cref{lem:inductive-alphaeq}, we first obtain the following properties of \(D\):

\begin{lemma}\label{lem:generated-language-prefix}
  Let \(t' \in \barNames^{n} \times X\), \(w' \in \Names^n \times X\), and \(a, b \in \Names\).
  \begin{enumerate}
    \item \label{itm:generated-language-prefix-free}
      \(bw' \in D([at']_\alpha)\) iff \(w' \in D([t']_\alpha)\) and \(a = b\).
    \item \label{itm:generated-language-prefix-bound-equal}
      \(aw' \in D([\newletter{a}t']_\alpha)\) iff \(w' \in D([t']_\alpha)\).
    \item \label{itm:generated-language-prefix-bound-different}
      If \(a \neq b\), then \(bw' \in D([\newletter{a}t']_\alpha)\) iff \(b \fresh [t']_\alpha\) and \(w' \in D([(a\ b)t']_\alpha)\).
  \end{enumerate}
\end{lemma}

\begin{proof}
  We will show each statement individually.
  \begin{enumerate}
    \item
      \emph{'If':}
      Assume \(w' \in D([t']_\alpha)\) and \(a = b\).

      Then there exists a \(\tilde{w}' \in [t']_\alpha\) with \(\ub(\tilde{w}') = w'\).
      Since \(\tilde{w}' \alphaeq t'\), we have \(a\tilde{w}' \alphaeq at'\) by \autoref{lem:inductive-alphaeq} \eqref{itm:inductive-alphaeq-free}.
      Thus, it follows that \(aw' = (a, \ub(\tilde{w}')) = \ub(a\tilde{w}') \in D([at']_\alpha)\).

      \emph{'Only if':}
      Assume \(bw' \in D([at']_\alpha)\).

      Then there exists a \(\tilde{w} \in [at']_\alpha\) with \(\ub(\tilde{w}) = bw'\).
      Since \(\tilde{w} \alphaeq at'\), we know from \autoref{lem:inductive-alphaeq} that \(\tilde{w} = a\tilde{w}'\) for some \(\tilde{w}' \in \barNames^n \times X\) with \(\tilde{w}' \alphaeq t'\).
      Thus, we have \(\tilde{w}' \in [t']_\alpha\).
      Furthermore, we have \((a, \ub(\tilde{w}')) = \ub(a\tilde{w}') = \ub(\tilde{w}) = bw'\), concluding that \(a = b\) and \(w' = \ub(\tilde{w}') \in D([t']_\alpha)\).
    \item
      Analogous to the above proof.
    \item
      \emph{'If':}
      Assume \(w' \in D([(a\ b)t']_\alpha\) and \(b \fresh [t']_\alpha\).

      Then there exists a \(\tilde{w}' \in [(a\ b)t']_\alpha\) with \(\ub(\tilde{w}') = w'\).
      Since \(\tilde{w}' \alphaeq (a\ b)t'\) and \(b \fresh [t']_\alpha\) with \(a \neq b\), it follows from \autoref{lem:inductive-alphaeq} \eqref{itm:inductive-alphaeq-bound} that \(\newletter{b}\tilde{w}' \alphaeq \newletter{a}t'\).
      Thus, we have
      \begin{equation*}
        bw' = (b, \ub(\tilde{w}')) = \ub(\newletter{b}\tilde{w}') \in D([\newletter{a}t']_\alpha).
      \end{equation*}

      \emph{'Only if':}
      Assume \(bw' \in D([\newletter{a}t']_\alpha)\).

      Then there exists some \(\tilde{w} \in [\newletter{a}t']_\alpha\) with \(\ub(\tilde{w}) = bw'\).
      Since \(\tilde{w} \alphaeq \newletter{a}t'\), we know from \autoref{lem:inductive-alphaeq} that \(\tilde{w} = \newletter{c}\tilde{w}'\) for some \(\tilde{w}' \in \barNames^n \times X\) and \(c \in \Names\).
      Since \((c, \ub(\tilde{w}')) = \ub(\newletter{c}\tilde{w}') = \ub(\tilde{w}) = bw'\), we know that \(c = b\) and \(\ub(\tilde{w}') = w'\).
      Since \(b \neq a\) and \(\newletter{b}\tilde{w}' \alphaeq \newletter{a}t'\), it follows from \autoref{lem:inductive-alphaeq} that \(b \fresh [t']_\alpha\) and \(\tilde{w}' \alphaeq (a\ b)t'\).
      Finally, it follows that \(w' = \ub(\tilde{w}') \in D([(a\ b)t']_\alpha)\).
  \end{enumerate}
\end{proof}

\begin{lemma}\label{lem:generated-language-structure}
  Let \(t, u \in \barNames^{n + 1} \times X\).
  If \(D([t]_\alpha) = D([u]_\alpha)\), then one of the following is true:
  \begin{enumerate}
    \item \label{itm:generated-language-structure-free}
      \(t = at'\) and \(u = au'\) for some \(a \in \Names\) and \(t', u' \in \barNames^n \times X\), or
    \item \label{itm:generated-language-structure-bound}
      \(t = \newletter{a}t'\) and \(u = \newletter{b}u'\) for some \(a, b \in \Names\) and \(t', u' \in \barNames^n \times X\).
  \end{enumerate}
\end{lemma}

\begin{proof}
  We will consider the following exhaustive list of cases for \(t, u\):
  \begin{itemize}
    \item \emph{For \(t = at'\) and \(u = bu'\) with \(t', u' \in \barNames^{n'} \times X\):}
      By definition, \(\ub(at') \in D([t]_\alpha) = D([u]_\alpha)\).
      It then follows by \cref{lem:generated-language-prefix} \eqref{itm:generated-language-prefix-free} that \(a = b\).
      Thus, \eqref{itm:generated-language-structure-free} holds.
    \item \textit{For \(t = \newletter{a}t'\) and \(u = bu'\) with \(t', u' \in \barNames^{n'} \times X\):}
      Pick a \(c \in \Names \setminus \supp(t', a, b)\).
      Then we have \(\newletter{a}t' \alphaeq \newletter{c} ((a\ c)t')\) and, by definition, \(\ub(c((a\ c)t')) \in D([\newletter{a}t']_\alpha) = D([bu']_\alpha)\).
      However, by \cref{lem:generated-language-prefix} \eqref{itm:generated-language-prefix-free}, this would imply \(c = b\), contradicting the assumption.
    \item \textit{For \(t = at'\) and \(u = \newletter{b}u'\) with \(t', u' \in \barNames^{n'} \times X\):}
      Analogous to the above case.
    \item \textit{For \(t = \newletter{a}t'\) and \(u = \newletter{b}u'\) with \(t', u' \in \barNames^{n'} \times X\):}
      Then \eqref{itm:generated-language-structure-bound} holds.
  \end{itemize}
\end{proof}

\begin{lemma}\label{lem:generated-language-injectivity}
  The operator \(D\) is injective for single pretraces: If \(D([t]_\alpha) = D([u]_\alpha)\) for \(t, u \in \barNames^n \times X\), then \([t]_\alpha = [u]_\alpha\).
\end{lemma}

\begin{proof}
  By induction on \(n\).
  \begin{itemize}
      \item \textit{For \(t = x\) with \(x \in X\):}
        Let \(u = y \in X\).
        By \cref{lem:inductive-alphaeq} \eqref{itm:inductive-alphaeq-variable}, we have
        \begin{equation*}
          D([t]_\alpha) = \{\ub(x)\} = \{x\} = \{y\} = \{\ub(y)\} = D([u]_\alpha).
        \end{equation*}
        Thus, we have \(x = y\).
      \item \textit{For \(t = at'\) with \(t' \in \barNames^{n'} \times X\):}
        We know from \cref{lem:generated-language-structure} that \(u = au'\) for some \(u'\).
        First, observe that \(D([t']_\alpha) = D([u']_\alpha)\):
        \begin{align*}
          &w' \in D([t']_\alpha) \\
          \Leftrightarrow~&aw' \in D([at']_\alpha) && \text{\cref{lem:generated-language-prefix} \eqref{itm:generated-language-prefix-free}} \\
          \Leftrightarrow~&aw' \in D([au']_\alpha) && \text{assumption} \\
          \Leftrightarrow~&w' \in D([u']_\alpha) && \text{\cref{lem:generated-language-prefix} \eqref{itm:generated-language-prefix-free}}.
        \end{align*}

        It then follows by inductive hypothesis that \(t' \alphaeq u'\) and thus \(t = at' \alphaeq au' = u\).
      \item \textit{For \(t = \newletter{a}t'\) with \(t' \in \barNames^{n'} \times X\):}
        We know from \cref{lem:generated-language-structure} that \(u = \newletter{b}u'\) for some \(u' \in \barNames^{n'} \times X\) and \(b \in \Names\).

        If \(a = b\), one can show that \(t = \newletter{a}t' \alphaeq \newletter{a}u' = u\) analogously to the above case.
        So assume \(a \neq b\).

        First note that \(\ub(\newletter{a}t') \in D([t]_\alpha) = D([\newletter{b}u']_\alpha)\).
        By \cref{lem:generated-language-prefix} \eqref{itm:generated-language-prefix-bound-different}, we have \(a \fresh [u']_\alpha\).
        By \cref{lem:inductive-alphaeq} \eqref{itm:inductive-alphaeq-bound} then suffices to show \([t']_\alpha = (a\ b) [u']_\alpha\) to conclude \(\newletter{a}t' \alphaeq \newletter{b}u'\).

        First, observe that
        \begin{align*}
          &w' \in D([t']_\alpha) \\
          \Rightarrow~&aw' \in D([\newletter{a}t']_\alpha) && \text{\cref{lem:generated-language-prefix} \eqref{itm:generated-language-prefix-bound-equal}} \\
          \Rightarrow~&aw' \in D([\newletter{b}u']_\alpha) && \text{assumption} \\
          \Rightarrow~&w' \in D([(a\ b)u']_\alpha) && \text{\cref{lem:generated-language-prefix} \eqref{itm:generated-language-prefix-bound-different}},
        \end{align*}
        and conversely,
        \begin{align*}
          &w' \in D([(a\ b)u']_\alpha) \\
          \Rightarrow~&(a\ b)w' \in D([u']_\alpha) && \text{equivariance} \\
          \Rightarrow~&b((a\ b)w')) \in D([\newletter{b}u']_\alpha) && \text{\cref{lem:generated-language-prefix} \eqref{itm:generated-language-prefix-bound-equal}} \\
          \Rightarrow~&b((a\ b)w') \in D([\newletter{a}t']_\alpha) && \text{assumption} \\
          \Rightarrow~&(a\ b)w' \in D([(a\ b)t']_\alpha) && \text{\cref{lem:generated-language-prefix} \eqref{itm:generated-language-prefix-bound-different}} \\
          \Rightarrow~&w' \in D([t']_\alpha) && \text{equivariance}.
        \end{align*}
        The statement then follows by inductive hypothesis.
  \end{itemize}
\end{proof}

\noindent As seen in \cref{thm:global-freshness-monad}, a set \(S \in M^\Br_n\) can be seen in \(T^\Br\) as a big join of all pretraces in \(S\).
By the join-semilattice structure of \(F' X\) and the interpretation of \(+\), we thus have
\begin{equation*}
  \overline{e_n(\kappa_n(\theta_n(S)))} = \bigcup_{[t]_\alpha \in S} \overline{e_n([t]_n)}.
\end{equation*}

As a prerequisite, we will characterize the behavior of the restriction operation for such terms:
\begin{lemma}\label{lem:generated-language-restriction}
  For any \(t \in \barNames^n \times X\) and \(b \in \FN(t)\), we have \(\res_{F'X, n}(\bind{b} e_n([t]_n)) = 0_{F'X, n}\).
\end{lemma}

\begin{proof}
  By induction on \(n\).
  \begin{itemize}
    \item \emph{For \(t = x \in X\):}
      By definition, we have
      \begin{align*}
        \res_{F'X, 0}(\bind{b} e_0([x]_0))
        &= \res_{F'X, 0}(\bind{b} \{x\}) \\
        &= \{x \in \{x\} \mid b \fresh x\} \\
        &= \emptyset = 0_{F'X, 0}.
      \end{align*}
    \item \emph{For \(t = at'\) with \(t' \in \barNames^{n'} \times X\):}
      We know that \(b \in \FN(at') = \FN(t') \cup \{a\}\).

      If \(b = a\), then
      \begin{align*}
        &\res_{F'X, n}(\bind{b} e_n([at]_n)) \\
        =~&\res_{F'X, n}(\bind{a} \pre_{F'X, n'}(a, e_{n'}([t']_{n'}))) && \text{definition} \\
        =~&0_{F'X, n} && \text{axioms}.
      \end{align*}

      If \(b \neq a\) (and thus \(b \in \FN(t')\)), we have
      \begin{align*}
        &\res_{F'X, n}(\bind{b} e_n([at]_n)) \\
        =~&\res_{F'X, n}(\bind{b} \pre_{F'X, n'}(a, e_{n'}([t']_{n'}))) && \text{definition} \\
        =~&\pre_{F'X, n'}(a, \res_{F'X, n'}(\bind{b} e_{n'}([t']_{n'}))) && \text{axioms} \\
        =~&\pre_{F'X, n'}(a, 0_{F'X, n'}) && \text{induction} \\
        =~&0_{F'X, n} && \text{axioms}.
      \end{align*}
    \item \emph{For \(t = \newletter{a}t'\) with \(t' \in \barNames^{n'} \times X\):}
      We know that \(b \in \FN(\newletter{a}t') = \FN(t') \setminus \{a\}\); i.e., \(b \in \FN(t')\) and \(b \neq a\).
      The argument is then analogous to the one in the previous case.
  \end{itemize}
\end{proof}

\noindent By definition of \(D\), it now suffices to show that \(\overline{e_n([t]_n)} = D([t]_\alpha)\) for \(t \in \barNames^n \times X\).

We will do this by induction on \(n\).
\begin{itemize}
  \item \emph{For \(t = x \in X\):}
    We know by \cref{lem:inductive-alphaeq} \eqref{itm:inductive-alphaeq-variable} that \([x]_\alpha = \{x\}\).
    It then follows that
    \begin{equation*}
      \overline{e_0([x]_0)} = \overline{\{x\}} = \{x\} = \{\ub(x)\} = D([x]_\alpha).
    \end{equation*}
  \item \emph{For \(t = at'\) with \(t' \in \barNames^{n'} \times X\):}
    \begin{align*}
      &\overline{e_n([at']_n)} \\
      =~&\{aw \mid w \in \overline{e_{n'}([t']_{n'})}\} && \text{definition of } \pre \\
      =~&\{aw \mid w \in D([t']_\alpha)\} && \text{induction} \\
      =~&D([at']_\alpha) && \text{\cref{lem:generated-language-prefix} \eqref{itm:generated-language-prefix-free}}.
    \end{align*}
  \item \emph{For \(t = \newletter{a}t'\) with \(t' \in \barNames^{n'} \times X\):}
    First, assume \(bv \in \overline{e_n([\newletter{a}t']_n)}\); i.e., \(v \in \overline{\abs_{F'X, n'}(\bind{a} e_{n'}([t']_{n'}))(b)}\).

    If \(a = b\), then we have
    \begin{align*}
      v &\in \overline{\abs_{F'X, n'}(\bind{a} e_{n'}([t']_{n'}))(a)} \\
      &= \overline{e_{n'}([t']_{n'})} && \text{\cref{lem:abs-eq}} \\
      &= D([t']_\alpha) && \text{induction}.
    \end{align*}
    By \cref{lem:generated-language-prefix} \eqref{itm:generated-language-prefix-bound-equal}, we thus have \(bv \in D([\newletter{a}t']_\alpha)\).

    If \(a \neq b\), we will first show that \(b \fresh e_{n'}([t']_{n'})\):
    Otherwise, we would have \(b \in \supp(e_{n'}([t']_{n'})) = \supp(D([t']_\alpha)) = \supp([t']_\alpha) = \FN(t')\) by inductive hypothesis and injectivity.
    Thus, for sufficiently fresh \(c\), we would obtain
    \begin{align*}
      v &\in \overline{\abs_{F'X, n'}(\bind{a} e_{n'}([t']_{n'}))(b)} \\
      &= \overline{\res_{F' X, n'}(\bind{c} (b\ c) \cdot (a\ c) \cdot e_{n'}([t']_{n'}))} && \text{definition} \\
      &= \overline{(b\ c) \cdot (a\ c) \cdot \res_{F' X, n'}(\bind{b} e_{n'}([t']_{n'}))} && \text{equivariance} \\
      &= \overline{(b\ c) \cdot (a\ c) \cdot 0_{F'X, n'}} && \text{\cref{lem:generated-language-restriction}} \\
      &= \overline{0_{F'X, n'}} = \emptyset.
    \end{align*}

    It then follows that
    \begin{align*}
      v &\in \overline{\abs_{F'X, n'}(\bind{a} e_{n'}([t']_{n'}))(b)} \\
      &= \overline{\abs_{F'X, n'}(\bind{b}(a\ b) \cdot e_{n'}([t']_{n'}))(b)} && \text{because } b \fresh e_{n'}([t']_{n'})  \\
      &= (a\ b) \cdot \overline{e_{n'}([t']_{n'})} && \text{\cref{lem:abs-eq}} \\
      &= (a\ b) \cdot D([t']_\alpha) && \text{induction} \\
      &= D([(a\ b) \cdot t']_\alpha) && \text{equivariance},
    \end{align*}
    and, by \cref{lem:generated-language-prefix} \eqref{itm:generated-language-prefix-bound-different}, \(bv \in D([\newletter{a}t']_\alpha)\).

    Conversely, assume \(bv \in D([\newletter{a}t']_\alpha)\).
    By \cref{lem:generated-language-prefix}, there exists some \(u' \in \barNames^{n'} \times X\) with \(v \in D([u']_\alpha)\) and \([\newletter{a}t']_\alpha = [\newletter{b}u']_\alpha\).
    It then follows that
    \begin{align*}
      v &\in D([u']_\alpha) \\
      &= \overline{e_{n'}([u']_{n'})} && \text{induction} \\
      &= \overline{\abs_{F'X, n'}(\bind{b} e_{n'}([u']_{n'}))(b)} && \text{\cref{lem:abs-eq}} \\
      &= \overline{\abs_{F'X, n'}(\bind{a} e_{n'}([t']_{n'}))(b)} && \text{assumption} \\
      &= \overline{e_n([\newletter{a}t']_n)(b)} && \text{definition}.
    \end{align*}
\end{itemize}\qed

\subsection*{Proof of \cref{thm:localfreshnessinjectivity}}

All of the following computations are done in \(T^\loc\) unless stated otherwise.

Given \(s, t \in \Term_{\Sigma^\Br, n}(X)\), we write \(X \vdash_n s \le t\) for \(X \vdash_n t + s = t\); intuitively, a term is "larger" than another one if adding that other term does not change anything.

\begin{lemma}\label{lem:derivable-inequality-equal}
  \(X \vdash_n s \le t\) and \(X \vdash_n t \le s\) iff \(X \vdash_n t = s\).
\end{lemma}

\begin{proof}
  \emph{'If':}
  By definition, we have \(X \vdash_n t + s = t\) and \(X \vdash_n s + t = s\).
  It then follows by symmetry of \(+\) that \(X \vdash_n t = s\).

  \noindent\emph{'Only if':}
  By idempotence of \(+\).

\end{proof}

For sum terms (with \(0\) being the neutral element), we have the following additional properties:

\begin{lemma}\label{lem:derivable-inequality-sums}
  Let \(T \in \powf(\Term_{\Sigma^\Br, n}(X))\) and \(s \in \Term_{\Sigma^\Br, n}(X)\). Then we have
  \begin{enumerate}
    \item\label{itm:derivable-inequality-sums-all}
      If \(X \vdash_n t \le s\) for every \(t \in T\), then \(X \vdash_n \sum_{t \in T} t \le s\).
    \item\label{itm:derivable-inequality-sums-some}
      If \(X \vdash_n s \le t\) for some \(t \in T\), then \(X \vdash_n s \le \sum_{t \in T} t\).
    \item\label{itm:derivable-inequality-sums-element}
      If \(s \in T\), then \(X \vdash_n s \le \sum_{t \in T} t\).
  \end{enumerate}
\end{lemma}

\begin{proof}
  By unfolding the definitions for \(\le\) and sums. \eqref{itm:derivable-inequality-sums-element} is a special case of \eqref{itm:derivable-inequality-sums-some} using \cref{lem:derivable-inequality-equal}.
\end{proof}

Furthermore, we have the following properties for (free and bound) prefixing:

\begin{lemma}\label{lem:derivable-inequality-prefix}
  Let \(t, s \in \Term_{\Sigma^\Br, n}(X)\) and \(u \in \Term_{\Sigma^\Br, n + 1}(X)\). Then we have
  \begin{enumerate}
    \item If \(X \vdash_n t \le s\) and \(X \vdash_{n + 1} as \le u\), then \(X \vdash_{n + 1} at \le u\).
    \item If \(X \vdash_n t \le s\) and \(X \vdash_{n + 1} \newletter{a}s \le u\), then \(X \vdash_{n + 1} \newletter{a}t \le u\).
  \end{enumerate}
\end{lemma}

\begin{proof}
  We will only show the first statement, the second one is completely analogous.
  By simple computation, we have
  \begin{align*}
    [u]_{n + 1}
    &= [u + as]_{n + 1} && \text{definition} \\
    &= [u + a(s + t)]_{n + 1} && \text{definition} \\
    &= [(u + as) + at]_{n + 1} && \text{axioms} \\
    &= [u + at]_{n + 1} && \text{definition}.
  \end{align*}
\end{proof}

The most important characterization, relating the generated language to terms, is the following one:

\begin{lemma}\label{lem:language-subset-inequality}
  Let \(w \in \barNames^n \times X\) and \(s \in \Term_{\Sigma^\Br, n}(X)\).
  If \(D([w]_\alpha) \subseteq \overline{e_n([s]_n)}\), then \(X \vdash_n w \le s\).
\end{lemma}

\begin{proof}
  Assume w.l.o.g. that \(s\) is in normal form; i.e., \(s = \sum_{v \in S} v\) for a set \(S \in \powf(\barNames^n \times X)\) of pretraces.

  Note that, by \cref{thm:languageinterpretationofpretraces} and the join-semilattice structure of \(F' X\), we have
  \begin{equation}\label{eq:interpretation-of-s}
    D([w]_\alpha) \subseteq \overline{e_n([s]_n)} = \bigcup_{v \in S} D([v]_\alpha).
  \end{equation}

  We will proceed by induction on \(n\).
  \begin{itemize}
      \item \emph{For \(w = x \in X\):}
        Since \(n = 0\), we know that \(S \subseteq X\).

        It follows from \cref{lem:inductive-alphaeq} that \(D([w]_\alpha) = \{ \ub(x) \} = \{x\}\).
        By \eqref{eq:interpretation-of-s}, it then follows that \(x \in D([v]_\alpha = \{v\})\) for some \(v \in S\); i.e., \(x \in S\).
        The statement then follows by \cref{lem:derivable-inequality-sums} \eqref{itm:derivable-inequality-sums-element}.
      \item \emph{For \(w = aw'\) with \(w' \in \barNames^{n'} \times X\):}
        We will define a new sum term
        \begin{equation*}
          s' :=
            \sum_{\substack{av' \in S}}^k v' +
            \sum_{\substack{\newletter{a}v' \in S}}^k v' +
            \sum_{\substack{\newletter{b}v' \in S \\ a \neq b, a \notin \FN(v')}}^k (a\ b)v'.
        \end{equation*}

        Note that, by \eqref{ax:distributivity-pre}, \(as'\) is derivably equal to a form where \(a\) is distributed over all terms of the sums.

        Next, we will show \(X \vdash_n as' \le s\) using \cref{lem:derivable-inequality-sums} \eqref{itm:derivable-inequality-sums-all} for all three sums in \(s'\).
        \begin{enumerate}
          \item \emph{For \(av' \in S\):}
            Then we already have \(X \vdash_n av' \le s\) by \cref{lem:derivable-inequality-sums} \eqref{itm:derivable-inequality-sums-element}.
          \item \emph{For \(\newletter{a}v' \in S\):}
            Analogous to the above case.
          \item \emph{For \(\newletter{b}v' \in S\) with \(a \neq b\) and \(a \notin \FN(v')\):}
            Since \(a \fresh [v']_\alpha\), we know that \(\newletter{b}v' \alphaeq \newletter{a}((a\ b)v')\) by \cref{lem:inductive-alphaeq} \eqref{itm:inductive-alphaeq-bound}.
            By \cref{lem:pretrace-alpha}, this equality also holds in \(T^\loc\).
            By \eqref{ax:loc}, we then have \(X \vdash_n a((a\ b)v') \le \newletter{a}((a\ b)v') = \newletter{b}v'\).
            Finally, we can conclude \(X \vdash_n a((a\ b)v' \le s\) using \cref{lem:derivable-inequality-sums} \eqref{itm:derivable-inequality-sums-some}.
        \end{enumerate}

        Furthermore, we will show that \(D([w']_\alpha) \subseteq \overline{e_{n'}([s']_{n'})}\).
        So let \(x \in D([w']_\alpha)\).
        By the join-semilattice structure of \(F' X\), it is enough to show that \(x \in D([u]_\alpha)\) for some summand \(u\) in \(s'\).
        By \cref{lem:generated-language-prefix} \eqref{itm:generated-language-prefix-free}, we know that \(ax \in D([aw']_\alpha)\) and, by \eqref{eq:interpretation-of-s}, \(ax \in D([v]_\alpha)\) for some \(v \in S\).
        We consider the cases for \(v\) as outlined in \cref{lem:generated-language-prefix}:

        If \(v = av'\) or \(v = \newletter{a}v'\), then \(x \in D([v']_\alpha)\) and, by definition, \(v'\) is a summand in \(s'\).

        If \(v = \newletter{b}v'\) with \(a \neq b\), then \(a \fresh [v']_\alpha\) and \(x \in D([(a\ b)v']_\alpha\).
        By definition, \((a\ b)v'\) is a summand in \(s'\).

        With this, we have \(X \vdash_{n'} w' \le s'\) by inductive hypothesis.
        Since \(X \vdash_n as' \le s\), we can then conclude \(X \vdash_n aw' \le s\) by \cref{lem:derivable-inequality-prefix}.
      \item \emph{For \(w = \newletter{a}w'\) with \(w' \in \barNames^{n'} \times X\):}
        Pick \(c \in \Names \setminus \FN(w')\) s.t. \(c \neq a\) and there is no \(v \in S\) with \(v = cv'\) or \(v = \newletter{c}v'\) (so there is no term prefixed with \(c\)).
        This is always possible because there are only \(k\) terms and thus only finitely many different prefixes used.
        Note that \(\newletter{a}w' \alphaeq \newletter{c}((a\ c)w')\).

        We will now define a new sum term
        \begin{equation*}
          s' := \sum_{\substack{\newletter{b}v' \in S \\ c \notin \FN(v')}}^k (b\ c)v'.
        \end{equation*}

        Note that, by \eqref{ax:distributivity-abs}, \(\newletter{c}s'\) is derivably equal to a form where \(\newletter{c}\) is distributed over all terms of the sum.

        Next, we will once again show \(X \vdash_n \newletter{c}s' \le s\) using \cref{lem:derivable-inequality-sums} \eqref{itm:derivable-inequality-sums-all} on \(s'\).
        So let \(\newletter{b}v' \in S\) with \(c \notin \FN(v')\), implying that \(c \fresh [v']_\alpha\).
        Since \(b \neq c\) by choice of \(c\), we have \(\newletter{b}v' \alphaeq \newletter{c}((b\ c)v')\).
        The inequality then follows by \cref{lem:derivable-inequality-equal} and \cref{lem:pretrace-alpha}.

        Furthermore, we will show that \(D([(a\ c)w']_\alpha) \subseteq \overline{e_{n'}([s']_{n'})}\).
        So let \(x \in D([(a\ c)w']_\alpha)\).
        By the join-semilattice structure of \(F' X\), it is once again enough to show that \(x \in D([u]_\alpha)\) for some summand \(u\) in \(s'\).
        By \cref{lem:generated-language-prefix} \eqref{itm:generated-language-prefix-bound-equal}, we know that \(cx \in D([\newletter{c}((a\ c)w')]_\alpha) = D([\newletter{a}w']_\alpha)\) and, by \eqref{eq:interpretation-of-s}, \(cx \in D([v]_\alpha)\) for some \(v \in S\).
        By \cref{lem:generated-language-prefix} and because there are no pretraces beginning with \(c\) or \(\newletter{c}\), we know that \(v = \newletter{b}v'\) with \(c \fresh [v']_\alpha\) and \(x \in D([(b\ c)v']_\alpha)\).
        By definition, \((b\ c)v'\) is a summand in \(s'\).

        With this, we have \(X \vdash_{n'} (a\ c)w' \le s'\) by inductive hypothesis.
        Since \(X \vdash_n \newletter{c}s' \le s\), we can then conclude \(X \vdash_n \newletter{a}w' = \newletter{c}((a\ c)w') \le s\) by \cref{lem:derivable-inequality-prefix}.
  \end{itemize}
\end{proof}

\noindent We now have to show that, for \(t, s \in \Term_{\Sigma^\Br, n}(X)\) with \(e_n([t]_m) = e_n([s]_n)\), we have \([t]_n = [s]_n\).
Assume w.l.o.g. that \(t = \sum_{w \in T} w\) for a set \(T \in \powf(\barNames^n \times X)\) of pretraces.
By \cref{lem:derivable-inequality-equal}, it suffices to show that \(X \vdash_n t \le s\) and \(X \vdash_n s \le t\).
We will only show the first inequality as the second one is completely symmetrical.
To achieve this, we will use \cref{lem:derivable-inequality-sums} \eqref{itm:derivable-inequality-sums-all}.

So let \(w \in T\).
By the join-semilattice structure of \(F' X\) and \cref{thm:languageinterpretationofpretraces}, we know that
\begin{equation*}
  D([w]_\alpha) = \overline{e_n([w]_n)} \subseteq \overline{e_n([t]_n)} = \overline{e_n([s]_n)}.
\end{equation*}
It then follows by \cref{lem:language-subset-inequality} that
\(X \vdash_n w \le rs\).  \qed

\subsection{Details for \cref{rem:namerestrictionmodel}}
\label{sec:tdrop-details}

Intuitively, we can think of $x|_N$ as the free restriction of $x$'s support to $N \subseteq \supp(x)$.
As such, we have the following properties:
\begin{lemma}
  \label{lem:restrictionproperties}
  Given a nominal set~$X$, let $x \in X$ and $N \subseteq \supp(x)$. Then we have:
  \begin{enumerate}
    \item \label{itm:restrictionsupp} The least support $\supp(x|_N)$ is exactly $N$.
    \item \label{itm:restrictionequivariant} For $\pi \in \Perm$, we have $\pi \cdot (x|_N) = (\pi \cdot x)|_{\pi \cdot N}$.
    \item \label{itm:restrictioneq} For $x' \in X$ and $N' \subseteq \supp(x')$, $x|_N = x'|_{N'}$ iff $N' = N$ and $x' = \pi \cdot x$ for some $\pi \in \Fix(N)$.
    \item \label{itm:restrictionsingle} We have $x|_N = \{y\}$ iff $x = y$ and $N = \supp(y)$.
  \end{enumerate}
\end{lemma}

\begin{proof}
  We show the statements individually:
  \begin{enumerate}
    \item Shown elsewhere \cite[Lemma 5.8]{SchroderEA17}.
    \item
      Let $\pi \cdot \tau \cdot x \in \pi \cdot (x|_N)$ for some $\tau \in \Fix(N)$.
      Then $\pi \cdot \tau \cdot x = \pi \cdot \tau \cdot \pi^{-1} \cdot \pi \cdot x \in (\pi \cdot x)|_{\pi \cdot N}$, because $\pi \cdot \tau \cdot \pi^{-1} \in \Fix(\pi \cdot N)$.

      Conversely, let $\tau \cdot \pi \cdot x \in (\pi \cdot x)|_{\pi \cdot N}$ for some $\tau \in \Fix(\pi \cdot N)$.
      Then $\tau \cdot \pi \cdot x = \pi \cdot \pi^{-1} \cdot \tau \cdot \pi \cdot x \in \pi \cdot (x|_N)$, because $\pi^{-1} \cdot \tau \cdot \pi \in \Fix(N)$.
    \item
      \emph{'If':} Obvious by definition of $x|_N$.

      \emph{'Only If':}
      By \eqref{itm:restrictionsupp}, we have $N = \supp(x|_N) = \supp(x'|_{N'}) = N'$.
      Furthermore, since $\pi = 1 \in \Fix(N')$, we have $x' \in x'|_{N'} = x|_N$.
      Thus, by definition, there exists a $\pi \in \Fix(N)$ with $x' = \pi \cdot x$.
    \item
      First, note that $y|_{\supp(y)} = \{y\}$, thus the 'if' case is immediately discharged.
      For 'only if', we also have $N = \supp(y)$ by \eqref{itm:restrictioneq}.
      Furthermore, for $\pi = \id \in \Fix(N)$, we have $x = \pi \cdot x \in x|_N = \{y\}$ and thus $x = y$.
  \end{enumerate}
\end{proof}

We can then make our claimed description into a $(T^\drop, 0)$-model $A$ with the operations defined as for $(F'X)_0$.
We can easily see that all operations preserve the requirement that the sets are closed under further name restriction.

Given a finite set $A = \{a_1, \dots a_n\} \subseteq \supp(x)$ where $a_1, \dots, a_n$ are pairwise distinct, we abbreviate the term
\begin{equation*}
  A \setminus x = \nu a_1.\res(\dots(\nu a_n.\res(x))),
\end{equation*}
where the order does not matter since we can swap name restrictions by axiom \eqref{ax:namerestriction}.

To prove the claimed description, first note that we can normalize terms as follows:
First, we can distribute name restrictions over joins using \eqref{ax:namerestriction} and \eqref{ax:namerestrictionsemilattice}.
We can then reduce every summand to a form $\nu a_1.\res(\dots (\nu a_n.\res(x)))$ where $a_1, \dots, a_n \in \supp(x)$ are pairwise distinct using \eqref{ax:namerestriction}.
Furthermore, we can add all further possible such restrictions for the summands using \eqref{ax:namerestrictionsemilattice}.

The result then follows using the following observation:
\begin{lemma}
  Let $x, y \in X$, $A \subseteq \supp(x)$, and $B \subseteq \supp(y)$.
  If $x|_{\supp(x) \setminus A} = y|_{\supp(y) \setminus B}$, then $X \vdash_0 A \setminus x = B \setminus y$ is derivable in $T^\drop$.
\end{lemma}
\begin{proof}
  We proceed by induction on $|A|$.
  \begin{itemize}
    \item
      For $A = \emptyset$:
      We have $y|_{\supp(y) \setminus B} = x|_{\supp(x)} = \{x\}$ and thus, by \cref{lem:restrictionproperties} \eqref{itm:restrictionsingle}, $y = x$ and $\supp(y) \setminus B = \supp(x)$ (i.e., $B = \emptyset$).
      The statement then follows by reflexivity.
    \item
      For $A = A' \cup \{a\}$ with $a\notin A'$: %$A = A' \cupdot \{a\}$:
      By \cref{lem:restrictionproperties} \eqref{itm:restrictioneq}, we know that $\supp(x) \setminus A = \supp(y) \setminus B$ and $y = \pi \cdot x$ for some $\pi \in \Fix(\supp(x) \setminus A)$.

      If $\pi(a) = a$, we have $a = \pi(a) \in \pi \cdot \supp(x) = \supp(y)$ by equivariance of $\supp$.
      Since $a \notin \supp(x) \setminus A = \supp(y) \setminus B$, we must have $a \in B$.
      Set $B' = B \setminus \{a\}$.
      Since $\supp(x) \setminus A' = \supp(y) \setminus B'$ and $\pi \in \Fix(\supp(x) \setminus A')$, we have $x|_{\supp(x) \setminus A'} = y|_{\supp(y) \setminus B'}$ by \cref{lem:restrictionproperties} \eqref{itm:restrictioneq}.
      It then follows by inductive hypothesis and the derivation rules that
      \begin{equation*}
        X \vdash_0 \nu a.\res(A' \setminus x) = \nu a.\res(B' \setminus y)
      \end{equation*}
      is derivable.

      If $\pi(a) \neq a$, then $\pi(a) \notin \supp(x) \setminus A = \supp(y) \setminus B$.
      Since $a \in A \subseteq \supp(x)$, we have $\pi(a) \in \pi \cdot \supp(x) = \supp(y)$ by equivariance of $\supp$, and thus $\pi(a) \in B$.
      Set $B' = B \setminus \{\pi(a)\}$.

      We are done once we have shown that
      \begin{equation*}
        X \vdash_0 \nu a.\res(A' \setminus x) = \nu \pi(a).\res(B' \setminus y).
      \end{equation*}
      Since $a \notin \supp(x) \setminus A = \supp(y) \setminus B$, we already have $a \notin \supp(y) \setminus B'$.
      Using the inductive hypothesis, it then suffices to show that
      \begin{equation*}
        x|_{\supp(x) \setminus A'} = ((a\ \pi(a)) \cdot y)|_{(a\ \pi(a)) \cdot (\supp(y) \setminus B')}
      \end{equation*}

      First, note that
      \begin{align*}
        &\supp(x) \setminus A' \\
        =~&(\supp(x) \setminus A) \cup \{a\} && \text{assumption} \\
        =~&(a\ \pi(a)) \cdot (\supp(x) \setminus A) \cup \{a\} && \text{by } a,\pi(a) \notin \supp(x) \setminus A \\
        =~&(a\ \pi(a)) \cdot ((\supp(x) \setminus A) \cup \{\pi(a)\}) && \text{equivariance} \\
        =~&(a\ \pi(a)) \cdot ((\supp(y) \setminus B) \cup \{\pi(a)\}) && \text{assumption} \\
        =~&(a\ \pi(a)) \cdot (\supp(y) \setminus B') && \text{assumption}.
      \end{align*}
      Furthermore, for $\tau = (a\ \pi(a)) \cdot \pi$, we have
      \begin{align*}
        &(a\ \pi(a)) \cdot y \\
        =~&(a\ \pi(a)) \cdot \pi \cdot x && \text{assumption} \\
        =~&\tau \cdot x && \text{definition of } \tau,
      \end{align*}
      and, since $\tau \in \Fix(\supp(x) \setminus A')$, the statement follows by \cref{lem:restrictionproperties} \eqref{itm:restrictioneq}.
  \end{itemize}
\end{proof}

\subsection{Details for \cref{sec:games}}

The functor $\mbar$ is described in purely categorical terms by taking
$\mbar(A,a)$, for an $M_0$-algebra $(A,a)$ with structure map
$a\colon M_0A\to A$, to be the coequalizer
\begin{equation*}
  \begin{tikzcd}
    M_1M_0A \arrow[shift left,"M_1a",r] \arrow[shift right,"\mu^{10}_A" below,r] & M_1 A\arrow[r] &  \mbar(A,a)
  \end{tikzcd}\qquad\text{(in the category of $M_0$-algebras).}
\end{equation*}
Equivalently, $\mbar (A,a)$ is described as follows. We have forgetful
functors $(-)_0$, $(-)_1$ from the category of $M_1$-algebras to the
category of $M_0$-algebras, sending an $M_1$-algebra
$((B_i),(b_{ij}))$ to $(B_0,b_{00})$ and $(B_1,b_{01})$,
respectively. Then, the free $M_1$-algebra $((B_i),(b_{ij}))$ w.r.t.\
$(-)_0$ over an $M_0$-algebra $(A,a)$ has $B_0=A$ and $b_{00}=a$;
applying $(-)_1$ to $((B_i),(b_{ij}))$ yields
$\mbar(A,a)$~\cite{FordEA22,ForsterEA25a}.

\subsection{Details for \cref{exp:m1-bar}}
\label{sec:m1-bar-details}

\noindent\emph{Global freshness:} Since, at depth 0, we have the
join-semilattice axioms, an \(M_0\)-algebra $A$ has an equivariant
join-semilattice structure.  Using the distributivity axioms and
(perm), depth-1 terms over $A$ can thus be normalized to the form
\(a_1x_1 + \dots + a_nx_n + \newletter{c}y\) for some sufficiently
fresh \(c \in \Names\) and \(x_1, \dots, x_n, y \in A\), requiring
that \(x_i \neq 0\) for all \(i\).  It follows that
\(\mbar A = A^{\Names}_\omega \times [\Names]A\).

\noindent\emph{Local feshness:} We prove the claimed description of \(\mbar\)
for local freshness semantics with name-dropping. As this description
makes~$\mbar$ a subfunctor of the functor $\Names \to_\fs (-)$, it
follows that~$\mbar$ preserves monos.

Let \((A, h_A)\) be an \(M_0\)-algebra.  First note that we can define
a \((\Sigma^\drop, 1)\)-model with \(A\) at depth~\(0\) and
\(\mbar(A, h_A)\) at depth \(1\).  The proof that this model actually
satisfies the axioms is the same as for
\cref{thm:semanticsmodelismodel}, observing that the proof for
\((F'X)_{n+1}\) does not depend on the actual structure of \((F'X)_n\)
and only uses the fact that \((F'X)_n\) is an \(M_0\)-algebra.
Since~$M_0A$ and $M_1A$ form the free $M_1$-algebra over the set~$A$,
we obtain in particular an $M_0$-homomorphism
\(m\colon M_1A \to \mbar(A,h_A)\) that acts by term evaluation.

Let \((Q, h_Q)\) be another \(M_0\)-algebra, and let \(q\colon M_1A \to Q\) be an $M_0$-homomorphism such that \(q \cdot M_1h_A = q \cdot \mu^{10}_A\).
We define \(m'\colon \mbar(A, h_A) \to Q\) by
\begin{equation*}\textstyle
  m'(f) = q\left( \left[ \sum_{a \in \supp(f)} a f(a) + \newletter{c} f(c) \right] \right) \quad \text{for } c \fresh f.
\end{equation*}
This is indeed an \(M_0\)-homomorphism: We will show that \(m'(f_1 + f_2) = m'(f_1) + m'(f_2)\), the proof for name restriction is largely analogous.
Since $q$ is an $M_0$-algebra morphism and thus distributes over $+$, it suffices to show that
\begin{gather*}
  \sum_{a \in \supp(f_1 + f_2)} a (f_1 + f_2)(a) + \newletter{c} (f_1 + f_2)(c) \\
  = \sum_{a \in \supp(f_1)} a f_1(a) + \newletter{c} f_1(c) + \sum_{a \in \supp(f_2)} a f_2(a) + \newletter{c} f_2(c)
\end{gather*}
is derivable from the axioms and the
assumption on $q$ that $q \cdot M_1h_A = q \cdot \mu^{10}_A$ which, algebraically speaking, allows us to collapse terms at depth $0$ into a single variable. For example, the term $x+y$ with variables $x,y \in A$ can be collapsed into a single variable $x+y \in A$, utilizing the join-semilattice structure of $A$.
Under these additional equalities, we will once again use \cref{lem:derivable-inequality-sums} and show that the
individual summands on each side are contained in the other side.  For
the bound summands, this is clear because $+$ distributes over
$\newletter{c}$.  For free summands, this reduces to the following cases:
\begin{itemize}
  \item For terms \(af_1(a)\) with \(a \in \supp(f_1) \cap \supp(f_1 + f_2)\), we have
    \begin{align*}
      af_1(a)
      &\le a(f_1(a) + f_2(a)) && \text{axioms} \\
      &= a(f_1 + f_2)(a) && \text{assumption on $q$}.
    \end{align*}
  \item For terms \(af_1(a)\) with \(a \in \supp(f_1) \setminus \supp(f_1 + f_2)\) and sufficiently fresh \(c\), we have
    \begin{align*}
      &af_1(a) \\
      \le~&a(f_1 + f_2)(a) && \text{axioms, assumption on $q$} \\
      =~&a(\nu c.\res((f_1 + f_2)(a))) && \text{axioms, } c\fresh(f_1 + f_2)(a) \\
      =~&a(\nu c.\res(((a\ c) \cdot (f_1 + f_2))(a))) && \text{because } a,c\#(f_2 + f_2) \\
      =~&a(\nu c.\res((a\ c) \cdot (f_1 + f_2)(c))) && \text{definition} \\
      =~&a((a\ c) \cdot \nu a.\res((f_1 + f_2)(c))) && \text{equivariance} \\
      \le~&\newletter{a}((a\ c) \cdot \nu a.\res((f_1 + f_2)(c))) && \text{axioms} \\
      =~&\newletter{c}(\nu a.\res((f_1 + f_2)(c))) && \text{derivation rules} \\
      \le~&\newletter{c}(f_1 + f_2)(c) && \text{axioms},
    \end{align*}
    notably using axiom \eqref{ax:namerestrictionsemilattice} in the last step.
  \item For terms \(a(f_1 + f_2)(a)\) with \(a \in \supp(f_1 + f_2) \setminus \supp(f_1)\) (and thus \(a \in \supp(f_2)\)) and sufficiently fresh \(c\), we have
    \begin{align*}
      &a(f_1 + f_2)(a) \\
      =~&af_1(a) + af_2(a) && \text{axioms, assumption on $q$} \\
      =~&a(\nu c.\res(f_1(a))) + af_2(a) && \text{axioms, }  c\fresh f_1(a) \\
      =~&a(\nu c.\res(((a\ c) \cdot f_1)(a))) + af_2(a) && \text{because } a,c\fresh f_1 \\
      =~&a(\nu c.\res((a\ c) \cdot f_1(c))) + af_2(a) && \text{definition} \\
      =~&a((a\ c) \cdot \nu a.\res(f_1(c))) + af_2(a) && \text{equivariance} \\
      \le~&\newletter{a}((a\ c) \cdot \nu a.\res(f_1(c))) + af_2(a) && \text{axioms} \\
      =~&\newletter{c}(\nu a.\res(f_1(c))) + af_2(a) && \text{derivation rules} \\
      \le~&\newletter{c}(f_1(c)) + af_2(a) && \text{axioms}.
    \end{align*}
  \item The cases for \(f_2\) are completely analogous.
\end{itemize}

To show that \(m'\) is unique, it then suffices to show that, for \(f \in \mbar(A, h_A)\) and sufficiently fresh \(c\),
\begin{equation*}\textstyle
  m\left( \left[ \sum_{a \in \supp(f)} a f(a) + \newletter{c} f(c) \right] \right) = f.
\end{equation*}
This is clear from the definition of \(\pre\) and \(\abs\), as well as the additional requirement for \(f\) that traces for all atoms include those from bound atoms.

\subsection*{Categorical Phrasing of the Equivalence Game}

The categorical formulation of the game is as follows.

\begin{defn}\label{def:game}
  Let $(\monad,\beta)$ be a graded semantics for a functor~$F$
  on~$\Nom$, and let $\gamma\colon C\to FC$ be a $F$-coalgebra. For an
  equivariant subset $Z\subseteq M_0C\times M_0C$, we define $c_Z,C_Z$
  as the coequalizer
  \begin{equation*}
    \begin{tikzcd}
      M_0Z\arrow[r,"l_0^*",shift left]\arrow[r,"r_0^*" below,shift right]
      & M_0C \arrow[r,"c_Z"] & C_Z
    \end{tikzcd}
  \end{equation*}
  in~$\Nom$, where $l,r$ are the left and right projection maps
  $Z\to M_0C$ (recall that~$(-)_0^*$ is part of the \emph{graded
    Kleisli star} of~$\monad$; cf.~\cref{def:graded-monad}). We then
  have an equivariant map
  \begin{equation}\label{eq:overline-Z}
    \overline Z =(M_0 C \xrightarrow{(\beta\cdot\gamma)_0^*}M_1C=\mbar M_0C\xrightarrow{\mbar c_Z}\mbar C_Z).
  \end{equation}

  The \emph{$n$-round $(\monad,\beta)$-behavioural equivalence game}
  $\game_n(\gamma)$ on~$(C,\gamma)$, played by Spoiler  and
  Duplicator, has \emph{configurations}
  $(s,t)\in M_0C\times M_0C$. The game proceeds in~$n$ rounds. A round
  starting in configuration $(s,t)$ is played as follows:
  \begin{enumerate}
  \item Duplicator picks an equivariant subset
    $Z\subseteq M_0C\times M_0C$ such that
    $\overline Z(s)=\overline Z(t)$ (for $\overline Z$ as per
    \eqref{eq:overline-Z}).
  \item Spoiler picks $(s',t')\in Z$.
  \end{enumerate}
  The next configuration is then $(s',t')$. Any player who is unable
  to move on their turn loses. After~$n$ rounds have been played,
  ending in configuration $(s,t)$, Duplicator wins if
  $M_0!(s)=M_0!(t)$; otherwise, Spoiler wins. This final
  check is referred to as \emph{calling the bluff}.
\end{defn}
This formulation is analogous to the one in the set-based
case~\cite{FordEA22}, from which it is obtained essentially by
replacing the base category~$\Set$ with~$\Nom$ and (correspondingly)
requiring~$Z$ to be an equivariant subset, i.e.\ a subobject.

When~$\monad$ is induced by a graded nominal theory~$T$, equivalence
of this formulation of the game to the algebraic formulation given in
the main body of the paper is immediate from the way~$\monad$ is
constructed from~$T$ as per \cref{sec:graded-algebra}. In more detail,
the correspondence works as follows. \lsnote{More details are
  currently still in the main body; transfer here.}

As indicated in the main body, we regard $Z\subseteq M_0C\times M_0C$
as a set of equalities, understanding $(s,t)\in Z$ as a depth-$0$
equation $C\vdash_0s=t$. Requiring~$Z$ to be equivariant in the
categorical game corresponds to allowing equations in~$Z$ to be used
in renamed form in algebraic derivations. The algebraic game indeed
effectively restricts to orbit-finite~$Z$; this is justified by the
fact that by \cref{rem:ax}, the system can be made finitary, so that
logical consequence is compact, i.e.~any equation derivable from a
given set~$Z$ of equalities follows already from a finite subset. The
coequalizer~$c_Z$ as per \cref{def:game} merges all pairs $(s,t)$ of
terms in $M_0C$ such that $Z\centails C\vdash_0 s=t$. As indicated in
the main body, we view the map $\beta\cdot\gamma\colon C\to M_1C$ as a
uniform-depth substitution~$\sigma$. The map
$(\beta\cdot\gamma)^*_0\colon M_0C\to M_1 C$ applies~$\sigma$ to
depth-$0$ terms over~$C$, and the map
$\mbar c_z\colon M_1C=\mbar M_0C\to\mbar C_Z$ identifies two depth-$1$
terms $u,v$ over~$C$ if $Z\centails C\vdash_1 u=v$. Thus,~$Z$ is a
legal move for Duplicator in configuration $(s,t)$ if
$Z\centails C\vdash s\sigma=t\sigma$, so the notion of admissible move
in the two games is the same. Finally, it is easy to see that the two
notions of calling the bluff agree.

\subsection*{Proof of  \cref{thm:games}}

We conduct the proof using the categorical formulation of the game. It
does not actually require the graded monad~$\monad$ to be presented by
a graded nominal theory; it does however need this property for the
monad~$M_0$. We introduce corresponding terminology:
\begin{defn}
  A monad on~$\Nom$ is \emph{regularly algebraic} if it is presented
  by a depth-$0$ nominal theory as per \cref{sec:graded-algebra}.
\end{defn}

\noindent The proof then needs the following fact:
\begin{lemma}\label{lem:kernel-monic}
  Let $M$ be a regularly algebraic monad on~$\Nom$. Then the following
  hold.
  \begin{enumerate}
  \item\label{item:coeq-surj} Coequalizers in the category of
    $M$-algebras are surjective.
  \item\label{item:kernel-monic} Let $f\colon A\to B$ be a morphism of
    $M$-algebras. Let $l,r\colon K\to A$ denote the kernel of~$f$, and
    let $c\colon A\to C$ be the coequalizer of~$l,r$. Then the unique
    factorizing morphism $m\colon C\to B$ such that $m\cdot c=f$ is
    monic.
  \end{enumerate}
\end{lemma}
\begin{proof}
  Claim~(\ref{item:coeq-surj}): Since operations in our nominal theories
  are total, the coequalizer of $M_0$-homomorphisms $l,r\colon A\to B$
  just quotients~$B$ by the equivariant congruence generated by the
  relation~$\{(l(x),r(x)\mid x\in A\}$.

  Claim~(\ref{item:kernel-monic}): Let
  $m(x)=m(y)$. By~(\ref{item:coeq-surj}), there are $x',y'\in B$ such
  that $c(x')=x$, $c(y')=y$. Then $f(x')=m(c(x'))=m(c(y'))=f(y')$,
  i.e.~$(x',y')$ is in the Kernel of~$f$. Thus,
  \mbox{$x=c(x')=c(y')=y$}.
\end{proof}

% \begin{proof}[Proof (Sketch)]
%   (\ref{item:coeq-surj}): Immediate from the totality of operations in
%   our nominal theories.~(\ref{item:kernel-monic}): Straightforward
%   using~(\ref{item:coeq-surj}).
% \end{proof}
\begin{proof}[Proof (\cref{thm:games})] The proof for the
  corresponding result over $\Set$~\cite[Theorem~6.6]{FordEA22} is in
  fact formulated in categorical generality. Specifically, the base
  category is only assumed to be concrete over~$\Set$ and to have
  certain limits and colimits (kernels and coequalizers), conditions
  which both hold for $\Nom$. Additionally, the proof uses a property
  that over $\Set$ follows from regularity of the category of
  $M_0$-algebras \cite[full version, Remark A.3]{FordEA22}, which
  may fail over~$\Nom$; however, we have established the requisite
  property for regularly algebraic monads on~$\Nom$ directly as
  \cref{lem:kernel-monic}.\ref{item:kernel-monic}.
\end{proof}

\subsection*{Details for \cref{exp:explicit-games}}

\noindent\emph{Global freshness:}
Terms~$t$ at depth~$1$ can be brought into the form
\begin{equation*}
  t = \sum_{a \in S} at_a + \newletter{c}t_{\scriptnew{c}},
\end{equation*}
where $c \in \Names$ is sufficiently fresh and $S \in \powf(\Names)$ can be extended to any finite superset $S' \supseteq S$ (by setting any new~$t_a$ to~$0$).
Given a configuration $(s,t)$ where w.l.o.g. $(\beta\cdot\gamma)^*_0(s)$ and $(\beta\cdot\gamma)^*_0(t)$ are viewed as terms as described above, a move~$Z$ is then admissible iff $Z \centails C \vdash_0 s_\rho = t_\rho$ is derivable for every $\rho \in S \cup \{\newletter{c}\}$, where the `only if' direction follows from the above description of $\mbar$.
More explicitly, let~$A$ be the $M_0$-algebra obtained by identifying terms in~$M_0C$ that are equal under~$Z$.
By the coequalizer property of~$\mbar$ (and the fact that equalities in~$Z$ cannot be used in substituted form), we then know that $Z \centails C \vdash_1 u = v$ iff~$u$ and~$v$ are equal in~$\mbar A$.

Duplicator may now choose to only play inequalities, using the fact that, over a join-semilattice, $u = v$ iff $u \le v$ and $v \le u$.
Treating elements of $M_0C$ as finite sets of states, we have $u \le v$ iff $x \le v$ for all $x \in u$ and, by the above discussion, Duplicator can thus limit themselves to playing only one inequality $x \le t'$ with $t' \subseteq t_\rho$ for every $x \in s_\rho$ (and symmetrically).
The game can then be rearranged as follows:
In configuration $(s,t)$, Spoiler begins by picking a $\rho \in S \cup \{\newletter{c}\}$ and $x \in s_\rho$ (or symmetrically, $y \in t_\rho$).
Duplicator then responds with a finite set $t' \subseteq t_\rho$ to play the inequality $x \le t'$ (i.e., configuration $(x + t',t')$).
This also means that Spoiler cannot switch sides after the initial move.

Finally, the given description of $(\beta \cdot \gamma)^*_0$ stated in the main body follows directly by computation using the description of $\mbar$.

\noindent\emph{Local freshness:}
Terms~$t$ at depth~$1$ can also be brought into the above form, additionally requiring that $C \vdash (a\ c) \cdot \nu a.\res(t_{\scriptnew{c}}) \le t_a$ for all $a \in S$.
Once again,~$S$ can be extended to finite supersets $S' \supseteq S$, now setting any new $t_a = (a\ c) \cdot \nu a.\res(t_{\scriptnew{c}})$.
Admissible moves can be described as above, using the additional requirement of the normal form to show the `only if' direction for $a \in S$.
As detailed in \cref{rem:namerestrictionmodel}, the elements of $M_0C$ can be thought of as sets of name restrictions closed under further restriction.
We can thus rearrange the game so Spoiler can move first, now picking a possibly restricted $x|_N \in s_\rho$ (or symmetrically), to which Duplicator responds with a finite set $t' \subseteq t_\rho$ of possibly restricted names closed under further restrictions, playing the inequality $x|_N \le t'$.

We will also show the explicit description of $(\beta \cdot \gamma)^\star_0$ by computation using the term representation outlined in \cref{sec:m1-bar-details,sec:tdrop-details}.
Note that, with this description, the case where Spoiler picks $\rho = \newletter{c}$ is subsumed by simply picking a sufficiently fresh $c$ by \cref{lem:abs-eq}.

We will once again use the notation
\begin{equation*}
  A \setminus x = \nu a_1.\res(\dots(\nu a_n.\res(x)))
\end{equation*}
for a finite set $A = \{a_1, \dots a_n\} \in \powf(\Names)$.
By the description given in \cref{sec:tdrop-details}, a term $(\supp(x) \setminus N) \setminus x$ then corresponds to a name restriction $x|_{\supp(x) \cap N}$.

We will first prove the following helper lemma:
\begin{lemma}
  \label{lem:explicit-kleisli-star-helper}
  Let $(C,\gamma)$ be a $H_\omega$-coalgebra, $x, y \in C$, and $N \subseteq \supp(x)$.
  \begin{enumerate}
    \item\label{itm:explicit-kleisli-star-free-cond}
      If $x \xrightarrow{a} y$, then $a \notin \supp(x) \setminus N$ iff $a \in N$.
    \item\label{itm:explicit-kleisli-star-free-eq}
      If $x \xrightarrow{a} y$, then $C \vdash_0 (\supp(x) \setminus N) \setminus y = (\supp(y) \setminus N) \setminus y$ is derivable.
    \item\label{itm:explicit-kleisli-star-bound-eq}
      If $x \xrightarrow{\scriptnew{c}} y$, then $C \vdash_0 (\supp(x) \setminus (N \cup \{c\})) \setminus y = (\supp(y) \setminus (N \cup \{c\})) \setminus y$ is derivable.
  \end{enumerate}
\end{lemma}
\begin{proof}
  We will prove the statements individually:
  \begin{enumerate}
    \item
      Since $x\xrightarrow{a}y$, we have $\{a\} \subseteq \{a\} \cup \supp(y) \subseteq \supp(x)$.
      Thus, $a \in \supp(x)$ holds and the statement follows immediately.
    \item
      Since \eqref{ax:namerestriction} (1) allows us to leave out name restrictions for names that are not in the support, it suffices to show $\supp(y) \cap (\supp(x) \setminus N) = \supp(y) \setminus N$.
      This follows from the fact that $\supp(y) \subseteq \{a\} \cup \supp(y) \subseteq \supp(x)$ since $x\xrightarrow{a}y$.
    \item
      Once again, it suffices to show $\supp(y) \cap (\supp(x) \setminus (N \cup \{c\})) = \supp(y) \setminus (N \cup \{c\})$.
      This follows from the fact that $\supp(y) \subseteq \supp(x) \cup \{c\}$ because $x \xrightarrow{\scriptnew{c}} y$.
  \end{enumerate}
\end{proof}

Let $s \in M_0C$ and fix some $c \fresh s$.
Choosing one representative for every element $x|_N \in s$, we obtain
\begin{align*}
  (\beta \cdot \gamma)^\star_0(s)
  &= (\beta \cdot \gamma)^\star_0\left( \left[ \sum_{x|_N \in s} (\supp(x) \setminus N) \setminus x \right]_0 \right) \\
  &= \left[ \sum_{x|_N \in s} (\supp(x) \setminus N) \setminus \beta(\gamma(x)) \right]_1 \\
  &= \left[ \sum_{x|_N \in s} (\supp(x) \setminus N) \setminus \left( \sum_{x \xrightarrow{a} y} ay + \sum_{x \xrightarrow{\scriptnew{c}} y} \newletter{c}y \right) \right]_1 \\
  &= \left[ \sum_{x|_N \in s, x \xrightarrow{a} y} (\supp(x) \setminus N) \setminus (ay) + \sum_{x|_N \in s, x \xrightarrow{\scriptnew{c}} y} (\supp(x) \setminus N) \setminus (\newletter{c}y) \right]_1.
\end{align*}

For the first sum, we have the derivable equality
\begin{align*}
  &\sum\limits_{\substack{x|_N \in s, \\ x \xrightarrow{a} y}} (\supp(x) \setminus N) \setminus (ay) \\
  =~&\sum\limits_{\substack{x|_N \in s, \\ x \xrightarrow{a} y, \\ a \in \supp(x) \setminus N}} (\supp(x) \setminus N) \setminus (ay) + \sum\limits_{\substack{x|_N \in s, \\ x \xrightarrow{a} y, \\ a \notin \supp(x) \setminus N}} (\supp(x) \setminus N) \setminus (ay) && \text{(case distinction)} \\
  =~&\sum\limits_{\substack{x|_N \in s, \\ x \xrightarrow{a} y, \\ a \notin \supp(x) \setminus N}} a((\supp(x) \setminus N) \setminus y) && \text{(axioms)} \\
  =~&\sum\limits_{\substack{x|_N \in s, \\ x \xrightarrow{a} y, \\ a \in N}} a((\supp(x) \setminus N) \setminus y) && \text{(\cref{lem:explicit-kleisli-star-helper} \eqref{itm:explicit-kleisli-star-free-cond})} \\
  =~&\sum\limits_{\substack{x|_N \in s, \\ x \xrightarrow{a} y, \\ a \in N}} a((\supp(y) \setminus N) \setminus y) && \text{(\cref{lem:explicit-kleisli-star-helper} \eqref{itm:explicit-kleisli-star-free-eq})} \\
\end{align*}

For the second sum, we have
\begin{align*}
  &\sum\limits_{\substack{x|_N \in s, \\ x \xrightarrow{\scriptnew{c}} y}} (\supp(x) \setminus N) \setminus (\newletter{c}y) \\
  =~&\sum\limits_{\substack{x|_N \in s, \\ x \xrightarrow{\scriptnew{c}} y}} \newletter{c}((\supp(x) \setminus (N \cup \{c\})) \setminus y) && \text{(axioms)} \\
  =~&\sum\limits_{\substack{x|_N \in s, \\ x \xrightarrow{\scriptnew{c}} y}} \newletter{c}((\supp(y) \setminus (N \cup \{c\})) \setminus y) && \text{(\cref{lem:explicit-kleisli-star-helper} \eqref{itm:explicit-kleisli-star-bound-eq})}.
\end{align*}

Using the model from \cref{sec:m1-bar-details}, we then obtain the claimed description.

\fi % \iffullversion

\end{document}